\documentclass[useAMS,fleqn,usenatbib]{mnras}
\usepackage{longtable}
\usepackage[normalem]{ulem}
\usepackage{graphicx}

\title[Physical parameters of close binary systems: VIII]
{Physical parameters of close binary systems: VIII}

\author[K.~Gazeas et al.]{K.~Gazeas$^{1}$
\thanks{e-mail:kgaze@physics.auth.gr, kgaze@phys.uoa.gr},
  S.~Zola$^{{2},{3}}$, A.~Liakos$^{4}$, B.~Zakrzewski$^{3}$, S.M.~Rucinski$^{5}$,
\newauthor  J.M. Kreiner$^{3}$, W. Ogloza$^{3}$, M. Drozdz$^{3}$, D. Koziel-Wierzbowska$^{2}$,
\newauthor  G.~Stachowski$^{3}$, M. Siwak$^{3}$, A. Baran$^{3}$, D. Kjurkchieva$^{6}$, D. Marchev$^{6}$,
\newauthor  A. Erdem$^{7}$ \& S. Szalankiewicz$^{3}$ \\
$^{1}$ Section of Astrophysics, Astronomy and Mechanics, Department of Physics, National and Kapodistrian University of Athens, \\
GR-15784 Zografos, Athens, Greece \\
$^{2}$ Astronomical Observatory, Jagiellonian University, ul. Orla 171, 30-244 Krakow, Poland\\
$^{3}$ Mt. Suhora Observatory, Pedagogical University, ul. Podchorazych 2, 30-084 Krakow, Poland\\
$^{4}$ Institute for Astronomy \& Astrophysics, Space Applications \& Remote Sensing, National Observatory of Athens, \\
Penteli, Athens, Greece\\
$^{5}$ Department of Astronomy and Astrophysics, University of Toronto, 50 St. George St., Toronto, Ontario, M5S 3H4, Canada \\
$^{6}$ Department of Physics, Shumen University, 9700 Shumen, Bulgaria \\
$^{7}$ Astrophysics Research Centre and Observatory, \c{C}anakkale Onsekiz Mart University, Terzio\u{g}lu Kamp\"{u}s\"{u}, \\
TR-17020 \c{C}anakkale, Turkey
}

\begin{document}

\date{Accepted ... Received ... in original form ...}

\pagerange{\pageref{firstpage}--\pageref{lastpage}} \pubyear{2020}

\maketitle

\label{firstpage}

\begin{abstract}
This paper presents the results of a combined spectroscopic and photometric study of 20 contact binary systems: HV~Aqr, OO~Aql,
FI~Boo, TX~Cnc, OT~Cnc, EE~Cet, RW~Com, KR~Com, V401~Cyg, V345~Gem, AK~Her, V502~Oph, V566~Oph, V2612~Oph, V1363~Ori, V351~Peg,
V357~Peg, Y~Sex, V1123~Tau and W~UMa, which was conducted in the frame of the {\it W~UMa Project}. Together with 51 already covered by the project and an additional 67 in the existing literature, these systems bring the total number of contact binaries with known combined spectroscopic and photometric solutions to 138. It was found that mass, radius and luminosity of the components follow certain relations along the MS and new empirical power relations are extracted. We found that 30~per~cent of the systems in the current sample show extreme values in their parameters, expressed in their mass ratio or fill-out factor. This study shows that, among the contact binary systems studied, some have an extremely low mass ratio ($q<0.1$) or an ultra-short orbital period ($P_{orb} <0.25$ d), which are expected to show evidence of mass transfer progress. The evolutionary status of these components is discussed with the aid of correlation diagrams and their physical and orbital parameters compared to those in the entire sample of known contact binaries. The existence of very short orbital periods confirms the very slow nature of the merging process, which seems to explain why their components still exist as MS stars in contact configurations even after several Gyr of evolution.
\end{abstract}



\section{Introduction}
This paper is the continuation of a series
(\citealt{kre2003},      
\citealt{bar2004},       
\citealt{zol2004},       
\citealt{gaz2005b},      
\citealt{zol2005},       
\citealt{gaz2006a},      
\citealt{zol2010},       
papers~I--VII, respectively), which presents studies of several dozen contact
and very close binaries carried out as part of the {\it W~UMa Project}. This was
aimed at the determination of accurate physical parameters of contact systems
in the vicinity of the Sun based on a homogeneous sample of more than a hundred
targets with combined spectroscopic and multicolour photometric data, analysed
with consistent methodology and under the same assumptions.


\begin{table*}

\begin{center}
\caption{Observation Log for 20 targets. For each system the corresponding
comparison star is named, as well as the detector and photometric filters used,
the observation period, number of observing nights and the observing site.}
\label{TabObsLog}
\begin{tabular}{llccccc}
\hline
System          &  Comp. Star       & Instrument    & Filters   & Dates    &   No of nights & site\\
\hline
HV Aqr          & GSC 5198-1221     & CCD       &$BVRI$   &Jul 23 -- 27, 2005      & 5    & UOAO \\
OO Aql          & GSC 1058-0409     & CCD       &$BVRI$   &Aug 31 -- Dec 4, 2006   & 27   & UOAO \\
FI Boo          & TYC 3488 0985     & PMT       &$BVR$    &May 13 -- Jul 16, 2005  & 9    & SUH  \\
TX Cnc          & GSC 1395-1070     & CCD       &$BVRI$   &Apr 1 -- 6, 2010        & 4    & UOAO \\
OT Cnc          & GSC 1387-942      & CCD       &$BVRI$   &Feb 26 -- 28, 2008      & 2    & SUH  \\
EE Cet          & GSC 0640-0134     & CCD       &$BVRI$   &Nov 23 -- Jan 10, 2007  & 4    & SUH  \\
RW Com          & GSC 1991-1695     & CCD       &$BVRI$   &Mar 22 -- Apr 11, 2011  & 3    & UOAO \\
KR Com          & HD 115981         & PMT       &$UBVR$   &Mar 26 -- Jun 8, 2003   & 4    & SUH  \\
V401 Cyg        & GSC 2654-1313     & CCD       &$BVRI$   &Jun 10 -- 13, 2005      & 4    & UOAO \\
V345 Gem        & GSC 2457-1613     & CCD       &$BVRI$   &Nov 26 -- 28, 2006      & 3    & UOAO \\
AK Her          & GSC 1536-1266     & CCD       &$BVRI$   &May 26 -- 27, 2010      & 2    & UOAO \\ 
V502 Oph        & GSC 0383-0999     & CCD       &$VRI$    &Apr 14 -- 17, 2005      & 4    & SAAO \\
V566 Oph        & GSC 0245-0022     & CCD       &$BVRI$   &May 28 -- 31, 2009      & 4    & UOAO \\ 
V2612 Oph       & GSC 0445-1293     & CCD       &$BVRI$   &Jun 14 -- 16, 2011      & 2    & UOAO \\
V1363 Ori       & GSC 4750-1148     & CCD       &$BVRI$   &Feb 7 -- Mar 26, 2003   & 6    & UOAO \\
V351 Peg        & GSC 1713-1171     & CCD       &$BVRI$   &Jul 31 -- Aug 18, 2005  & 8    & UOAO \\
V357 Peg        & TYC 2254 2520     & PMT       &$UBVR$   &Oct 17 -- 18, 2003      & 2    & SUH  \\
Y Sex           & HD 86901          & PMT       &$BVRI$   &Mar 25 -- 31, 2003      & 7    & SAAO \\
V1123 Tau       & TYC 1238 1126     & PMT       &$BVR$    &Nov 22 -- Dec 2, 2003   & 2    & SUH  \\
W UMa           & HD 83564          & PMT       &$BVRI$   &Feb 1 -- Feb 16, 2001   & 3    & SUH  \\
\hline
\end{tabular}
\end{center}
UOAO: University of Athens Observatory, SUH: Mt. Suhora Observatory,
SAAO: South African Astronomical Observatory
\end{table*}

The rationale for the project and the method of deriving physical parameters
of the components which were as accurate as possible, while avoiding
non-unique solutions, was described in detail in {Paper~I}. Subsequent changes,
modifications and improvements in the procedure were presented in {Paper~II}.
Each system was observed photometrically and spectroscopically, using the latest
and most accurate available techniques.  The spectroscopic part included
high-resolution spectroscopic measurements from David Dunlap Observatory (DDO)
in Toronto, Canada (Radial Velocity Programme: \citet{ruc2005} and \citet{ruc2013})
and was described in detail in \cite{pri2009b}, while the photometric programme
was described by \citep[][and references therein]{kre2003}.

In this paper, we present the results of our analysis of recent and unpublished
multicolour photometric observations of 20 previously-known contact binary systems,
some of which have been analysed individually in the past by other authors.
However, less than 40 per~cent of them have results based on very accurate multicolour
photometric observations combined with a spectroscopically-determined mass ratio.
Application of non-uniform methodology, different spectroscopic accuracy
and other limitations create an inhomogeneous sample unsuitable for reliable
statistical studies on the properties of the components of these systems.
The 20 systems analysed in this work will extend the 51 systems studied
previously within this Project (see papers I--VII) to a total of 71,
which is a significant increase in the overall sample.

The accurate determination of the physical parameters of binary systems can
be improved significantly, leading to a uniform sample suitable for statistical
analyses, when data is collected with the same instrument and the reduction
procedure is similar and well defined. The sample of contact binaries observed
as part of the {\it W~UMa Project} and the {\it Radial Velocity Project} \citep[][and
references therein]{ruc2005} offers such an opportunity, where the accuracy in the
solutions exceeds that of previous studies.

A study of a uniform sample of 112 systems was previously presented
in \citet{gaz2006b}, \citet{gaz2008} and \citet{ste2012}. It includes
solutions for the 51 systems available at that time utilizing spectroscopy
from the DDO {Radial Velocity Project}, together with an additional 61 systems
collected from literature which also have combined photometric and spectroscopic
solutions. The 20 systems analysed in this work, together with 6 further systems
which have since appeared in the literature expand this to a total of 138 systems.

The manuscript is organised as follows: In Section 2 we summarize the results
from previous investigations for each individual system, giving a brief historical
overview of the past studies.
In Section 3 we present the new photometric data acquisition and reduction, while in
Section 4 describe the procedure used for obtaining the models and interpreting the resulted uncertainties.
Section 5 presents the results on each individual target, in comparison to the studies obtained in the past.
Section 6 follows with a discussion on the evolutionary status of the sample under study and
finally Section 7 presents an overview of the physical properties of contact binaries resulting from the overall study in the {\it W~UMa Project} to date.


\section{Notes on individual systems}
In this section we summarize the results from previous investigations for each individual system, giving a brief historical overview of past studies.
\subsection{HV~Aqr} HV~Aqr (TYC 5198-659-1) is a magnetically active eclipsing binary system, discovered by \citet{hut1992} during a photometry survey aimed at detecting minor planets. The first dedicated photoelectric measurements were obtained by \citet{sch1992}. \citet{rob1992}, based on his CCD observations and assuming the effective temperature of both components to be 6000~K, presented a contact configuration model for HV~Aqr, with $q$ = 0.15--0.16, $i$ = 78$^\circ$ and $f$ between  48 and 49~per~cent. The most recent spectroscopic observations by \citet{ruc2000} provided a mass ratio of $q=0.145$ and a spectral type of F5V. Based on the spectroscopic data obtained at DDO, \citet{DAngelo2006} found the signature of a tertiary component and identified HV~Aqr as a triple system. This third component, of K2--3V spectral type, was confirmed by high resolution imaging with an Adaptive Optics (AO) technique \citep{ruc2007b}. The companion orbits the close pair at a distance of 74~AU with a period of $\sim$430 yr. Its mass was estimated to be 0.7 M$_{\sun}$. The mass ratio derived by \citet{ruc2000} was applied by \citet{liq2013} in their model of the system.
\subsection{OO~Aql} OO~Aql (HD 187183) was discovered by \citet{hof1932} who estimated its spectral type as G5. \citet{Binnendijk1968} was the first to present a complete light curve. There have been numerous subsequent photometric observations of this system, such as those by \citet{laf1985}, \citet{dem1981}, \citet{ess1992}, \citet{gur1994}, and \citet{dju1998}. Spectroscopic classification obtained by \citet{pri2007} gives a F9V spectral type when G-band (430 nm) is used, while the weak hydrogen lines indicate a later G8V type. Based on the broadening function (BF) method \citep{ruc2002a}, the mass ratio of this system was determined to high accuracy with a value of $q = 0.846 \pm 0.007$. \citet{ruc2007b} found that OO~Aql has a cool tertiary companion of M5V (or later) spectral type, orbiting the binary with a period of $\sim$420 yr. Recently, \citet{icl2013} examined the O$-$C diagram of this system in order to determine the parameters of the orbiting component. They presented a model  resulting in a quadruple configuration. \citet{moc1981} considered OO~Aql to be the prototype of a group of newly-formed contact binaries.  \citet{hri1989}  studied the system  through a combined radial velocity and light-curve analysis. His model indicated that the orbital inclination $i$ of the system is close to 90$^\circ$ and, accounting for proximity effects, the spectroscopic mass ratio turned out to be $q = 0.843 \pm 0.008$. This value is very close to one obtained by \citet{pri2007}, who utilized the BF method. The physical parameters have since been re-determined by \citet{deb2011}, who assumed the spectroscopic mass ratio  derived by \citet{pri2007}.
\subsection{FI~Boo} FI~Boo (HIP 75203) was discovered by the {\it HIPPARCOS} mission \citep{Perryman1997}. It was initially listed as a pulsating variable with a linear ephemeris corresponding to maximum light and half of the period of variation. \citet{due1997} was the first to include it in the list of possible contact binaries, followed by other observers who determined times of minima, resulting in an updated linear ephemeris. The spectroscopic orbit for FI~Boo was determined by \citet{lu2001}, resulting in a spectroscopic mass ratio of $q = 0.372$ with an estimated spectral type of G3V.  No third body has been definitively detected spectroscopically, however \citet{pri2006} suggested a ``stochastic'' detection of a third companion, possibly orbiting the contact binary in less than 3 yr. Nearly at the same time, \citet{DAngelo2006} announced the possible detection of a tertiary contributing only 1.2~per~cent to the total system luminosity. They estimated its effective temperature to be approximately $T_{3} = 3900$~K and its mass as $M_{3} = 0.52 $M$_{\sun}$.  \citet{ter2006} observed the system in the $UBVR_{c}I_{c}$ bands. They updated the linear ephemeris, classified the system as an A-type contact binary and performed a combined photometric and spectroscopic analysis. They arrived at a model with very shallow partial eclipses ($i$=43.1$^\circ$) of components with masses of $0.82 $~M$_{\sun}$ and $0.31 $~M$_{\sun}$.
An even shallower orbital inclination ($i$=38.05$^\circ\pm2.69$) was presented by \cite{chr2013}. Contrary to the results obtained by \cite{ter2006} they suggested that this was a W-type system. This result was obtained by comparing their light curve and the radial velocity curve given by \citet{lu2001}. The effective temperature $T_{1}=5939\pm35$ K was calculated using equation (3), presented in \cite{zwit2003}, and resulted in the following absolute parameters: $M_{h} = 0.40 \pm 0.05 M_{\sun}$, $M_{c} = 1.07 \pm 0.05 M_{\sun}$, and $R_{h} = 0.85 \pm 0.07 R_{\sun}$, $R_{c} = 1.28 \pm 0.07 R_{\sun}$, where 'h' and 'c' denote the 'hot' and 'cool' stars, respectively, in their work. Study of the O-C diagram confirms the presence of a third body with a mass estimated to be approximately 0.45 $M_{\sun}$.
\subsection{TX~Cnc} TX~Cnc (TYC 1395-907-1) is a system of particular interest as it belongs to the Praesepe open cluster, one of the youngest (900~Myr) open clusters containing such binaries \citep{ruc1998}. The Praesepe cluster membership of TX~Cnc was confirmed by numerous astrometric and parallactic observations (\cite{ros1988}, \cite{van1999}, \cite{jon1991}, \cite{hog2000}) and it is also in a perfect agreement with the absolute magnitude estimation of \citet{ruc1997} ($M_{V}$ = 3.60~mag, for $(B-V)_{0}$ = 0.54~mag, corresponding to a spectral type of F8V). The photometric variability of TX~Cnc was discovered by \citet{haf1937}, while the first complete photoelectric light curve was obtained by \citet{yam1972}. \cite{pri2006b} presented the most recent spectroscopic study, where they determined the system mass ratio $q$ = 0.455 $\pm$ 0.011. The authors found no third light contamination and confirmed the spectral type to be indeed F8V. Based on the $V$-band photometric data and the spectroscopic mass ratio from \citet{pri2006b}, \citet{liu2007} obtained the absolute parameters of the components of TX~Cnc are: $M_{c}$ = 1.32~M$_{\sun}$, $M_{h}$ = 0.60~M$_{\sun}$, $R_{c}$ = 1.28~R$_{\sun}$ and $R_{h}$ = 0.91~R$_{\sun}$, utilizing the same definition for the 'hot' and 'cool' components, as in \cite{zwit2003}. Their O$-$C analysis suggests that a third body orbits the contact binary with a period of 26.6 yr, however with a quite large uncertainty. \citet{zha2009} obtained accurate photometric observations in $BVI$, analysed the O$-$C diagram, improved its linear ephemeris and pointed out a possible quadratic term indicating a period increase. Re-analysing all available data, they obtained a comparable value for the system  mass ratio of $q$ = 0.450. Contrary to \citet{liu2007}, they found no evidence for a tertiary companion either from the O$-$C analysis or from the light curve modelling.  
\subsection{OT~Cnc} OT~Cnc (GSC~1387-0475, ASAS~083128+1953.1, TYC~1387-475-1) was discovered by the All Sky Automated Survey ($ASAS$) \citep{poj1997, poj2004, poj2005, pac2006}. OT~Cnc has one of the shortest orbital periods ($P_{orb}$ = 0.2178~d) among contact binaries. Spectroscopic observations by \citet{ruc2008b} showed a K3V spectral type, which we adopt in this work. The system was first modelled by \citet{ruc2008b} using the V and I-band $ASAS$ photometric data and the  spectroscopically determined mass ratio of $q$ = 0.474 $\pm$ 0.008. \citet{yan2010} presented a model with higher quality photometric observations in $BVR$-bands. Their model resulted in a somewhat higher value for the orbital inclination ($i = 49.9^\circ$) compared to that derived by \citet{ruc2008b} ($i = 42^\circ$) and a third light contribution of about 13--14~per cent.
\subsection{EE~Cet} EE~Cet (ADS 2163 B) is the southern (slightly fainter) component of the visual binary WDS~02499+0856 \citep{Mason2001}. It was discovered by the {\it HIPPARCOS} mission \citep{Perryman1997}, by noticing the variability of the combined light of both visual components. \citet{Lam2001} performed photometric measurements of the visual pair and gave (but only for one epoch) the following values $V(A)$ = 9.47~mag and $V(B)$ = 9.83~mag. \citet{pri2006} lists the orbital parameters (orientation and separation) of  WDS~02499+0856 and gave $\theta$ = $194^\circ$, $\rho$ = 5.66~arcsec and magnitude difference $\Delta$$V$ = 0.07~mag (the magnitude difference can be as large as $\Delta$$V$ = 0.36~mag, due to photometric variability of the eclipsing binary). WDS~02499+0856 turned out to be a quadruple system, when the northern component  was found to be a double-lined (SB2) binary from the DDO spectroscopic observations \citep{pri2006}. \citet{DAngelo2006} re-confirmed the multiplicity of the system, and listed it among the contact binaries with additional components. Radial velocity observations from \citet{ruc2002b} resulted in a well defined circular orbit of the contact binary, with $K_{1}$ = 84.05 km s$^{-1}$, $K_{2}$ = 266.92 km s$^{-1}$ ($q = 0.315$) and a F8V spectral type. \citet{kar2007} using their own velocity curve analysis method, arrived at almost identical results for the mass ratio. \citet{dju2006} presented the first model, resulting in orbital inclination of $i=78.5^\circ$ and a fill-out factor of $f = 32.69$~per~cent, $T_{2}$ = 6314~K and $T_{1}$ = 6095~K, when spots were added. Their no-spot  model resulted in very close value for the fill-out factor  but slightly different geometrical and orbital parameters. The physical parameters derived in this study were: $M_{1}$ = 1.37~M$_{\sun}$, $M_{2}$ =0.43~M$_{\sun}$ and mean radii $R_{1}$ =1.35~R$_{\sun}$, $R_{2}$ = 0.82~R$_{\sun}$. It is worth noting here that the light curves analysed by these authors included the visual component in the photometric aperture with a contamination of about 54~per~cent.
\subsection{RW~Com} RW Com (HIP 61243) was discovered almost a century ago \citep{jor1923}. Evidence of magnetic activity in this system was found spectroscopically by \cite{str1950}, who detected Ca II emission lines, and classified the spectral type of the system as G2. Since then, photometric observations have confirmed that the system is indeed magnetically active, exhibiting asymmetries in its light curve \citep{oco1951, mil1987}. The first radial velocity (RV) observations by \citet{mil1985} gave: $(M_1+M_2) \sin^3 i = 0.69 \pm 0.03$ $M_{\sun}$, $M_1/M_2$ = 2.9  $\pm$ 0.2. 
\citet{sri1987} studied the O$-$C diagram of RW Com and found a cyclic variation with a period of about 16 yr and suggested that a third body orbits the system. RV observations obtained by \citet{pri2009a} gave contradictory results. They estimated the spectral type to be K2V and obtained a system mass ratio  $q = 0.471 \pm 0.006$, which leads to a  much larger total mass: $(M_1+M_2) \sin^3 i = 1.052 \pm 0.013$ $M_{\sun}$. Their spectra did not show any trace of a third star, counter to an earlier suggestion based on the cyclic variation of the orbital period. \citet{dju2011} performed a combined photometric and spectroscopic analysis, using the mass ratio of \citet{pri2009a}. They found the two components to have a shallow contact (approximately 6~per~cent) and having larger masses. Asymmetries seen in the light curve were explained as being due to cool and hot spots on each component's surface.
A recent study by \cite{ozav2020} presented the orbital period variation using data from over 100 years. Their results indicate the presence of two low mass bodies, physically bound with the contact binary. The orbital period and mass of the third and fourth bodies are, respectively,
$P_{3}=101.03\pm4.78$ yr, $P_{4}=9.72\pm0.11$ yr, $M_{3}=0.298\pm0.006$ $M_{\sun}$, and $M_{4}=0.123\pm0.006$ $M_{\sun}$.
\subsection{KR~Com} KR~Com (HIP 65069) was discovered by the {\it HIPPARCOS} satellite \citep{Perryman1997}. It is a member of a very close visual binary with a separation of only 0.12 arcsec and 0.60 mag difference between the components, with the contact binary being brighter. The rather large amplitude of the RV curves and the observed low-amplitude photometric variability is most likely due to the combination of a low-inclination orbit and light contamination by the visual companion. The third light contribution was estimated by \cite{ruc2002b} to be very large ($l_{3} = 0.56 \pm 0.04$) and the spectral type of the close pair as G0IV. The Tycho-2 colour index of $B-V$ = 0.52~mag, corresponding to an earlier F8V spectral type, is also likely due to the light contamination from the visual companion. KR~Com has one of the smallest mass ratios known among contact binaries ($q = 0.091 \pm 0.002$). The minimum total mass was found  by \cite{ruc2002b} to be $0.517 \pm 0.008$ M$_{\sun}$. The photometric light curve of this system was initially analysed by \citet{zas2010}, who arrived at a model with a contact configuration, a temperature difference between the components of $\Delta$$T$ = 523 $\pm$ 364~K and an orbital inclination of $i$ = 52.1$^\circ$. They also performed a study of the circumbinary component using the available astrometric data and concluded that the third object is orbiting the contact binary in a highly eccentric orbit ($e=0.934$) with a period of 10.98 yr. Using the above parameters, one can derive the masses of the binary system components as $M_{1}$ = 1.42 $M_{\sun}$, $M_{2}$ = 0.129 $M_{\sun}$, and a minimum mass of $M_{3}$ = 1.60 $M_{\sun}$ for the companion.
\subsection{V401~Cyg} V401~Cyg (HIP 95816) was discovered by \citet{hof1929}. It was initially classified as a pulsating variable of RR~Lyr type with a period of 0.2895~d. \citet{lur1947} was the first who recognized its real nature as an eclipsing binary and assigned the correct orbital period. The first photoelectric observations were obtained by \citet{spi1959} who confirmed the binary nature of V401~Cyg. Photoelectric photometry was later obtained by \citet{pur1964}, who noticed a remarkable increase of its orbital period. A similar study of the O$-$C diagram was performed by \citet{her1993}, who also pointed out that the orbital period was increasing. \citet{wolf2000} performed a light-curve analysis and calculated a photometric mass ratio of $q_{ph} = 0.3$. Detailed spectroscopic observations were performed by \citet{ruc2002b}, who found the spectroscopic mass ratio to be $q_{sp} = 0.290 \pm 0.011$. They also noticed a small contribution of a third light (about 3~per~cent). \citet{DAngelo2006} confirmed the existence of a third body from the spectroscopic data and estimated the effective temperature of the tertiary to be about 4700 K, while the mass ratio of $M_{3}/(M_{1}+M_{2}) = 0.29$, leading to a value of $M_{3}$=0.73 $M_{\sun}$.
\subsection{V345~Gem} V345 Gem (HIP 37197) was considered for a long time to be a single star of F0V spectral type \citep{can1919}, the fainter component of the visual binary WDS~07385+3343 \citep{Mason2001}.  Its variability was discovered by the {\it HIPPARCOS} mission \citep{Perryman1997} and it was initially classified as a pulsating variable with a period of 0.137389~d. The small amplitude of the light variations was correctly considered to be due to the contribution from the fainter visual companion. The accurate photometric study performed by \citet{gom2003} showed that the variable could be a contact binary with twice the period (0.2747736~d) and an amplitude of 0.07~mag. \citet{pri2006} calculated $\theta$ = 352$^\circ$, $\rho$ = 3.19~arcsec and $\Delta$$V$ = 1.40~mag. They also noted the very large proper motion errors in the {\it TYCHO} observations, confirming that the eclipsing system and the visual companion are gravitationally bound. \cite{pri2007} performed spectroscopic observations, obtaining a small mass ratio of $q = 0.142$ and assigned a later spectral type of F7V.
\citet{yan2009} performed a combined photometric and spectroscopic solution of the system and derived a contact configuration, with a high fill-out factor of $f=72.9$~per~cent, an orbital inclination of $i$=73.3$^\circ$ and a third light contribution of $l_{3}=0.20$. The $l_3$ is attributed to a third body orbiting the contact binary with a period of $P_{3}=646.7 \pm 0.7$ d.
\subsection{AK~Her} AK~Her (HIP 84293) is the brighter component of the visual binary ADS~10408 and was the subject of investigation by several observers \citep[][and references therein]{bin1961}. Spectroscopic observations obtained by \citet{pri2006b} determined a spectroscopic mass ratio of $q = 0.277$. \citet{pri2006b} estimated an earlier spectral type of F4 than the one (F8) given by \citet{san1934},  still not consistent with the Tycho-2 colour index of $B-V$ = 0.49~mag. They detected no third light contamination but in the notes they stated that the fainter ADS~10408 component could not be the cause of LITE changes and suggested the possibility that AK Her could be a quadruple system. \citet{cal2014} published an analysis of multicolour light curves, which resulted in a model with spots, an orbital inclination ($i$=81.7$^\circ$), and fill-out factor of $f=33.2$~per~cent. \citet{sam2010} obtained a set of UBV observations, and combining them with the radial velocity data from \citet{san1934} they were able to calculate the physical parameters of the components.

Detailed discussion of the orbital period changes was performed by \cite{ozav2020}. The first order changes of the orbital period arise from mass transfer between the component within the system. The periodic variation seen in the O-C diagram is due to the presence of a third component, whose mass upper limit is estimated to be $0.15\pm0.01$ $M_{\sun}$ and its orbital period is equal to $P_{3}=71.28$ yr. The residuals from the above two effects also show another cyclic variation on the O-C diagram. \cite{ozav2020} proposed either a low mass fourth body or magnetic activity in the system, as a possible cause. The existence of the fourth body in this system was also mentioned by \citet{pri2006b}, while changes of orbital period based on magnetic effects were described by \citet{sam2010}.

\subsection{V502~Oph} The eclipsing nature of V502~Oph (HIP 81703) was discovered by \citet{hof1935}, while the first ephemeris, based on visual observations, was given by \citet{lau1937}. The light curve of the system exhibits intrinsic variations, and also its orbital period has been varying over the years \citep{bin1969}. Spectroscopic observations and radial velocity curves of V502~Oph have been published by several investigators \citep{str1948, str1959, kin1984}. The spectra of the components were classified as G2V for the primary and F9V for the secondary by \citet{str1959}. \citet{pyc2004} argued for a spectral type of G0V, and determined the system mass ratio to be $q = 0.335\pm0.009$. V502~Oph is a contact binary for which there is evidence for the presence of a third companion. \citet{der1992} examined the O$-$C diagram and found that the arrival times of the minima showed a modulation with a period of about 35 yr. This was later confirmed spectroscopically by \citet{hen1998}, who found evidence of a tertiary component in the spectra. \citet{yuc2006} studied the secular variation of the orbital period and found a periodic variation of 57.88 yr in the O$-$C diagram. Based ond these facts, \citet{pri2006} listed V502~Oph in their list of contact binaries with additional components, however, \citet{DAngelo2006} found no evidence of a tertiary component. Several photometric observations and analyses for this system have been published by \cite{kwe1958, kwe1968a}, \citet{hin1960}, \citet{wil1967}, \citet{bin1969}, \citet{pol1975}, \citet{mac1982}, \citet{zol1988} and \citet{rov1988}. Observations with the VLA revealed that it is a binary radio source \citep{hug1984}. W~UMa-type systems usually show low radio activity \citep{ruc1995} and radio emission from this system may suggest that a companion could  still be undetected \citep{hug1984}. Detailed solutions of the system was constructed recently by \citet{zhou2016}. Using the spectroscopic mass ratio from Pych et al. (2004) and fixing $T_{1}=6140$ K after \citet{rov1988}, they obtained the following absolute parameters: $M_{1} = 0.46\pm0.02$ $M_{\sun}$, $M_{2} = 1.37\pm0.02$ $M_{\sun}$, $R_{1} = 0.94\pm0.01$ $R_{\sun}$, $R_{2} = 1.51\pm0.01$ $R_{\sun}$. A tertiary component is also detected around the contact binary system, whose mass is estimated to have a lower limit of 0.16 $M_{\sun}$.
\subsection{V566~Oph} V566~Oph (HIP 87860) is a bright ($V_{max} = 7.46$) system discovered by \citet{hof1935}, initially described as an Algol-type binary. \citet{fre1954} noticed shape variations and asymmetries in the light curves, and calculated the orbital period $P = 0.4096623(4)$~d and an orbital inclination of 72.3$^\circ$. The light curve shows a flat-bottom secondary minimum. V566~Oph was the subject of several photometric and spectroscopic studies \citep[][and references therein]{twi1979, pri2006b}. Spectroscopic studies were started in the mid 1960s by \citet{hea1965}, later \citet{mcl1983a} and \citet{hil1989} repeated the spectroscopic observations and determined the radial velocities using the Cross-Correlation Function (CCF) method. The results from these studies are consistent with the most recent one performed by \citet{pri2006b}, who found a mass ratio of $0.263 \pm 0.012$  and estimated the system spectral type to be F4V. No third light was detected. These results agree with the photometric model obtained by \citet{moc1972} and other studies based only on photometric data which estimated a mass ratio in the range: $0.23 < q_{ph} < 0.24$. \citet{eaton1986} used ultraviolet observations from the IUE satellite \citep{bog1978}, finding $i = 80.5^\circ$ and $f = 55$~per~cent. Based on modelling of $UBVR$ observations, \citet{deg2006} gave $i = 80.8^\circ$, while \citet{lia2010} obtained  $i = 80.7^\circ$ and $f = 36.7$~per~cent from analysis of $BVRI$ data. Both studies fixed the mass ratio at the value obtained by \citet{pri2006b}. The most recent study was presented by \citet{sel2018} using a fixed value for the effective temperature $T_{1} = 6456$  K, according to the spectral type F4/F5V \citet{abt2009}. As a result they found $i = 80.4\pm0.2$ and $f = 34.1\pm1.7$, similar to values found by \citet{lia2010}. They also resulted in the following absolute parameters: $M_{1} = 1.50\pm0.01$ $M_{\sun}$, $M_{2} = 0.38^\circ\pm0.01$ $M_{\sun}$, $R_{1} = 1.49\pm0.01$ $R_{\sun}$, $R_{2} = 0.81\pm0.01$ $R_{\sun}$. The periodic changes visible in the O-C diagram (after removing the parabolic trend) are explained by both the presence of a third body and the magnetic activity in the components. \citet{sel2018} also discussed the evolutionary status of V566~Oph and they suggested that the system is on the first stage of A-type evolution.
\subsection{V2612~Oph} V2612~Oph (HD 170451) was suspected to be a variable star by \citet{hil1958}. Several years later, $V$ band photometry by \citet{kop2002} showed it was an eclipsing binary. The light curve shows an asymmetric shape, possibly the result of magnetic activity. \citet{yan2005} analysed the light curve of \citet{kop2002} and determined a photometric mass ratio $q = 0.323$, orbital inclination $i = 65.7^\circ$ and  fill-out factor $f = 23.0$~per~cent. \citet{pri2007} performed spectroscopic observations and estimated the spectral type of the system as F7V and a lower mass ratio $q = 0.286$. A recent study by \citet{ozd2014} used seven years of photometric observations gathered between 2003 and 2009. They showed that the light curve of V2612~Oph undergoes seasonal and long term variations, which were attributed to solar-type magnetic activity. They also estimated the distance of this system to be 140~pc, confirming that the binary is not a member of the NGC~6633 open cluster, disputed earlier by \citet{hid2005} and \citet{kha2005}. Based on the $BVRI$ photometric observations and the spectroscopic mass ratio from \citet{pri2007}, \citet{cal2014} presented a new model including magnetic spots for this system, suggesting that strong magnetic activity occurs in both stars.
\subsection{V1363~Ori} V1363~Ori (HIP 23809) was discovered by the {\it HIPPARCOS} mission \citep{Perryman1997} and listed as a contact binary with an orbital period of 0.431915~d. Its spectral type, as listed in the Henry Draper Extension Charts, is F5 \citep{nes1995}. The {\it TYCHO} catalogue provides a colour index of $B-V$ = 0.56~mag, which corresponds to a later spectral type of F9. \cite{gom1999} performed the first ground-based observations and provided improved orbital elements and photometric parameters. The spectroscopic study by \citet{pyc2004} resulted in a mass ratio of $q = 0.205$ and the spectral type was estimated as early to mid F. The {\it HIPPARCOS} mission reported a poor parallax measurement, which might indicate a spectroscopically undetectable companion. Furthermore, \citet{pri2006} speculated that the large parallax errors might indicate multiplicity. However, their search for a tertiary component turned out to be negative. Subsequent searches for an additional component by \citet{DAngelo2006} using spectroscopy and by \citet{ruc2007b} through AO observations were also unsuccessful. \citet{gaz2005a}  presented the first combined photometric and spectroscopic model for this system and calculated the physical parameters of the system as: $M_{1}$ = 1.314 $\pm$ 0.599 M$_{\sun}$, $M_{2}$ = 0.269 $\pm$ 0.123 M$_{\sun}$, $R_{1}$ = 1.665 $\pm$ 0.087 R$_{\sun}$, $R_{2}$ = 0.886 $\pm$ 0.046 R$_{\sun}$, $L_{1}$ = 5.193 $\pm$ 0.026 L$_{\sun}$, $L_{2}$ = 1.505 $\pm$ 0.032 L$_{\sun}$. The absolute bolometric magnitudes of the components are: $M_{bol,1}$ = 2.960~mag, $M_{bol,2}$ = 4.305~mag.
\subsection{V351~Peg} The variability of the V351~Peg (HIP 115627) binary system was first noticed in the {\it HIPPARCOS} satellite data \citep{Perryman1997}. It is listed in the {\it HIPPARCOS} catalog as a pulsating variable with half the actual period (0.29665~d is given, instead of 0.59330~d). The system was apparently mistaken for a pulsating variable due to the equal depths of the minima. V351~Peg was subsequently observed photometrically by \citet{gom1999}, who re-classified it as a contact binary and improved the system ephemeris. Spectroscopic observations obtained by \citet{ruc2001} resulted in a mass ratio of $q = 0.360$ and a spectral type of A8V. \citet{ruc2007b}  detected no additional component orbiting the binary, neither was one found spectroscopically by \citet{DAngelo2006}. \citet{alb2005} obtained photometric data in the $BVR$ bands and published a model which included spots. A low orbital inclination was found ($i = 63^\circ$), the fill-out factor of $f = 20.64$~per~cent and very similar effective temperatures of the components.
\subsection{V357~Peg} V357~Peg (HIP 117185) is a contact binary system, discovered by the {\it HIPPARCOS}  mission. \citet{yas2000} performed $BVR$ photometry and noted that the secondary minima are deeper than the primary ones, but without obvious asymmetries in the maxima. The first radial velocity curve was measured by \citet{ruc2008}, who calculated the mass ratio of the system as $q = 0.401 \pm 0.004$ and estimated the spectral type to be F2V, which we adopt in this work. Recently, \citet{deb2011} presented their results for this system, using the V-band observations from the $ASAS-3$ project \citep{poj1997}. They found it to have an orbital inclination of $i = 73.2^\circ$, a contact configuration with  fill-out factor of $f = 10$~per cent and a temperature difference between the components $\Delta$$T = 562$~K. The resulting physical parameters were: $M_{1}=1.720 \pm 0.015 $M$_{\sun}$ and $M_{2} = 0.690 \pm 0.013 $M$_{\sun}$. \citet{ekm2012} used their $BVR$ band photometric data and the spectroscopic mass ratio from \citet{ruc2008} and arrived at a model with a similar inclination but a higher fill-out factor (31.2~per cent).  The temperatures of the components differ by 300~K. A somewhat different absolute parameters than those in \citet{deb2011} were determined: $M_{1} = 0.85 \pm 0.03 M_{\sun}$ and $M_{2} = 0.34 \pm 0.02 M_{\sun}$. \citet{pri2006} and \citet{ruc2007b} found no evidence for an additional component orbiting the binary, which is consistent with the absence of  variations in the O$-$C diagram.  
\subsection{Y~Sex} Y~Sex (HIP 49217) is the brighter member of the visual binary system WDS~10028+0106 ($\theta = 154^\circ$, $\rho = 0.50$ arcsec, $V_{1}$ = 10.08~mag and $V_{2}$ = 12.70~mag). The variability of Y~Sex was first noticed by \citet{hof1934}. \citet{hil1979} analysed its light curve and confirmed a marginal contact configuration, adding that it undergoes total eclipses. He determined a photometric mass ratio of $q = 0.175$, and found that the  orbital inclination is $i = 76.8^\circ$. Y~Sex was first observed spectroscopically by \citet{mcl1983}, who found a mass ratio of $q = 0.18\pm0.03$, confirming the photometric estimation. Despite contamination of the spectra by the very close visual companion, \citet{pri2009b}  determined an improved mass ratio of $q = 0.195 \pm 0.008$, and estimated the light contribution of the visual companion at the level of $ l_3/(l_1 +l_2) = 0.11 \pm 0.03$. Classification based on David Dunlap Observatory (DDO) spectra \citep{pri2009b} gave a spectral type of F5/6. Y~Sex has been extensively observed for its orbital period variations \citep{her1993, qia2000, wolf2000, heq2007}. \citet{wolf2000} found that the cyclic variation of the orbital period can be interpreted as the presence of an invisible third body in a $57.6$ yr eccentric orbit ($e = 0.52$), estimating its spectral type as M4/5 and mass as $M_{3} = 0.3$M$_{\sun}$, while \citet{yan2003} derived only a small value of a third light $l_{3} = 0.0064 \pm 0.0008$ in their modelling. The orbital period of this system was constant over the years 1953-79 \citep{her1993} but observations from 1982 onward confirm a secular decrease of the period at a rate of $-5.5 \times 10^{-8}$d~yr$^{-1}$ \citep{qia2000}, while the overall photometric and time travel phenomena were presented by \citet{dry2006}. Attributing such a variation to the visual companion seems to be very likely and therefore,  \citet{pri2006} included Y~Sex in the catalog of contact binaries with additional components. \citet{deb2011} gave a combined model for the system, based on the spectroscopic observations of \citet{pri2009b} and the photometric data from the $ASAS$ survey. They derived the following physical parameters: $M_{1} = 1.471 $M$_{\sun}$, $M_{2} = 0.287 $M$_{\sun}$, $R_{1} = 1.568 $R$_{\sun}$, $R_{2} = 0.795 $R$_{\sun}$.
\subsection{V1123~Tau} The eclipsing binary V1123~Tau (TYC~1238-831-1) was discovered to be variable by {\it HIPPARCOS} and, based on the colour index, it was classified as a $\beta$ Lyr-type eclipsing binary of G0V spectral type. Later, \citet{kaz1999} proposed it was a contact binary. V1123~Tau is a member of the WDS~03350+1743 visual binary, accompanied by a fainter star ($\theta$ = 136$^\circ$, $\rho$ = 4.3 arcsec and $\Delta$$V$ = 1.77). \citet{ruc2008} performed spectroscopic observations excluding additional light contribution (or with only a marginal contribution of 3~per~cent during periods of poor seeing) by positioning the third component away from the spectrograph slit. They determined the mass ratio of the contact binary as $q = 0.279$ and estimated its spectral type to be G0V, confirming the classification given by \citet{kaz1999}. \citet{gut2009} obtained the spectral types of both components of the visual binary and confirmed the contact binary G0V spectral type, and assigned a K0V type to the visual companion. \citet{ozd2006} presented the first ground-based photometry of the system and estimated the colour index as $B-V = 0.684$~mag. \citet{pri2006} confirmed the existence of an additional component orbiting the contact system by studying the large proper motion errors from {\it TYCHO} observations. The first combined photometric and spectroscopic model was given by \citet{deb2011}, using the V-band observations of $ASAS-3$ project \citep{poj1997} and the spectroscopic mass ratio from \citet{ruc2008}. According to their model, the contact binary has an orbital inclination of $i = 68.1^\circ$ and a fill-out factor of $f$ = 18~per cent. A similar study was also performed by \citet{zha2011}, based on their own multi-band observations. Their results are in good agreement with those presented by \citet{deb2011}. \citet{ekm2012} obtained new BVR photometric observations and derived a somewhat larger orbital inclination of $i = 74^\circ$, which resulted in slightly smaller masses than those given by \citet{deb2011} and \citet{zha2011}.

\subsection{W~UMa} W~UMa (HIP 47727) is the prototype system of the entire category of
contact binaries. It was discovered in 1903 \citep{mul1903} and since then has been intensively
studied by many observers. It is the brighter component of the visual pair ADS~7494 (WDS J09438+5557)
with a separation of 6.4~arcsec \citep{Mason2001}. However, it is still not certain whether the two
visual components are gravitationally bound. \citet{whe1974} explained the periodic shape of O$-$C
diagram by the presence of a third body orbiting the close binary system. \citet{pri2006} confirmed
its existence and gave astrometric parameters. W~UMa also shows irregular orbital period changes,
which probably indicate non-uniform mass and angular momentum transfer processes within the contact
system. The most recent spectroscopic study was made by \citet{pri2007}, who derived a mass ratio
of $q = 0.484 \pm 0.003$, and a spectral type of F5V. The first set of absolute parameters for W~UMa
was given by \citet{kal1985}, who calculated masses and radii as 1.15 and 0.56 $M_{\sun}$ and 1.11 and
0.76 $R_{\sun}$ for the primary and secondary component, respectively. A detailed light curve analysis
was performed by \cite{Linnell1991}. One of the main conclusions of his work is the explanation of
the O'Connell effect in this system by the presence of a pair of cool spots on the surface of the
more massive star. Apart from numerous photometric studies \citep[][and references therein]{Linnell1991},
the system has been observed several times spectroscopically (e.g. \cite{mcl1981, Rucinski1993}).
It is interesting that the absolute spectroscopic elements of the system are still not consistent,
with large differences in the centre-of-mass velocity, ranging from V$_{0}=0$ km s$^{-1}$ \citep{Binnendijk1966}
to $V_{0}=-43$ km s$^{-1}$ \citep{Popper1950} and $V_{0}=-50$ km s$^{-1}$ \citep{str1950b}.


\section{Photometric Observations and Data Reduction}
This study is based on new photometric observations collected within
the period 2001--2011, either with CCD cameras or photomultiplier
tube (PMT) photometers. The observing strategy for all systems is
the same as that described in \citet{zol2010} (Paper VII and references
therein). We gathered multicolour photometric data using only one instrument
for each target, in order to avoid any systematic effects due to instrumental
calibration or filters mismatch.
In addition, our goal was to obtain complete light curves within as short a time as possible to minimize
the effect of any intrinsic variability, e.g. due to magnetic activity, which might prevent the
combining of data from individual runs.
The orbital period of several  systems analyzed here is short enough to be covered
within one night. However, for some systems observations were continued for several days.
The purpose was to follow the variability over several orbital cycles to make sure that the
timescale of any intrinsic variability is significantly longer than the duration of our
observations.  In this way, the light curves modelled  in this manuscript (and in all
papers in the series) represent a snapshot of photometric behaviour during the
observational period given in Table \ref{TabEfem}.
In the case of OO~Aql, longer runs were also performed to check the stability of the light
curve. We did not detect any shape varation of the light curve during
the entire observing period which might have prevented all the segments of the light
curve which had been gathered from being combined.

Photometric observations of  OO~Aql, HV~Aqr, TX~Cnc, RW~Com, V401~Cyg,
V345~Gem, AK~Her, V566~Oph, V2612~Oph, V1363~Ori, and V351~Peg were obtained
at the University of Athens Observatory (UOAO) with the 40~cm telescope.
This telescope is equipped with a set of $UBVRI$ filters (Bessell
specifications) and three CCD cameras (ST8, ST8-XMEI and ST10-XME), which were
used in turn depending on availability.
The observing setup results in a field of view of 10$\times$15 arcmin for the ST8
and ST8-XMEI CCD cameras and 11$\times$16 arcmin for the ST10-XME CCD camera.
An f/6.3 focal reducer was used in some observations to increase the field
of view to 15$\times$24 arcmin and 17$\times$26 arcmin for the ST8 and ST10,
respectively.


\begin{table*}
\begin{center}
\caption{Linear ephemerides used for phasing the observations and initial information derived
from spectroscopic data (DDO Program). The standard errors for each value are expressed in
parentheses, in units of last decimal places quoted.}
\label{TabEfem}
\begin{tabular}{llllccc}
\hline
System          & $T_{0}$ (HJD)         &  $P_{orb}$ (d)    &  sp. type     &  $T_{1}$ (K)  & mass ratio        &   Ref \\
\hline
HV Aqr	        &	2453579.40480(20)	&	0.3744574(4)	&	F5V	        &	6460(100)	&	0.145(50)		&	III	\\
OO Aql	        &	2454067.26641(11)	&	0.5067934(2)	&	F9V	        &	6000(100)	&	0.846(7)		&	XII	\\
FI Boo	        &	2453504.39378(11)	&	0.3899999(8)	&	G3V	        &	5650(100)	&	0.372(21)		&	IV	\\
TX Cnc	        &	2455289.34028(25)	&	0.3828824(2)	&	F8V	        &	6100(100)	&	0.455(11)		&	XI	\\
OT Cnc	        &	2454523.32792(14)	&	0.2178118(4)	&	K3/5V	    &	4500(200)	&	0.474(8)		&	(R)	\\
EE Cet	        &	2454460.39590(10)	&	0.3799230(8)	&	F8V	        &	6100(100)	&	0.315(5)		&	VI	\\
RW Com	        &	2455663.50137(11)	&	0.2373484(3)	&	K2/5V	    &	4600(200)	&	0.471(6)		&	XIV	\\
KR Com	        &	2452789.42334(28)	&	0.4079676(9)	&	G0IV        &	5920(100)	&	0.091(2)		&	VI	\\
V401 Cyg	    &	2453535.42230(17)	&	0.5827268(5)	&	F0V	        &	6980(100)	&	0.290(11)		&	VI	\\
V345 Gem	    &	2454068.53814(25)	&	0.2747744(1)	&	F7V	        &	6200(100)	&	0.142(3)		&	XII	\\
AK Her	        &	2455343.44153(17)	&	0.4215229(9)	&	F4V	        &	6530(100)	&	0.277(24)		&	XI	\\
V502 Oph	    &	2453475.58546(11)	&	0.4533863(2)	&	G0V	        &	5920(100)	&	0.335(9)		&	IX	\\
V566 Oph	    &	2454982.35897(53)	&	0.4096567(3)	&	F4V	        &	6530(100)	&	0.263(12)		&	XI	\\
V2612 Oph	    &	2455727.39631(23)	&	0.3753080(2)	&	F5V	        &	6460(100)	&	0.286(3)		&	XII	\\
V1363 Ori	    &	2453447.28240(70)	&	0.4319217(6)	&	F(early-mid)&	6700(250)	&	0.205(15)		&	IX	\\
V351 Peg	    &	2453591.56489(29)	&	0.5932972(5)	&	A8V	        &	7500(150)	&	0.360(6)		&	V	\\
V357 Peg	    &	2452931.24349(13)	&	0.5784516(4)	&	F2V	        &	6700(100)	&	0.401(4)		&	XIII\\
Y Sex	        &	2452724.30270(35)	&	0.4198193(2)	&	F5/6V	    &	6400(100)	&	0.195(8)		&	XV	\\
V1123 Tau	    &	2452975.40516(12)	&	0.3999463(7)	&	G0V	        &	5920(100)	&	0.279(4)		&	XIII\\
W UMa	        &	2451952.34017(14)	&	0.3336352(2)	&	F5V	        &	6450(100)	&	0.484(2)		&	XII	\\
\hline
\end{tabular}
\end{center}
\begin{small}
References: III: \citet{ruc2000},  IV: \citet{lu2001},  V: \citet{ruc2001}, VI: \citet{ruc2002b},  IX: \citet{pyc2004}, XI: \citet{pri2006b}, XII: \citet{pri2007}, XIII: \citet{ruc2008}, XIV: \citet{pri2009a}, XV: \citet{pri2009b}, (R): \citet{ruc2008b}
 \end{small}
\end{table*}


FI~Boo, KR~Com, V357~Peg, V1123~Tau, and W~UMa were observed at the Mt. Suhora observatory
(SUH)  with the 60~cm telescope and a PMT photometer, while EE~Cet and OT~Cnc
were observed with the same telescope and, respectively, an Apogee Alta U47 or a SBIG ST10-XME CCD
camera. The PMT photometer was equipped with a set of $UBVRI$ filters (Bessell
specification) and includes a set of circular apertures ranging between 10 and 40 arcsec.
Both CCD cameras also have sets of Bessell-specification $UBVRI$ filters. The focal length
of this telescope is 750~cm (at Cassegrain focus) and provides a field of
view of $6.1 \times 6.1$ arcmin and $4.6 \times 6.8$ arcmin, for the Apogee and
SBIG cameras, respectively.

Observations of V502~Oph were obtained with the 75~cm telescope at the South African
Astronomical Observatory (SAAO) with the UCT CCD camera, which has a 576$\times$420 pixel
sensor with 22 $\mu$m pixel size. It is equipped with a set of $UBV(RI)_{C}$ filters
(Johnson-Cousins specification) and has a field of view of approximately 1.5$\times$2.5 arcmin.
Observations of Y~Sex were obtained with the same telescope  and a PMT photometer,
with a similar set of $UBV(RI)_{C}$ filters.
CCD images were reduced using the aperture photometry method, while PMT data were
extracted during observations, utilizing the circular apertures of the instrument.
All photometric data were corrected to heliocentric time (HJD) and for the differential term of
atmospheric extinction. Colour extinction effects were compensated by careful selection
of comparison stars with colours close to those of the targets and therefore was
not accounted for separately.

A detailed observation log is presented in Table \ref{TabObsLog}.

\section{Light curve modelling}
The light curve of each system consists of between a few hundred and a few thousand
individual measurements per filter. In order to speed up computations, we calculated
100--200 mean points in each filter for running the Monte Carlo (MC) code. We phased the
data using the ephemerides listed in Table \ref{TabEfem}, which were calculated on
the basis of new minima times from our observations and additional ones gathered
from \citet{kre2004}\footnote{http://www.as.up.krakow.pl/ephem/}. Graphical representations
of the resulting O$-$C diagrams for all systems analysed in this paper are also available
on this website. This database is continuously updated with new times of minima and therefore
the ephemerides are re-calculated and change over time.
Differential magnitudes were not converted to a standard system, as this is not required
for modelling, but were left in the instrumental system. The magnitudes were transformed
into flux units and normalized to 1 at the highest quadrature in phase diagrams.
We used the Wilson-Devinney (W-D) code \citep{Wilson1971, Wilson1990} appended
with the MC algorithm as the search procedure (see Papers I and III for details)  for
light curve modeling.


\begin{figure}
\includegraphics[height=6.8cm,scale=1.0,angle=0.0]{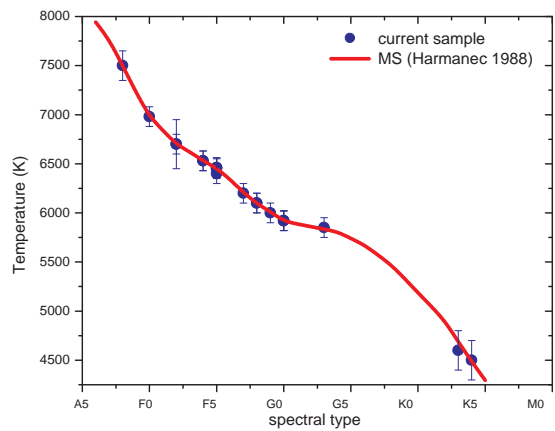}
 \caption{The effective temperature of primary components was estimated by determining
the spectral type from spectroscopic observations. The link between the spectral type
and effective temperature leads to a quite accurate results, following the empirical
model and calculated values given by \citet{har1988}. This empirical model is shown as continuous
line, while the adopted effective temperature for each system in the current
sample is shown in filled circles, together with their corresponding uncertainty.}
\label{FigTemp}
\end{figure}


The computations were made using the radial velocity measurements and the multicolour light curves
in  an iterative way. In the first step, the mass ratio parameter was kept fixed at the spectroscopic
value, given in Table \ref{TabEfem}. The mass ratio was reversed (i.e. the value $1/q$ was used in the
model) when needed, in order to have the deeper minimum of the light curve at
phase $\phi$=0.0 and thus phase the spectroscopic and photometric data with a single ephemeris,
given in Table \ref{TabEfem}. Thus the star eclipsed in the primary minimum is not always the more massive one,
contrary to the convention commonly used in spectroscopic papers. The effective temperature of the primary
component(the star eclipsed at phase 0) was fixed and its value was determined using
the \cite{har1988} calibration tables based on the system spectral type.
This method for spectral type estimation leads to an uncertainty of about 100~K for low-temperature
systems (F--M spectral types) and up to 150~K for the higher-temperature ones of the A spectral type.
An overview of the spectral types corresponding temperatures along with their uncertainties is given
in Table \ref{TabEfem}, and graphically shown in Figure \ref{FigTemp}.
The only exception applies to KR~Com, the spectral type  of which was determined to be G0IV \citep{ruc2002b}.
Since the components of contact binaries belong to the MS, for KR Com we adopted a temperature corresponding
to the G0V spectral type.

Albedo and gravity darkening coefficients were fixed at their theoretical values of $A = 0.5$ and
$g = 0.32$ for stars with convective envelopes ($T < 7200$~K) and $g = 1$ and $A = 1$ for stars with
radiative envelopes ($T > 7200$~K). The limb darkening coefficients were taken from the tables of
\cite{diaz95} and \cite{cla2011}, according to the effective temperature of the components and the
filters used. The effective temperature of the secondary component, potential, phase shift, and the
luminosity of the primary component were treated as free parameters. Luminosity of the secondary
components were computed from their temperatures and the system geometry.
If there were asymmetries visible in the light curves, a cool spot was also included,
which involves an additional four  parameters describing the spot. In some cases, we considered more than one spot,
but decided to keep only one if the improvement was not better than half the $\chi^2$ value of the single-spot model.
The same criterion was applied to no-spot and single-spot solutions.
The third light ($l_{3}$) parameter was also adjusted for systems which have confirmed or proposed  companions
(e.g. based on the O$-$C diagram shape).
It is worth noting that reliability of solutions decreases when spot(s) or third light are included in the
models. The parameters which would be most affected are orbital inclination and the temperature of the secondary.


\begin{figure*}
\includegraphics[height=7.8cm,scale=1.0,angle=270]{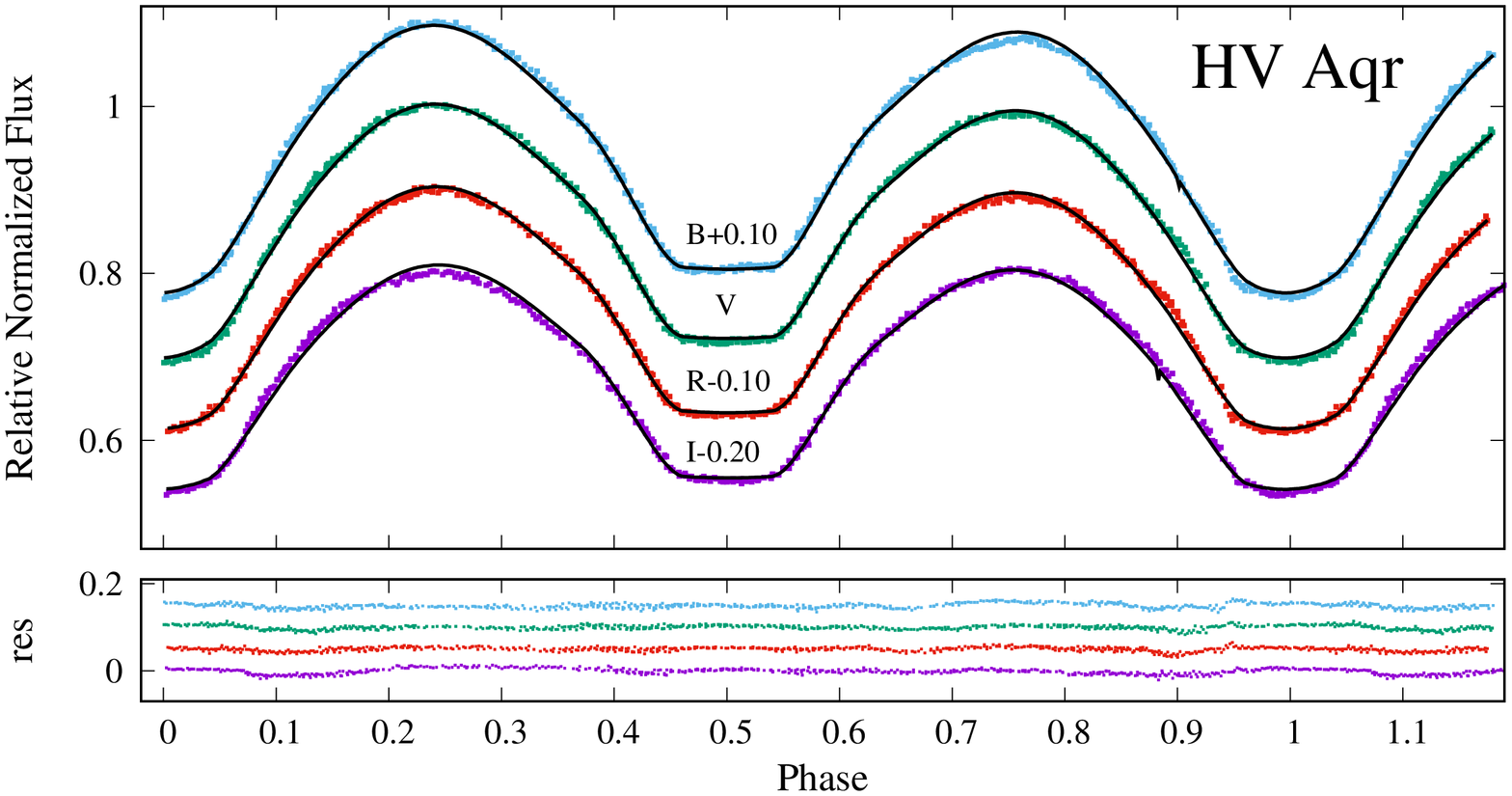}
\includegraphics[height=7.8cm,scale=1.0,angle=270]{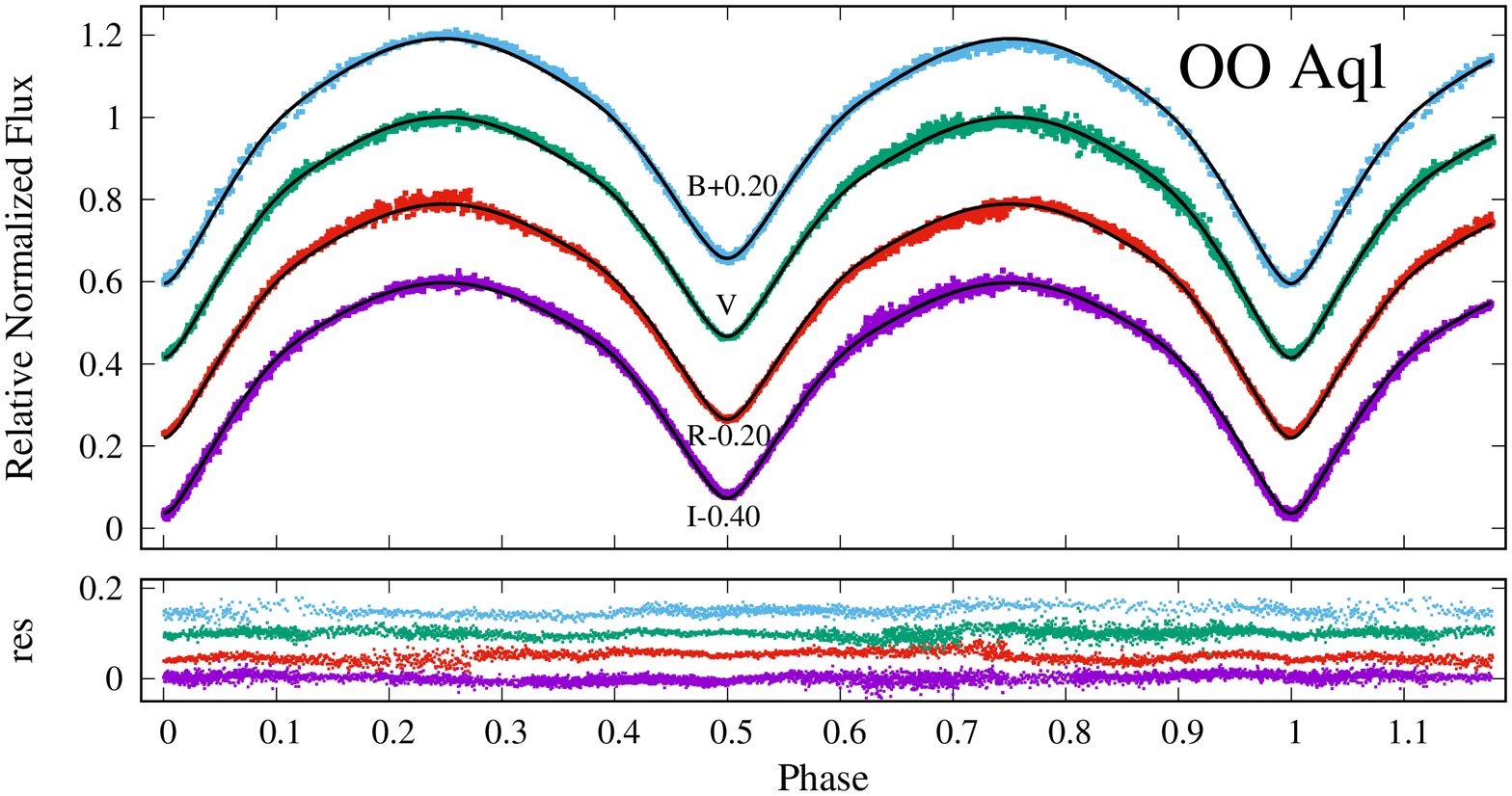}
\includegraphics[height=7.8cm,scale=1.0,angle=270]{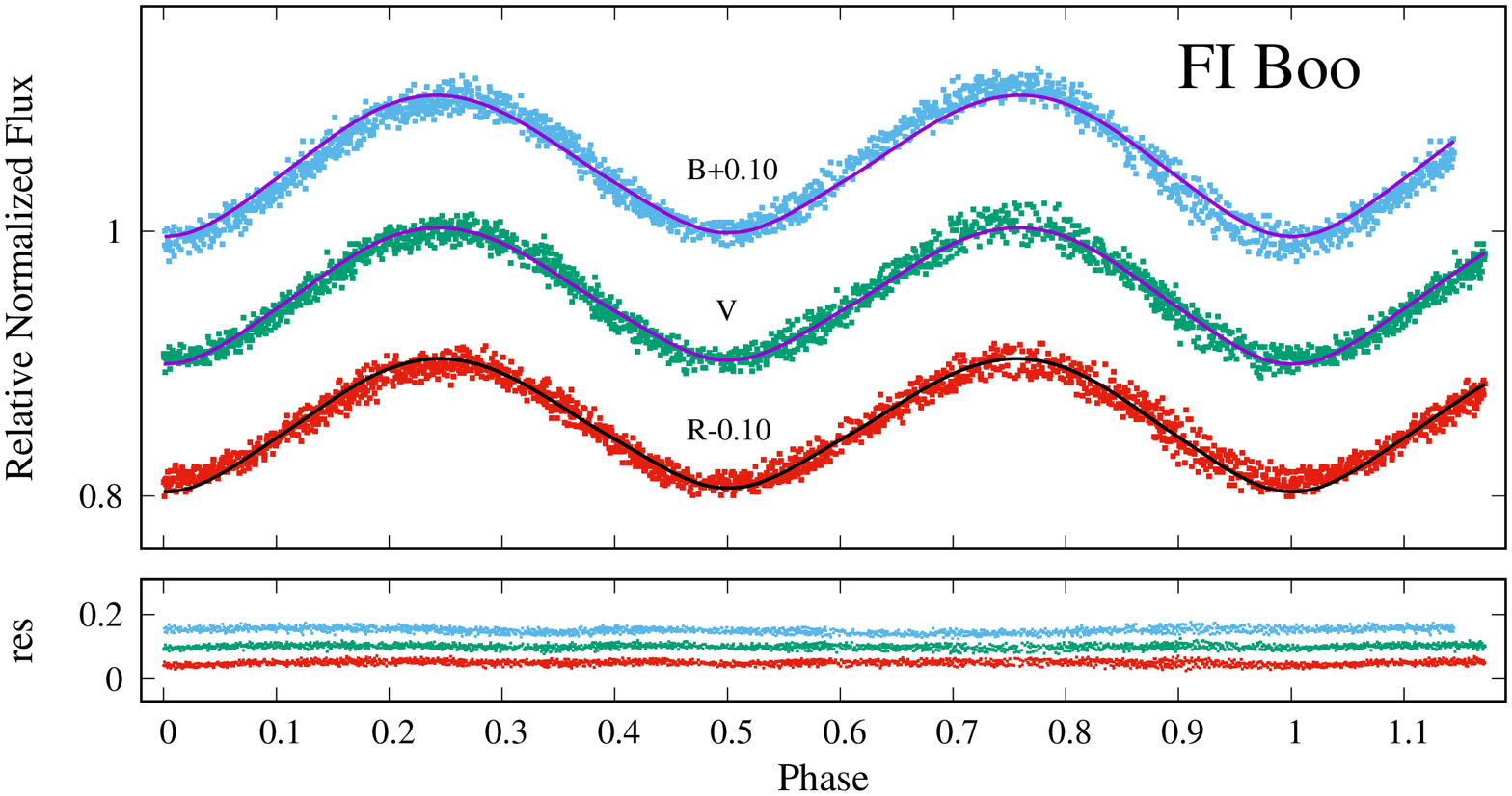}
\includegraphics[height=7.8cm,scale=1.0,angle=270]{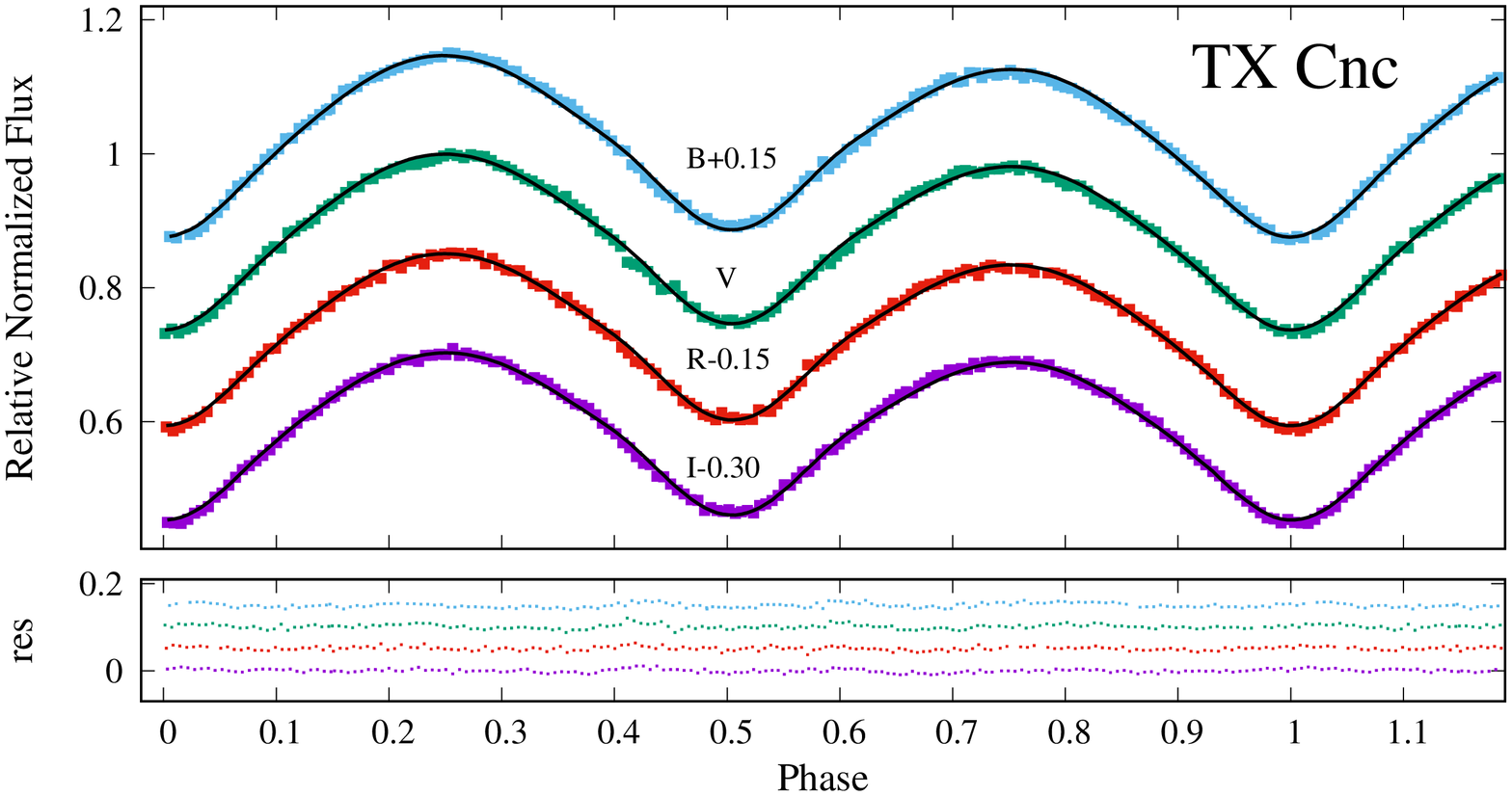}
\includegraphics[height=7.8cm,scale=1.0,angle=270]{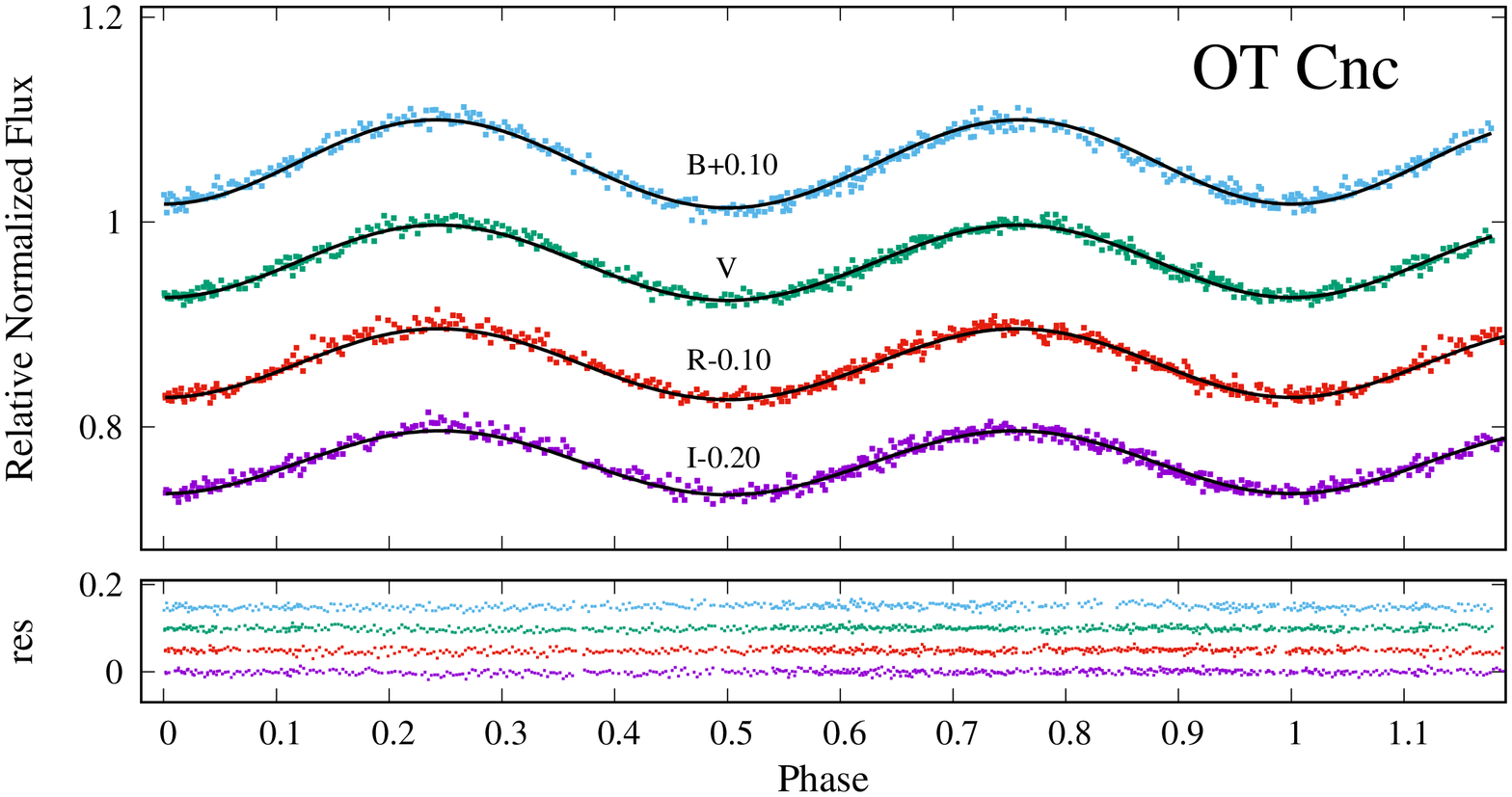}
\includegraphics[height=7.8cm,scale=1.0,angle=270]{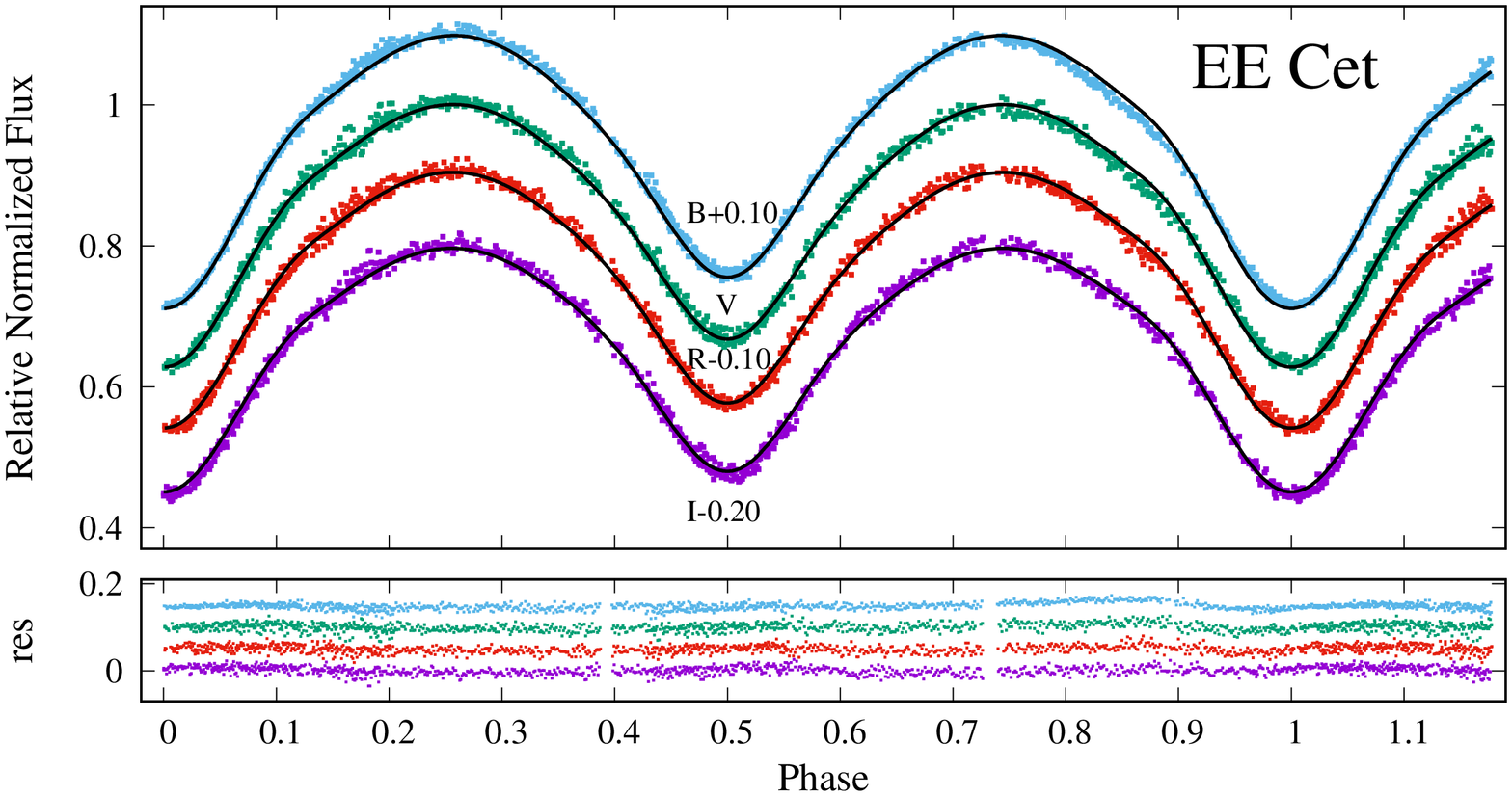}
\includegraphics[height=7.8cm,scale=1.0,angle=270]{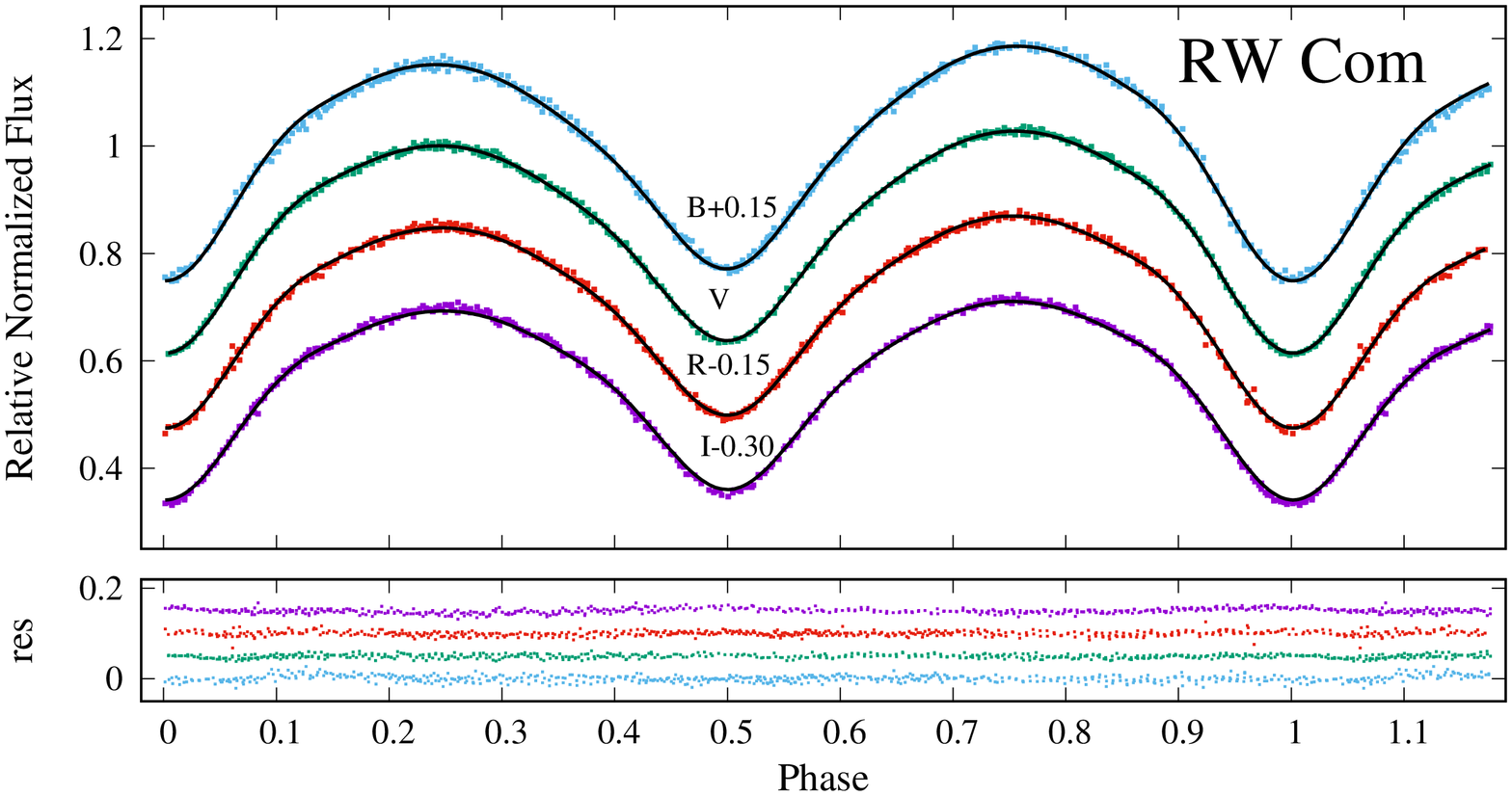}
\includegraphics[height=7.8cm,scale=1.0,angle=270]{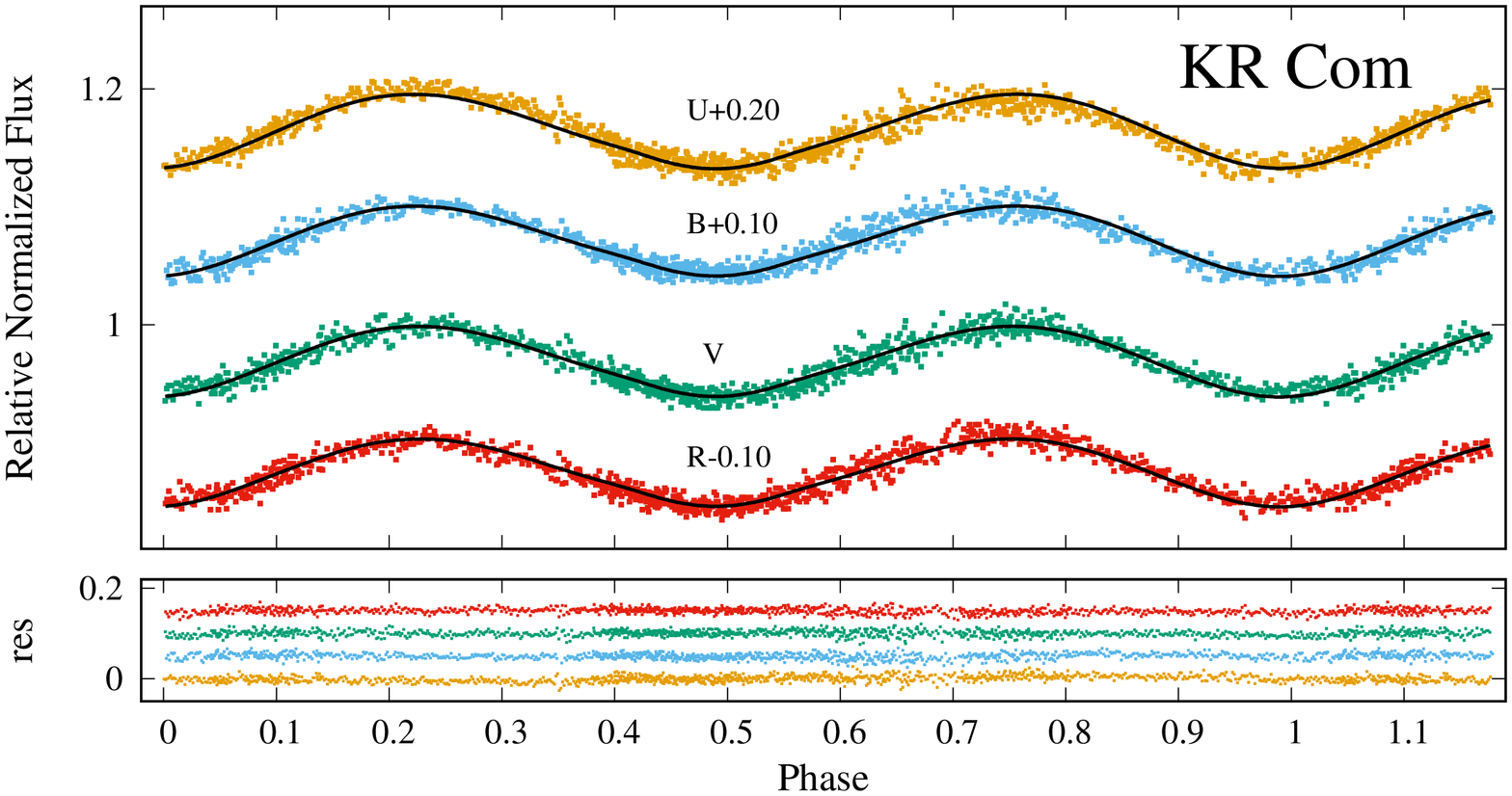}
\includegraphics[height=7.8cm,scale=1.0,angle=270]{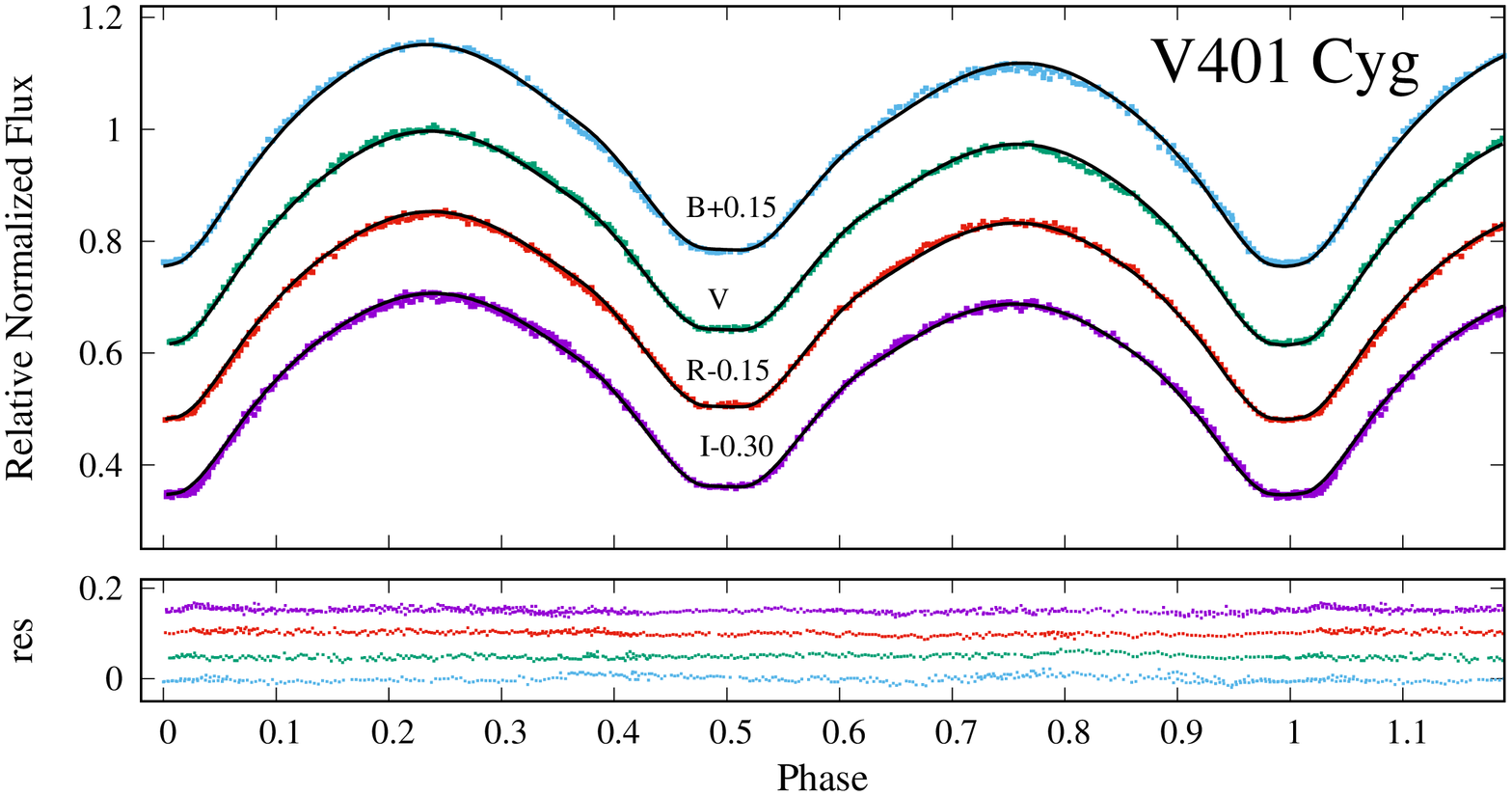}
\includegraphics[height=7.8cm,scale=1.0,angle=270]{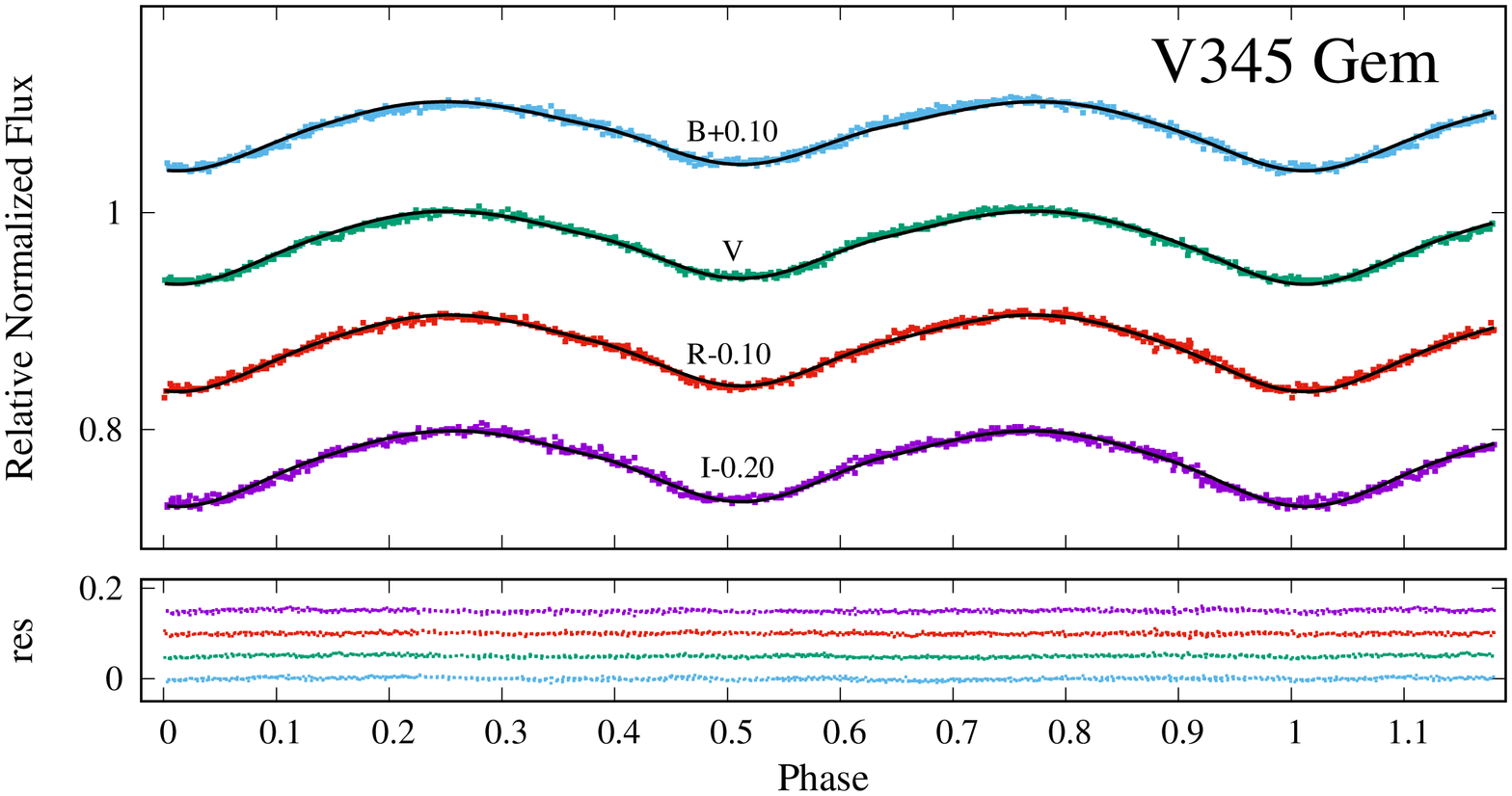}
 \caption{Comparison of theoretical and observed light curves
  of HV Aqr, OO Aql, FI Boo, TX Cnc, OT Cnc, EE Cet, RW Com, KR Com, V401 Cyg and V345 Gem. The values in y-axis
  are given in normalized flux for all the observed $UBVRI$ filters, shifted relative to each other for clarity.
  Symbols represent individual observations, while the theoretical light curves are shown as continuous lines.
  The residuals after fitting the model to the observed data are given for each plot, shifted
  by 0.05 with respect to each other for clarity.}
\label{FigLC1}
\end{figure*}

Details of the procedure used to derive the parameters and their uncertainties were outlined
in \citet{kre2003} (Paper~I) and in \citet{bar2004} (Paper~II). Light curve modelling of each system
was performed using the same methods and techniques as those in \citet{zol2004} (Paper III) and
other papers of the series.
The  applied Monte Carlo (MC) method  does not require providing starting values for the adjusted
parameters, instead it searches for the best solution within given ranges.
The code randomly generates a set of values for each of the free parameters representing the model of the system, and stores them in a search array
consisting of 3000 elements if there are no spots or third light (L3) in the model,
otherwise a larger set of 4000 elements is used. Once the search array has been filled,
the next set of values to be generated is compared with the models already stored and the worst
element (in the sense of highest weighted, reduced  {$\chi^2$} value) is replaced
if a better one has been found. Models with larger $\chi^2$ values are discarded. This procedure
is repeated until there is little difference between the best and worst elements in the search
array. Here we set up the difference limit to be 0.001.
When convergence is obtained there will be very little scatter of
the values of adjusted parameters, thus providing no reliable estimate of their uncertainties.
We determined the uncertainties of the free parameters in the $\chi^{2}$  minimization according to the
the method described in Numerical Recipes in Fortran (2nd edition, Section 15.6). This was done
by monitoring the search array during the computations and saving a copy of the array when the criteria of 1-3$\sigma$ were fulfilled. By analyzing the
intermediate search arrays we derived the adjusted parameter uncertainties at 68, 90 and 95 per~cent
confidence levels ($1\sigma$, $2\sigma$, and $3\sigma$, respectively). We chose to present
the values at the $2\sigma$ level.

\begin{figure*}
\includegraphics[height=7.8cm,scale=1.0,angle=270]{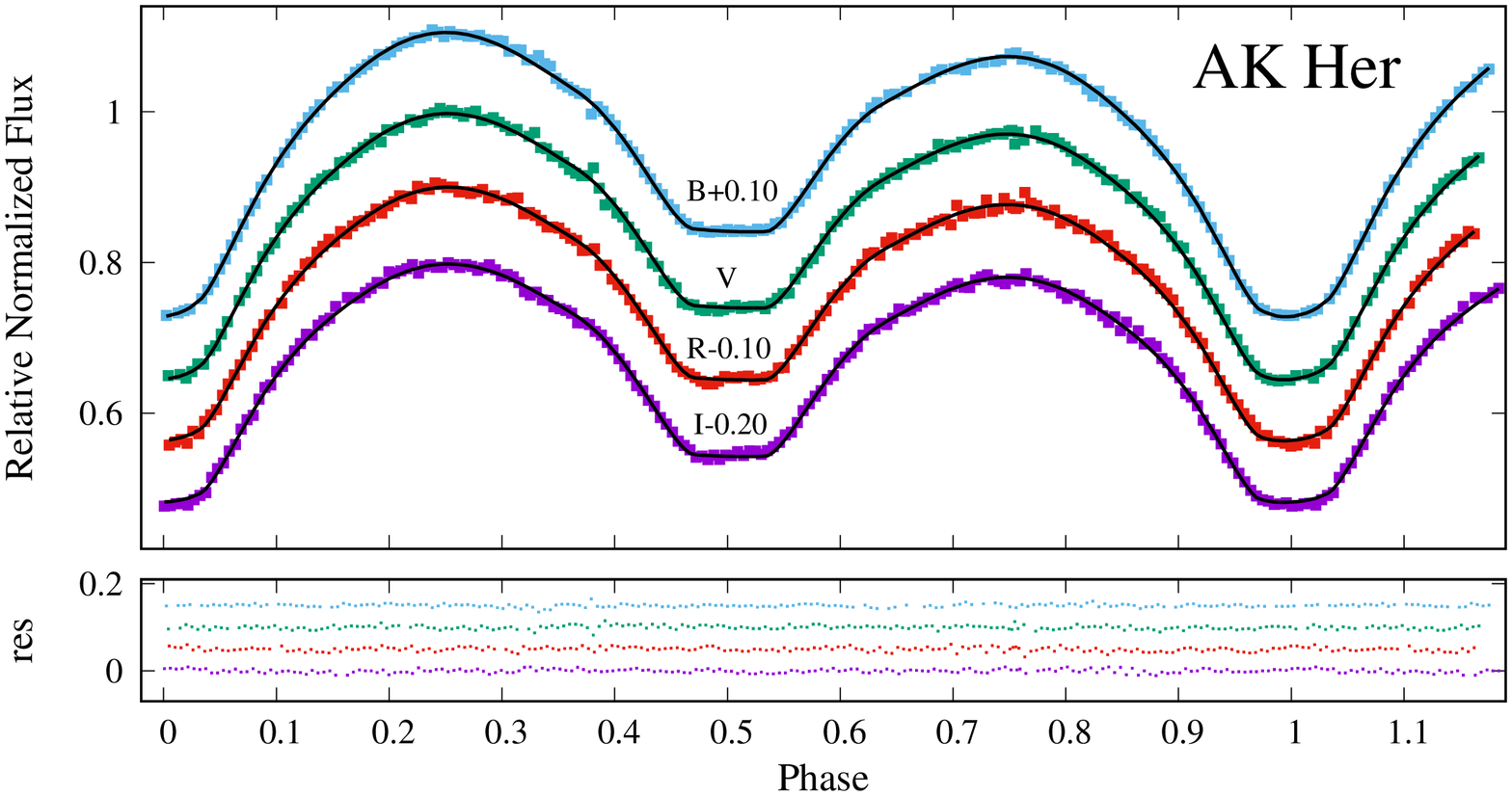}
\includegraphics[height=7.8cm,scale=1.0,angle=270]{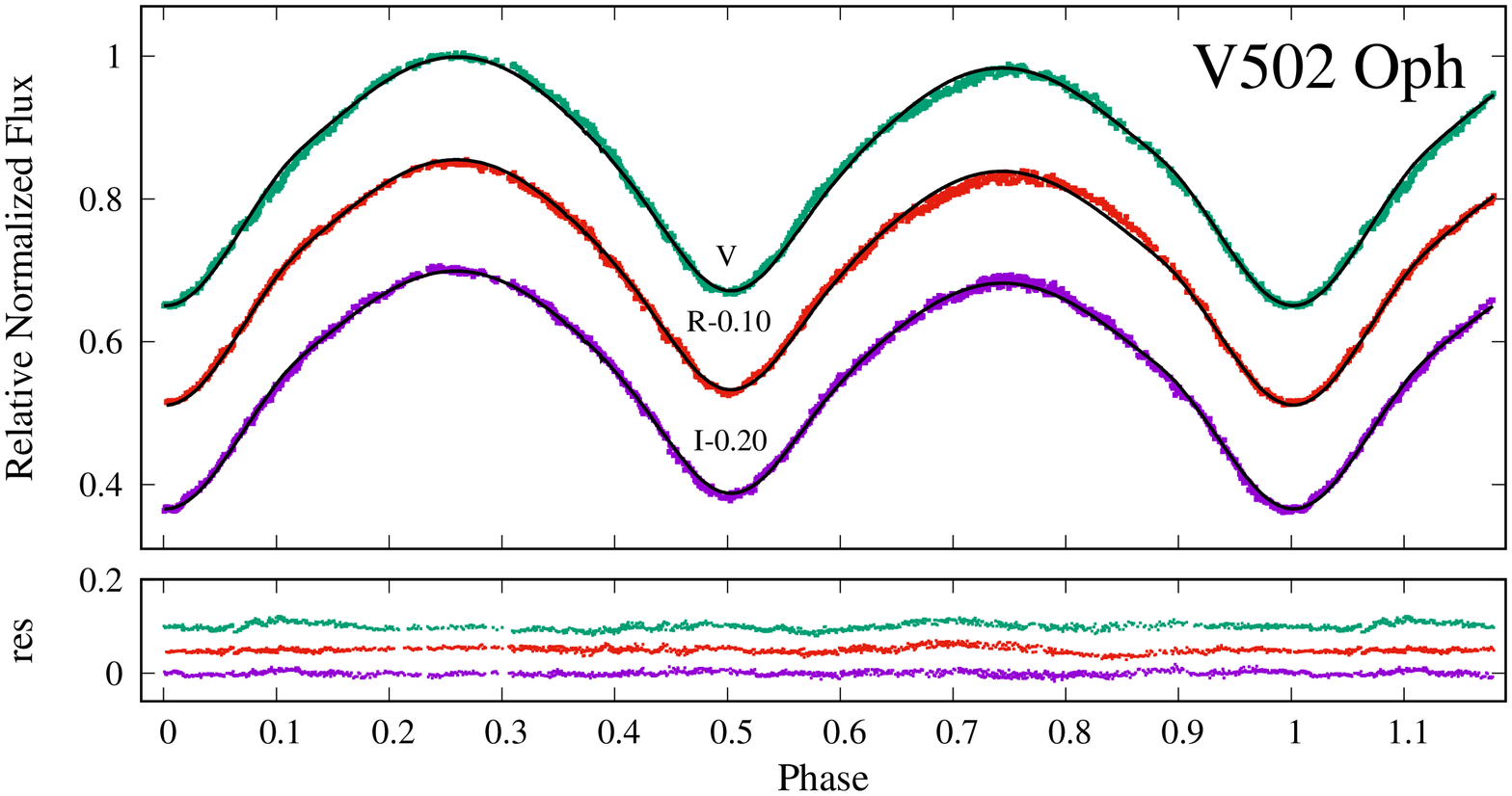}
\includegraphics[height=7.8cm,scale=1.0,angle=270]{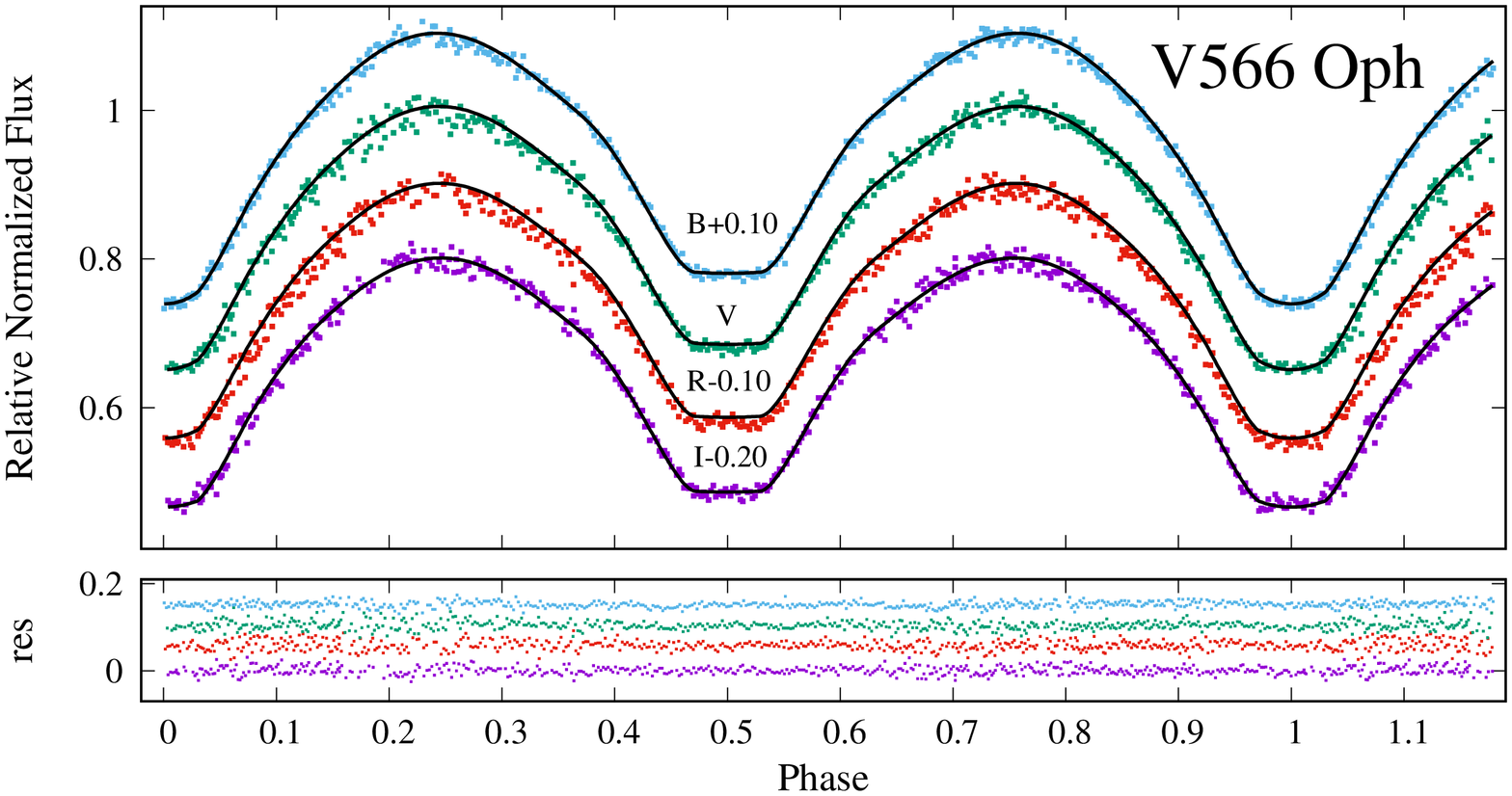}
\includegraphics[height=7.8cm,scale=1.0,angle=270]{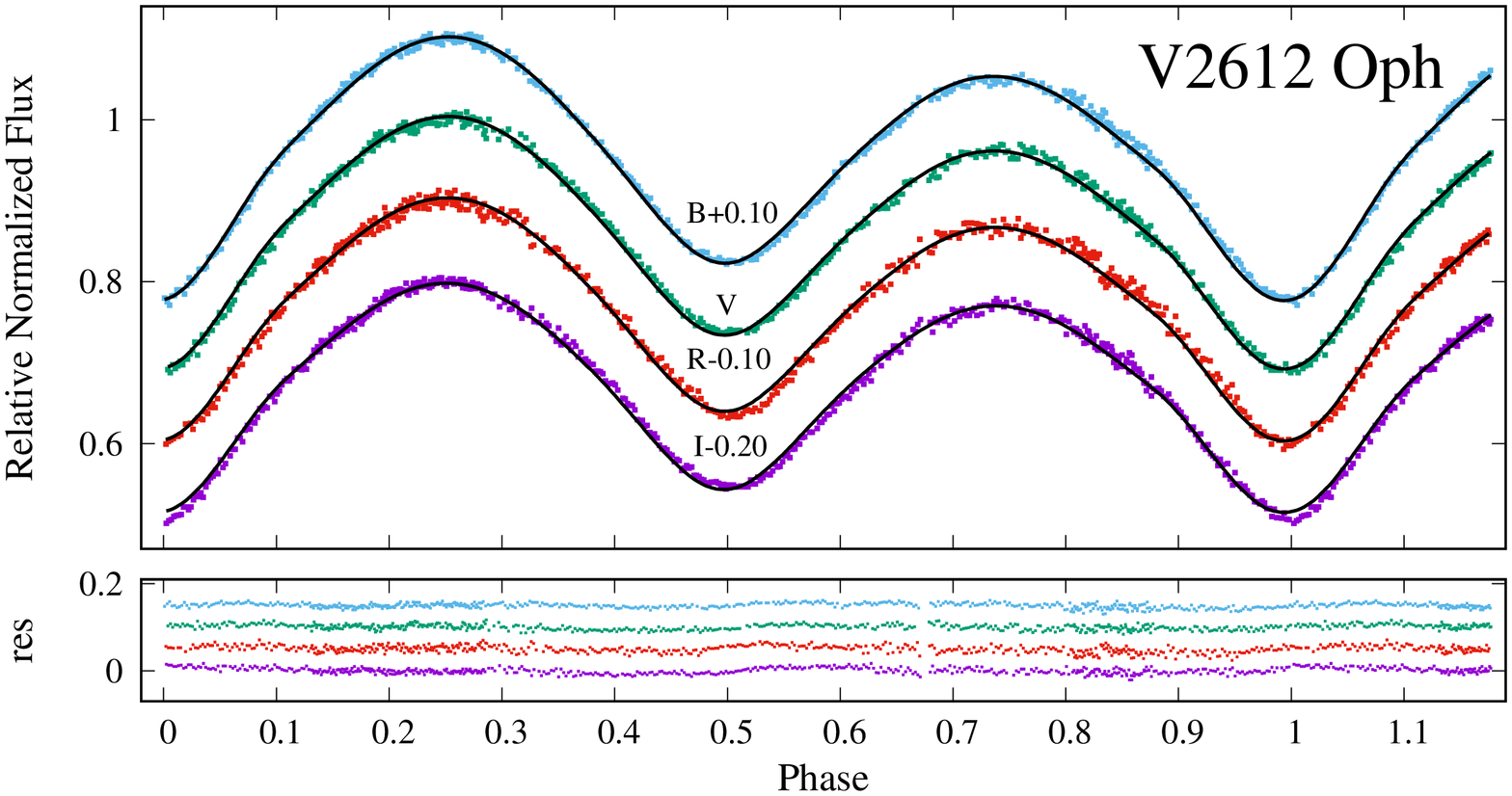}
\includegraphics[height=7.8cm,scale=1.0,angle=270]{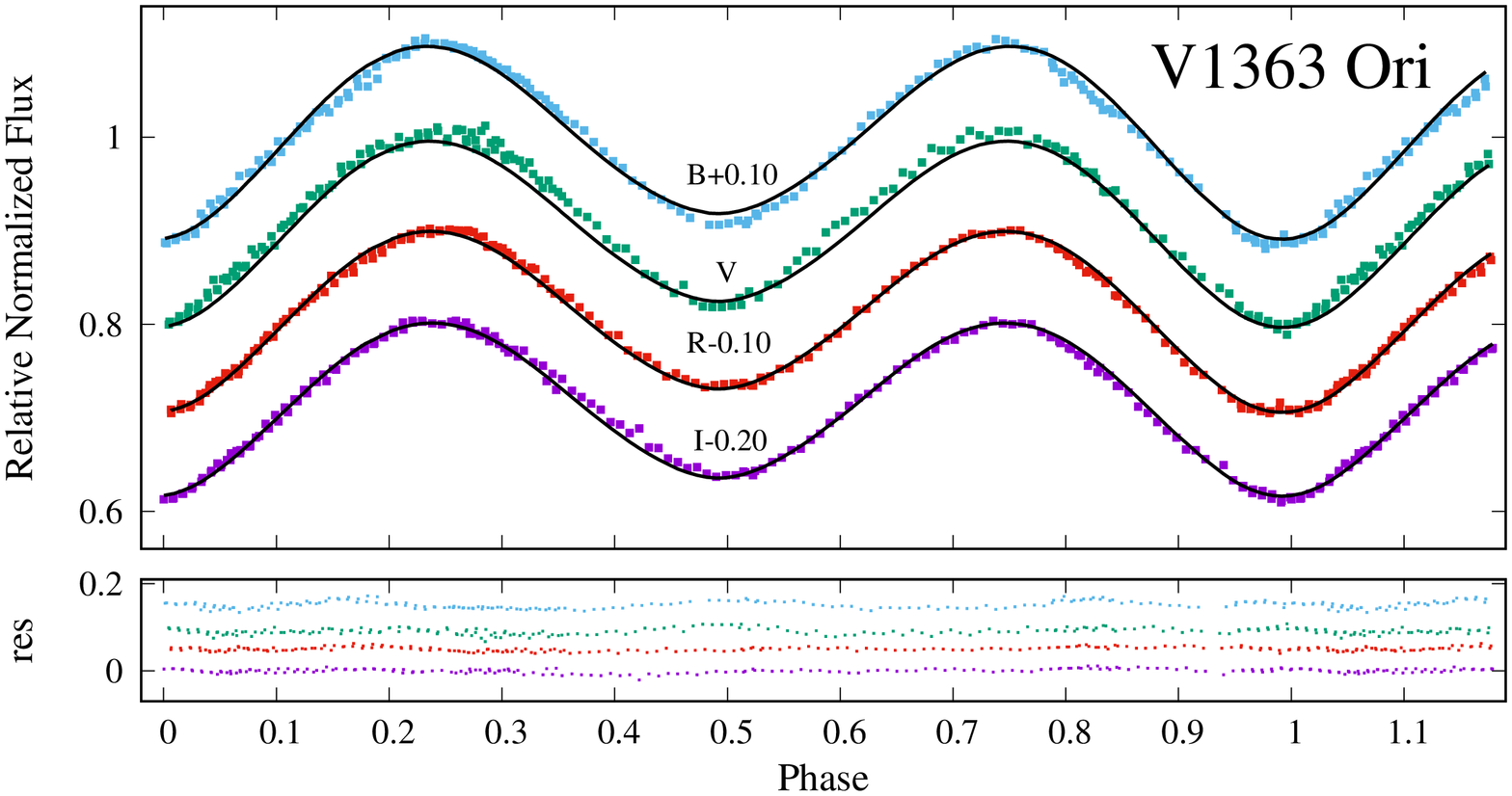}
\includegraphics[height=7.8cm,scale=1.0,angle=270]{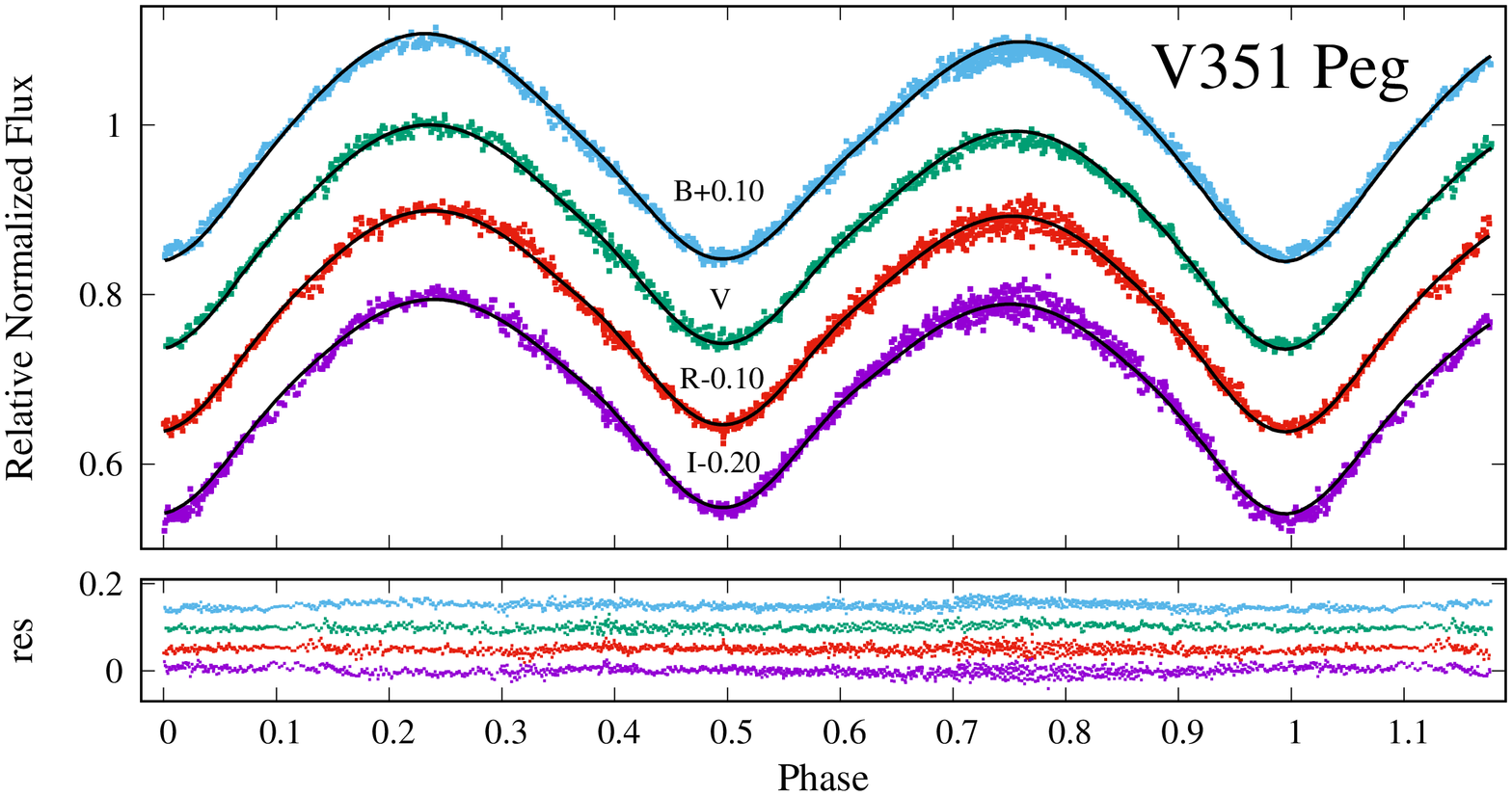}
\includegraphics[height=7.8cm,scale=1.0,angle=270]{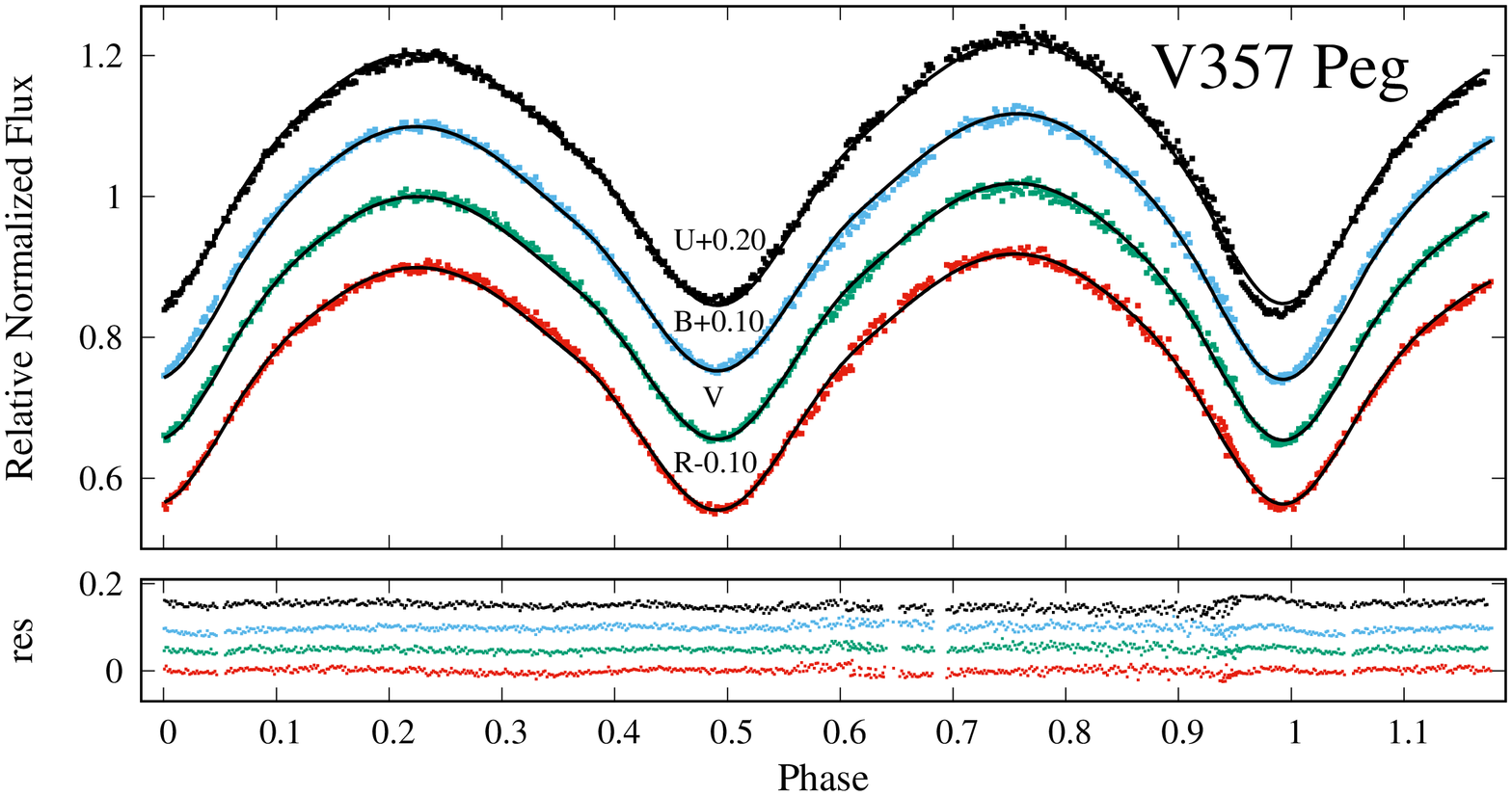}
\includegraphics[height=7.8cm,scale=1.0,angle=270]{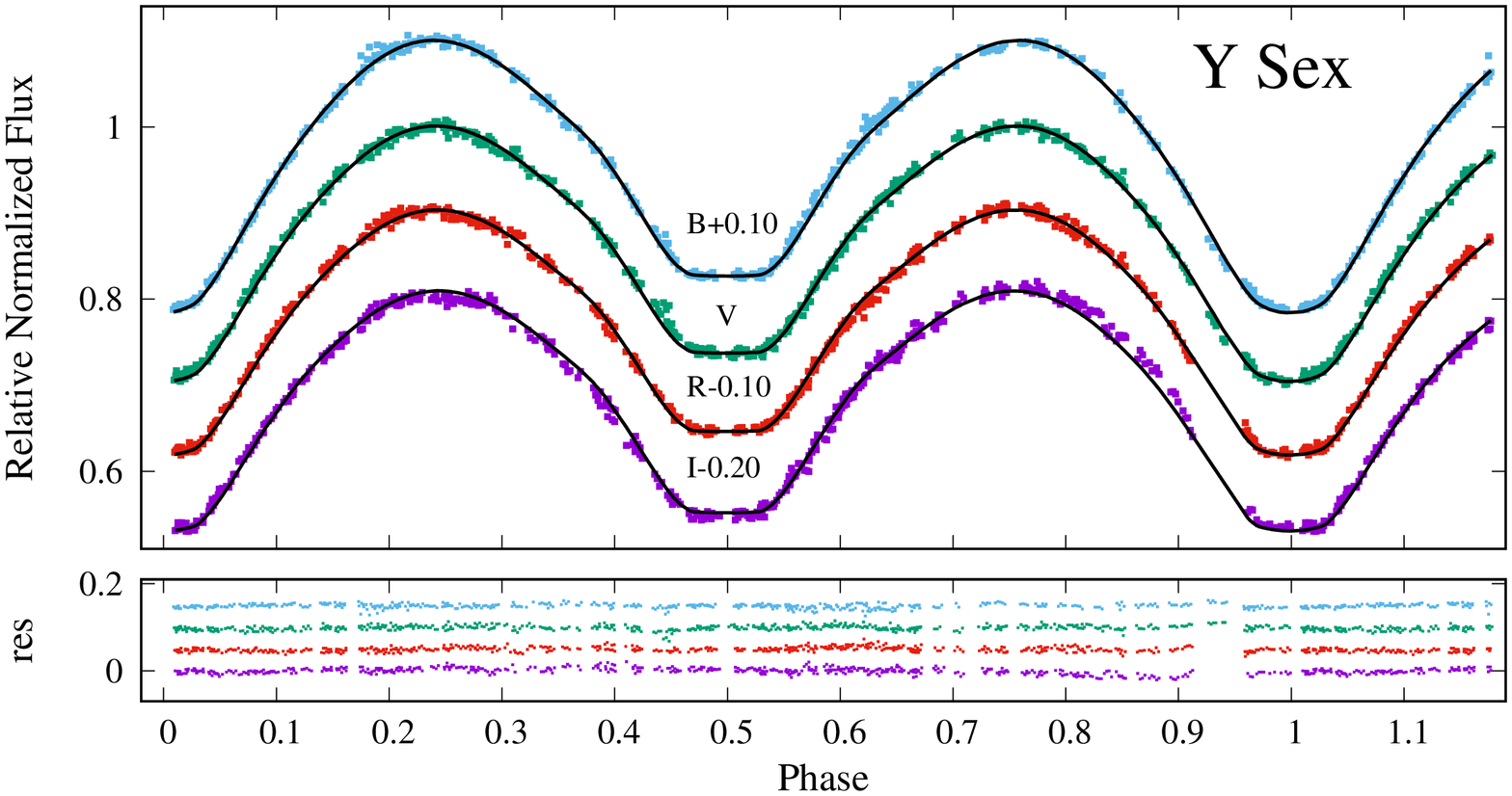}
\includegraphics[height=7.8cm,scale=1.0,angle=270]{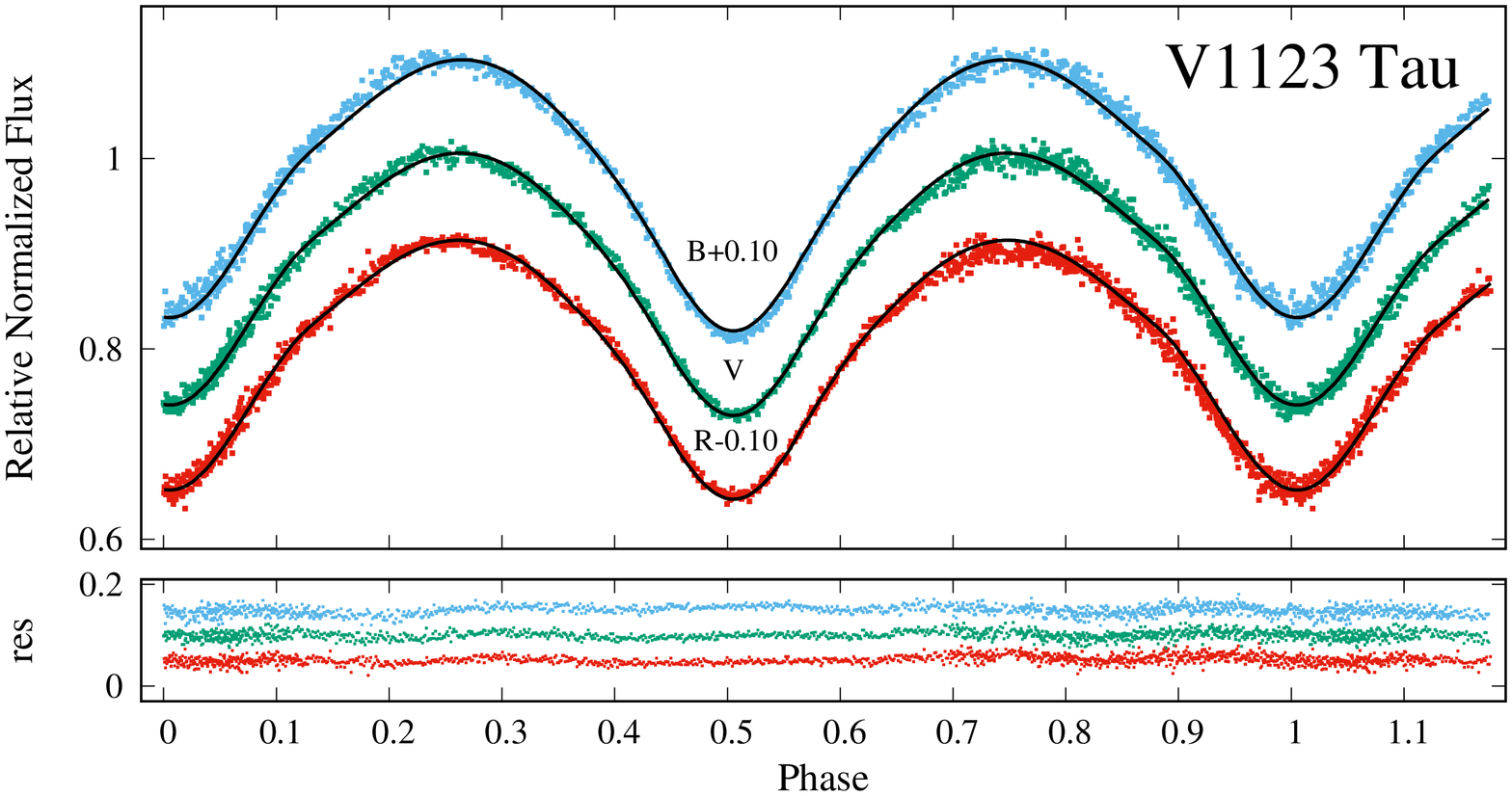}
\includegraphics[height=7.8cm,scale=1.0,angle=270]{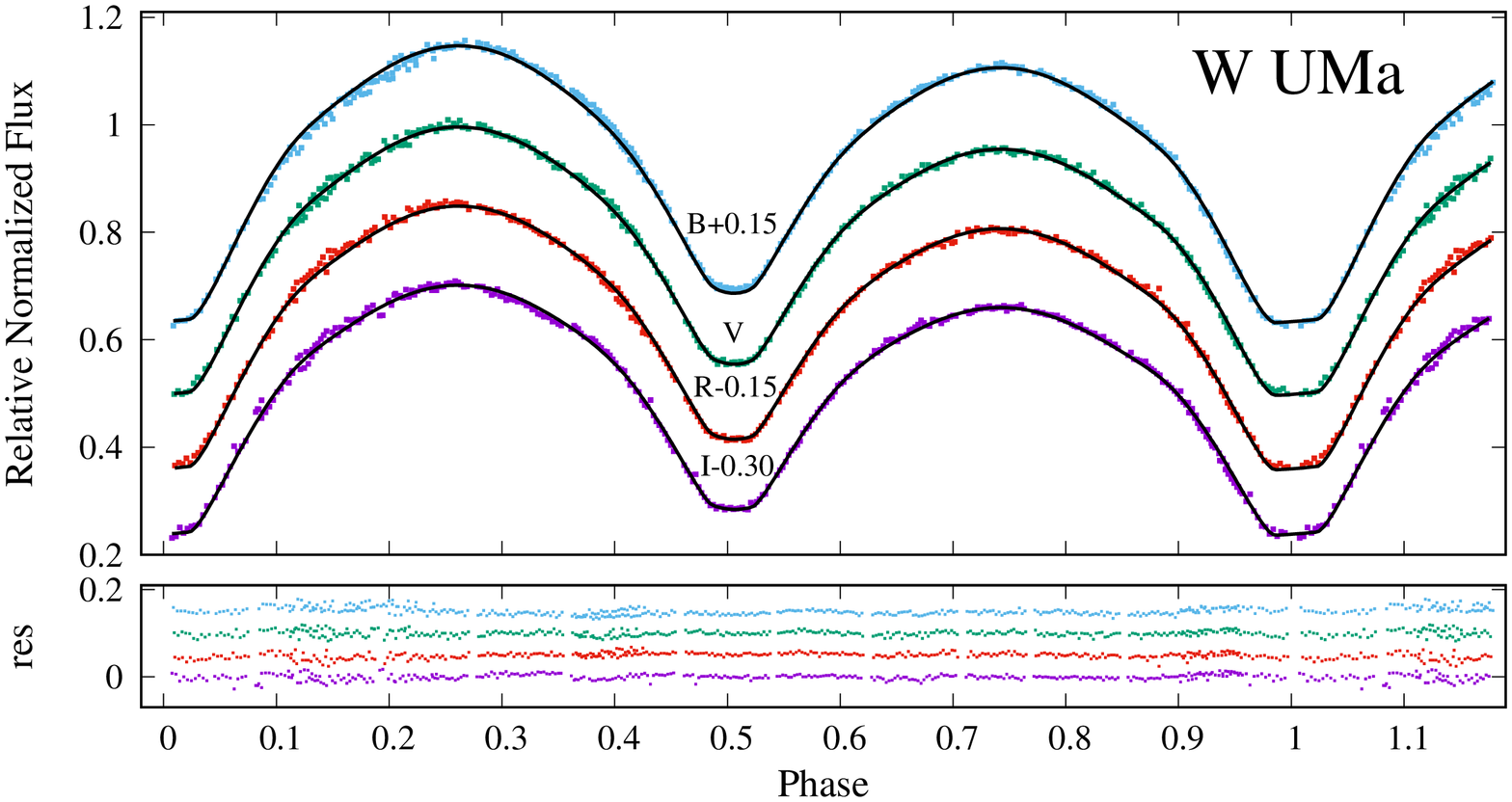}
 \caption{The same as in Fig. \ref{FigLC1} but for AK Her, V502 Oph, V566 Oph, V2612 Oph, V1363 Ori, V351 Peg, V357 Peg, Y Sex, V1123 Tau, and W UMa.}
\label{FigLC2}
\end{figure*}

\begin{figure*}
\includegraphics[height=5.8cm,scale=1.0,angle=270]{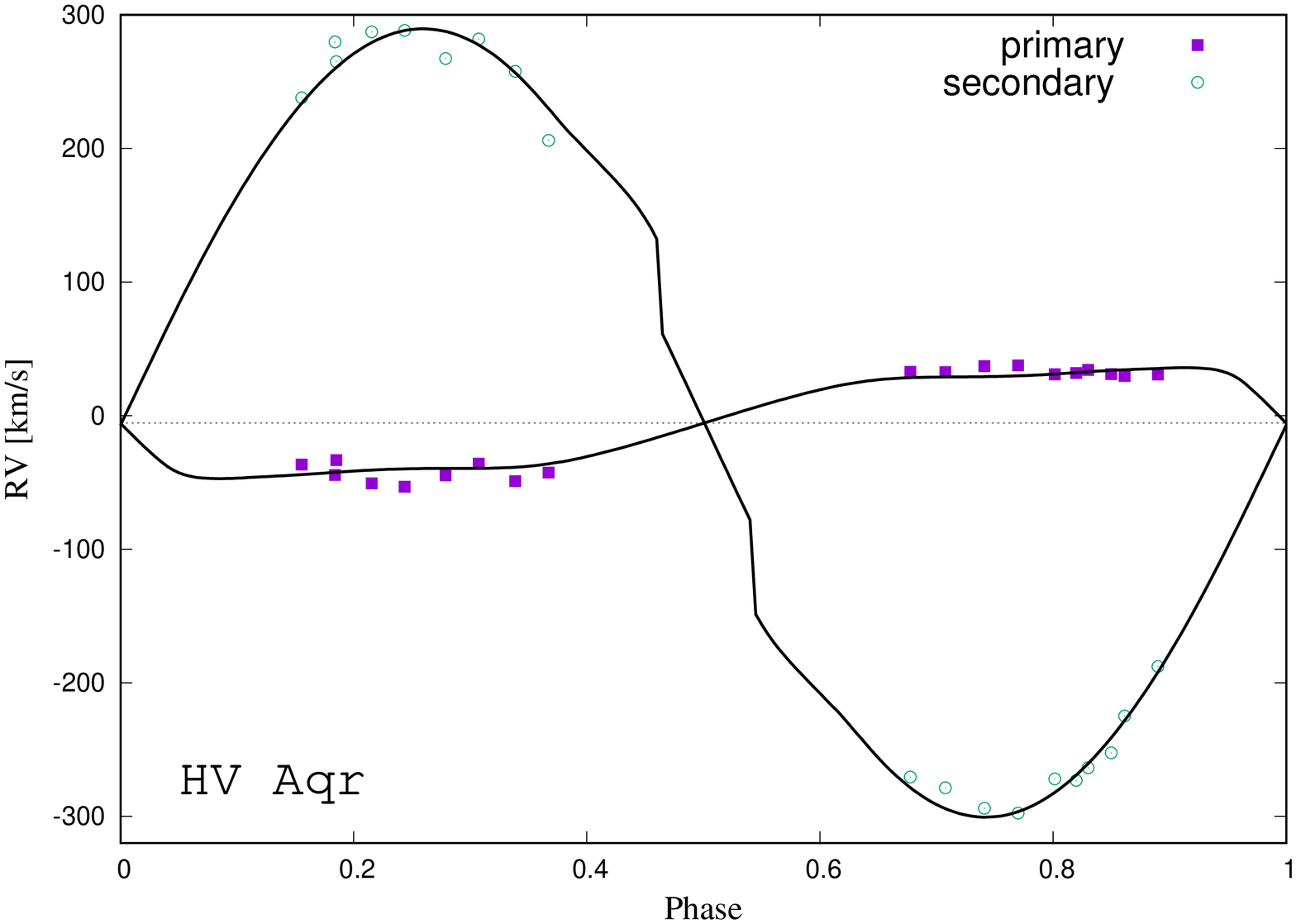}
\includegraphics[height=5.8cm,scale=1.0,angle=270]{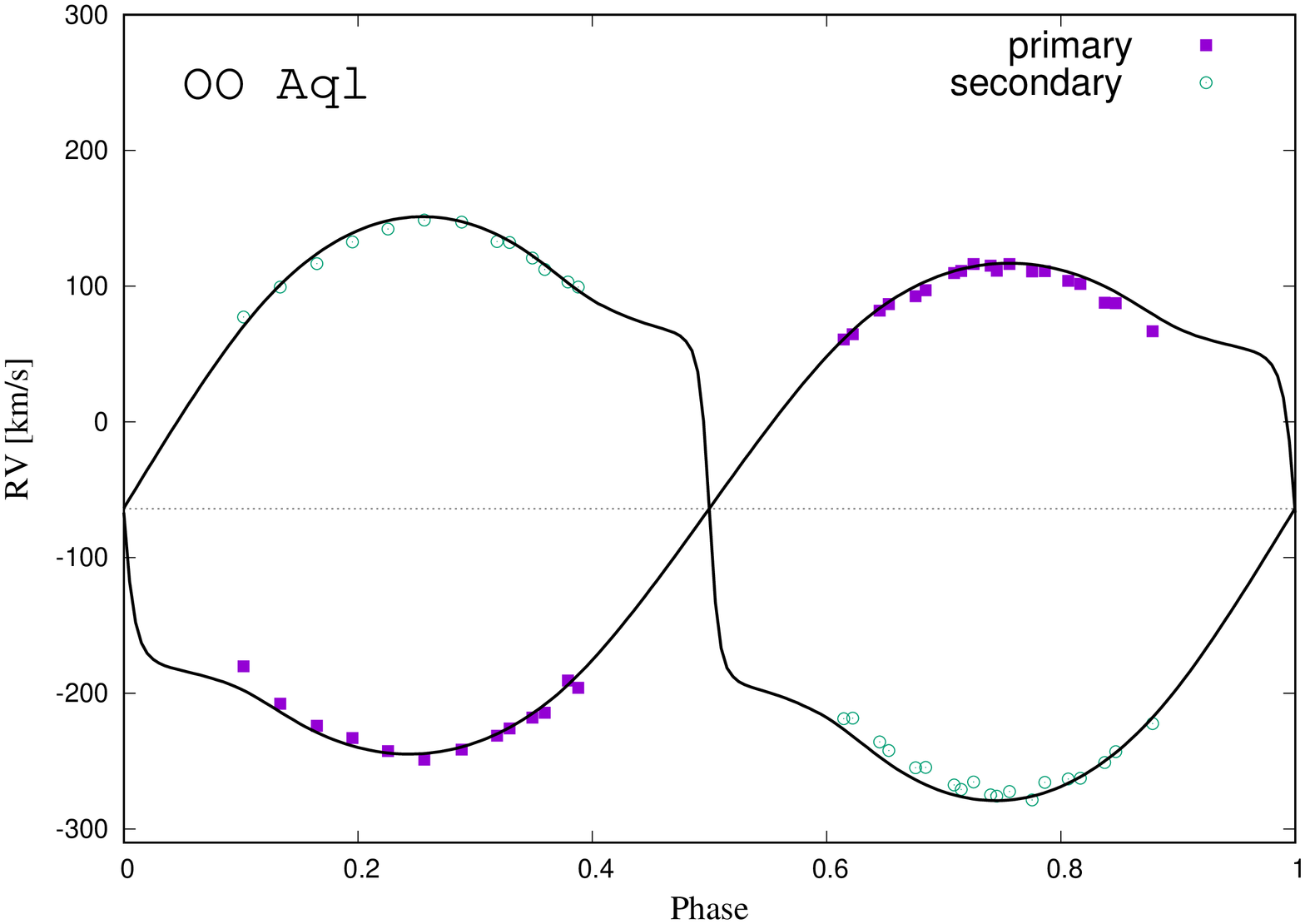}
\includegraphics[height=5.8cm,scale=1.0,angle=270]{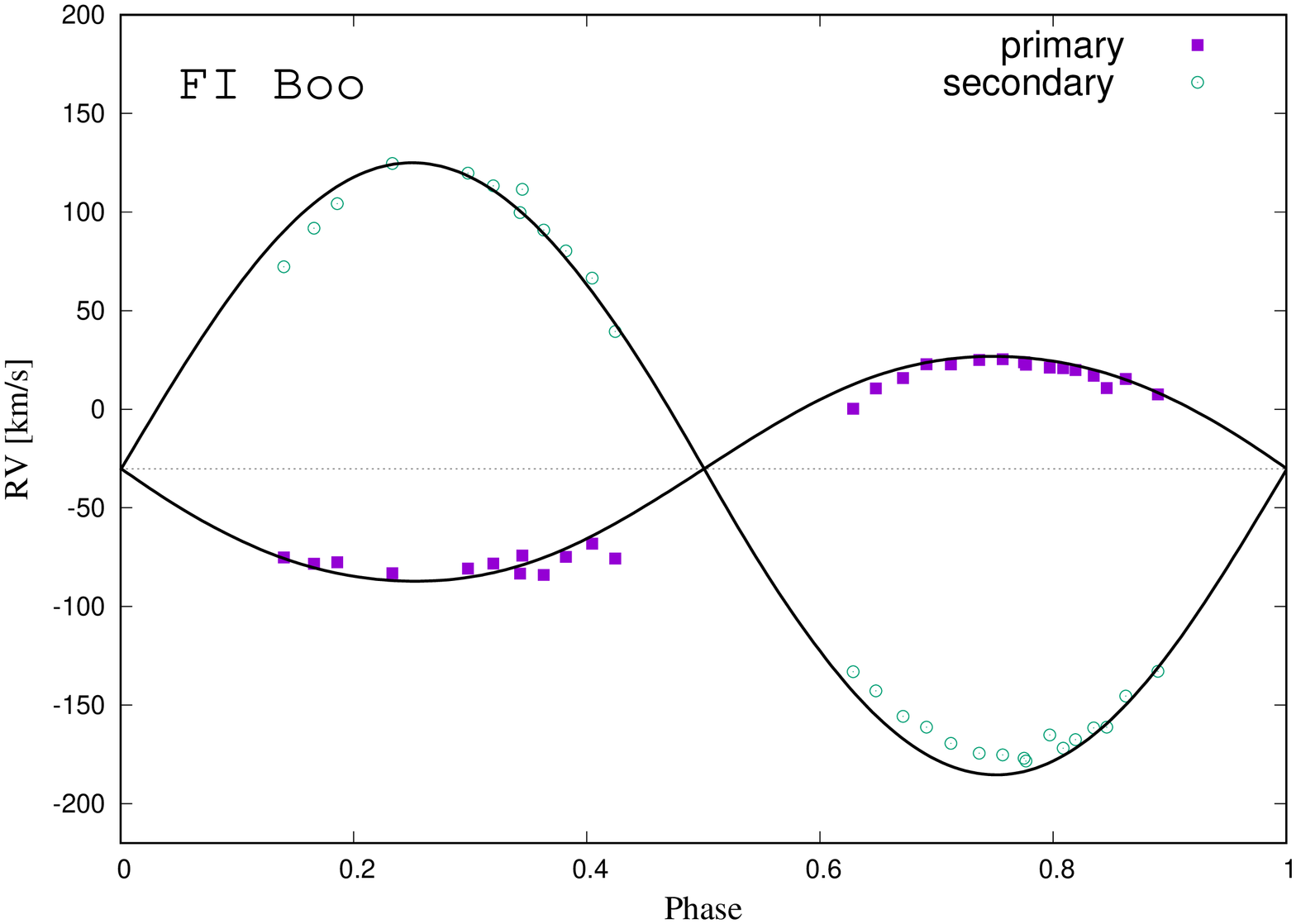}
\includegraphics[height=5.8cm,scale=1.0,angle=270]{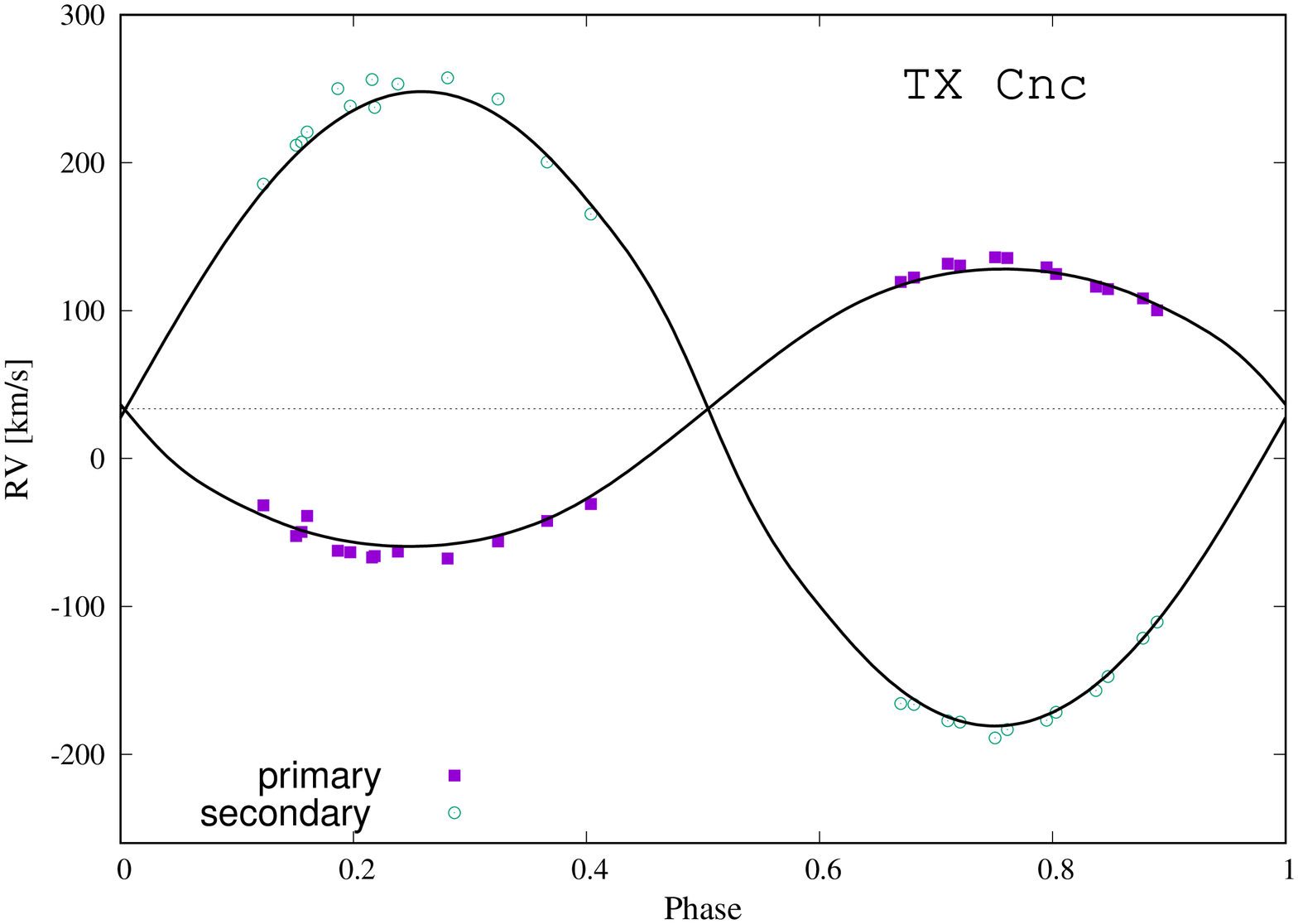}
\includegraphics[height=5.8cm,scale=1.0,angle=270]{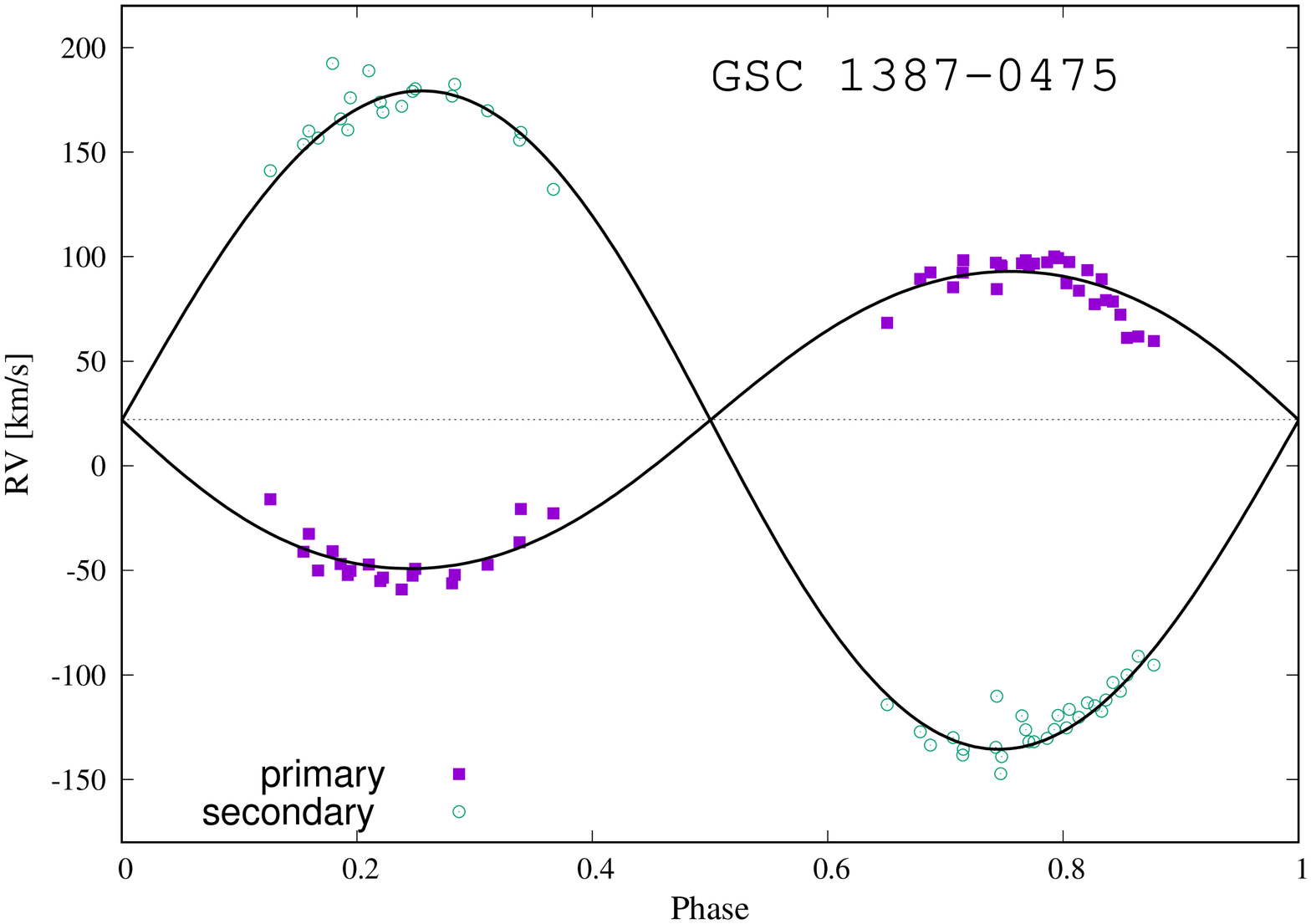}
\includegraphics[height=5.8cm,scale=1.0,angle=270]{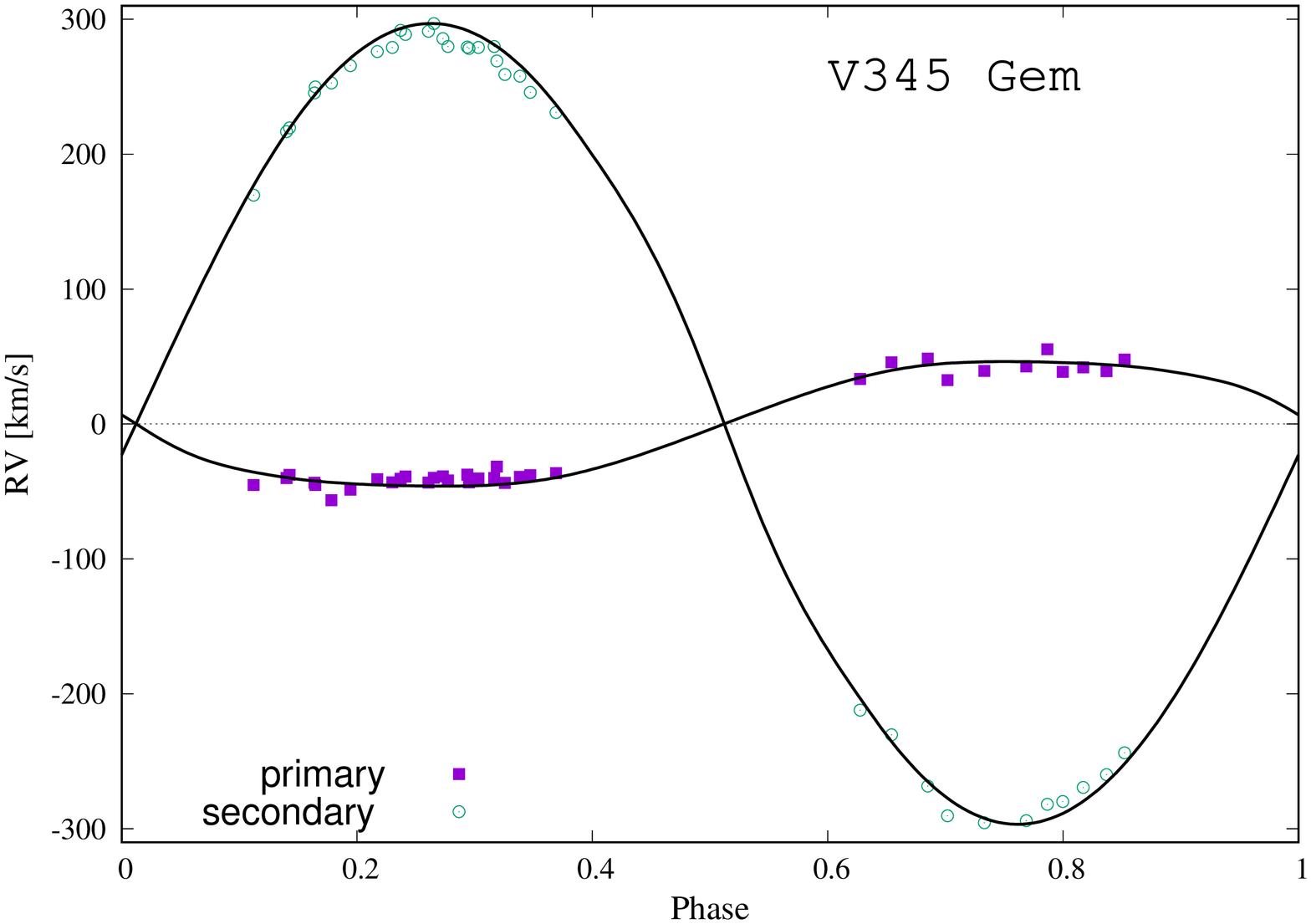}
\includegraphics[height=5.8cm,scale=1.0,angle=270]{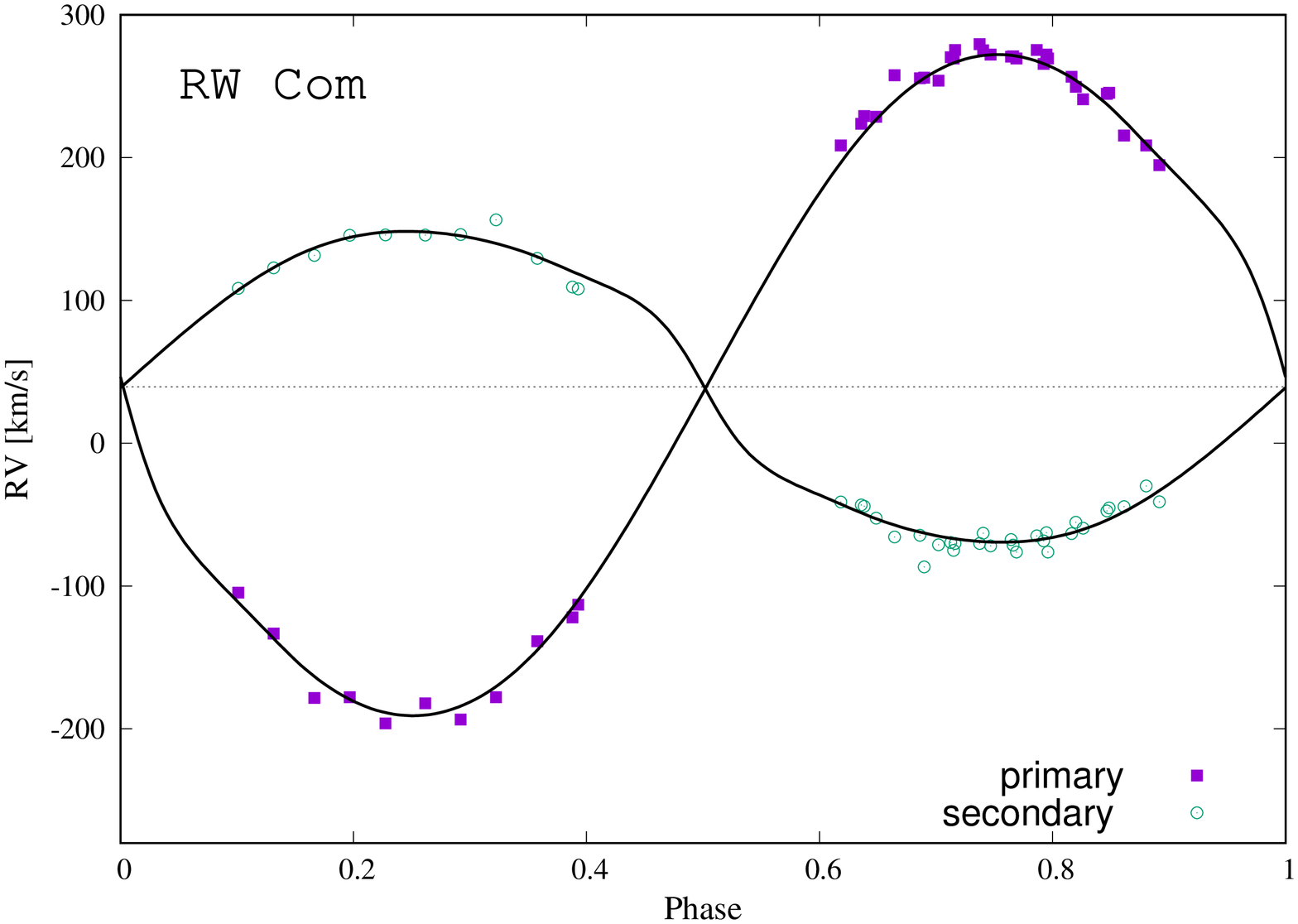}
\includegraphics[height=5.8cm,scale=1.0,angle=270]{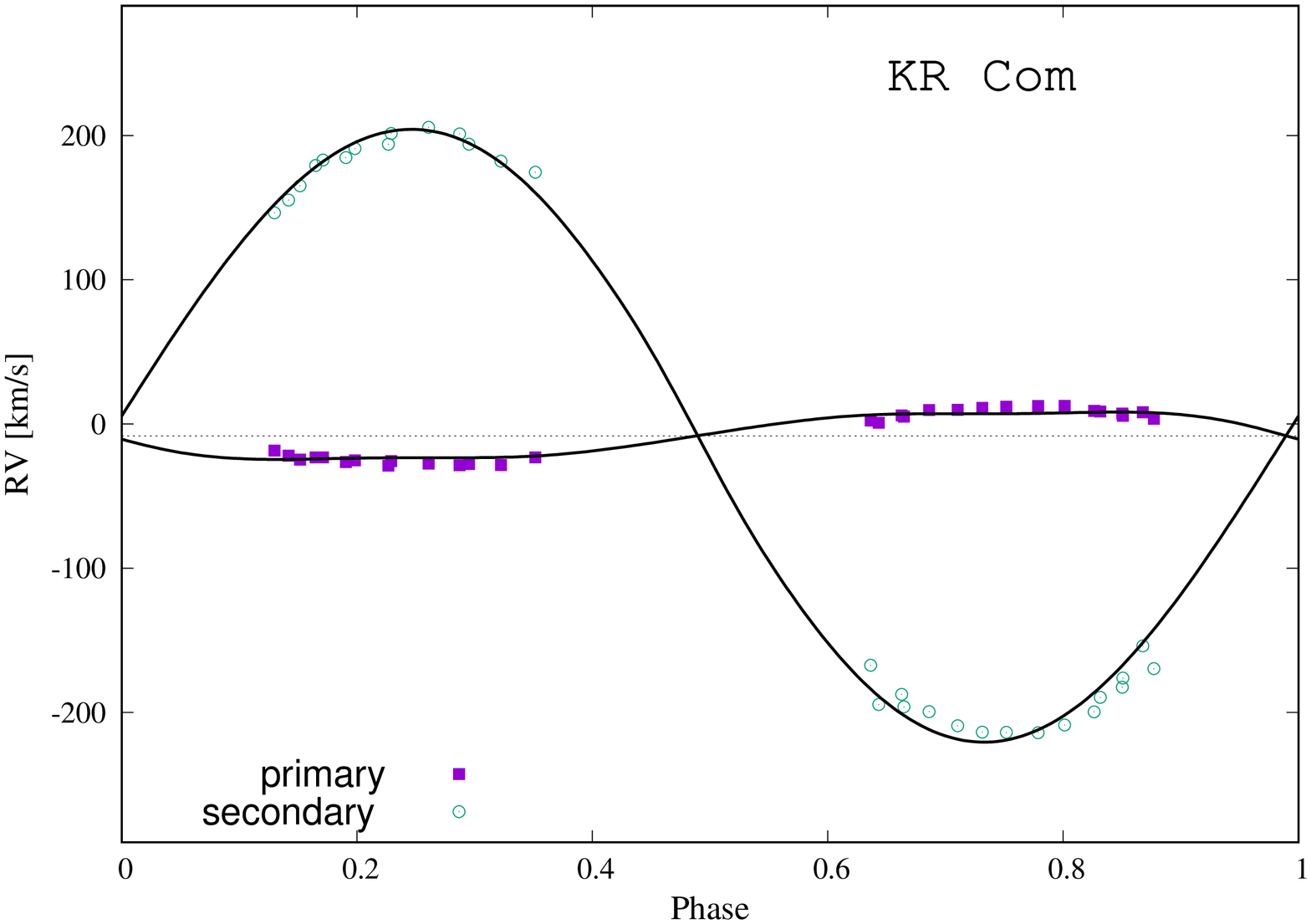}
\includegraphics[height=5.8cm,scale=1.0,angle=270]{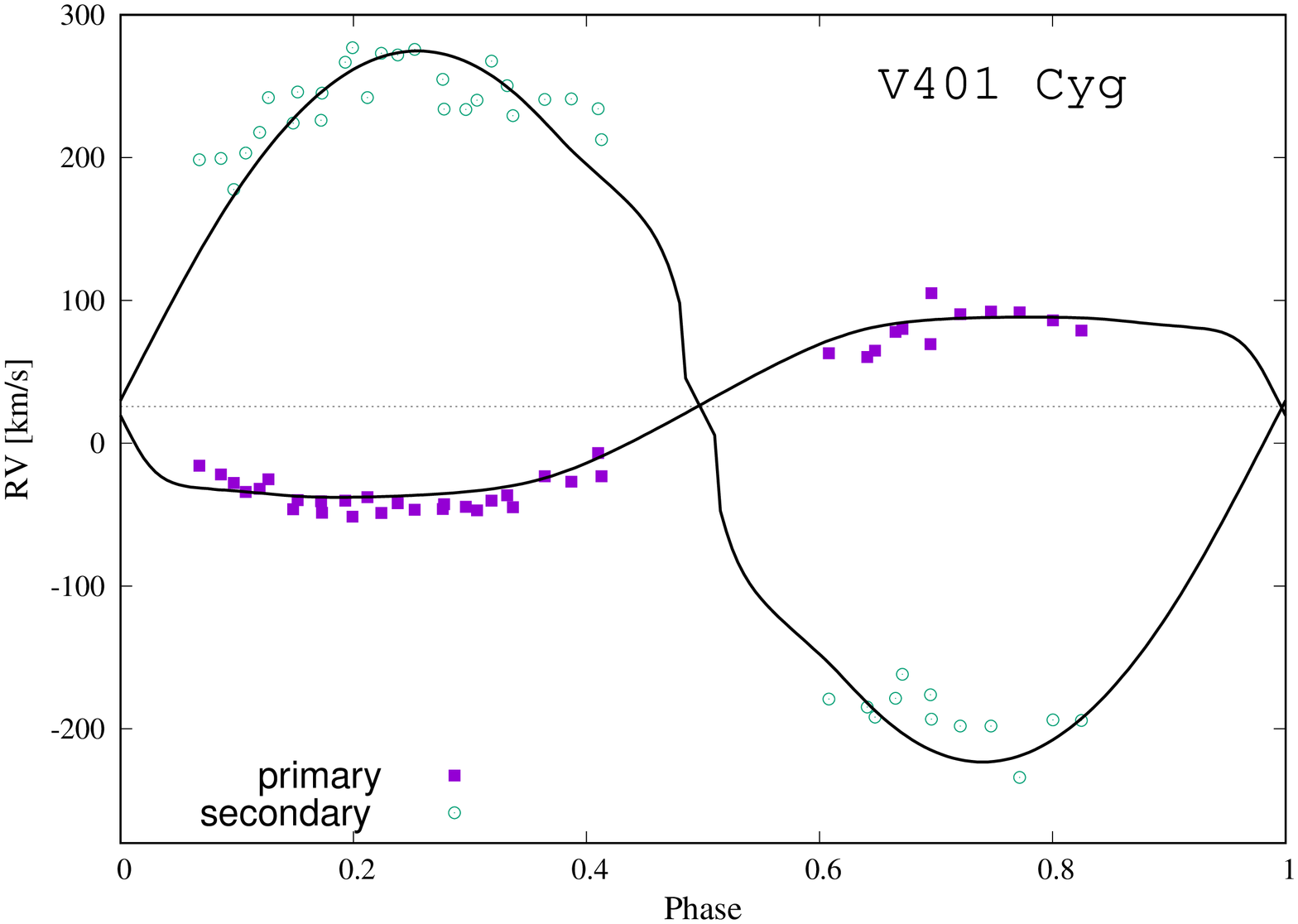}
\includegraphics[height=5.8cm,scale=1.0,angle=270]{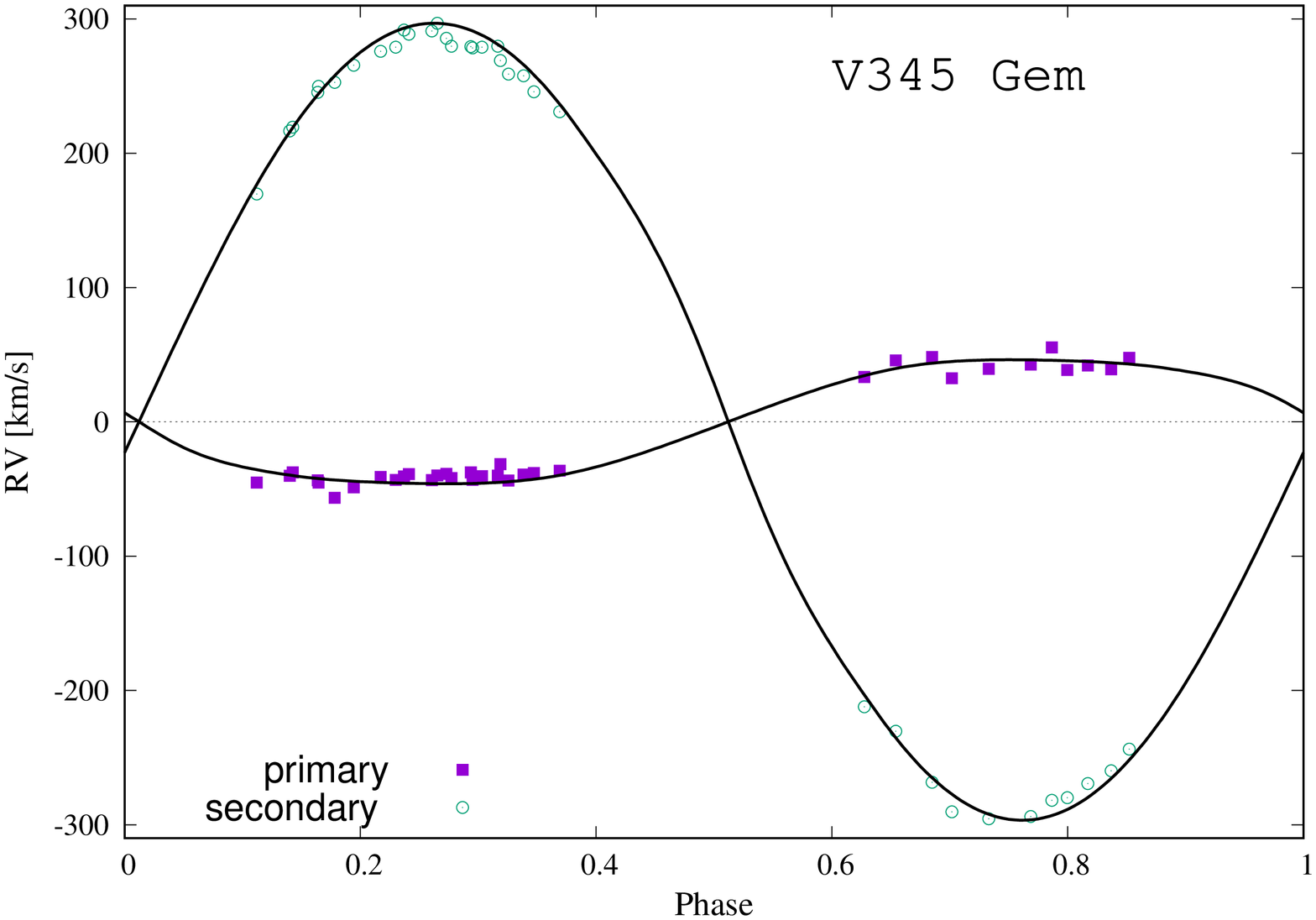}
\includegraphics[height=5.8cm,scale=1.0,angle=270]{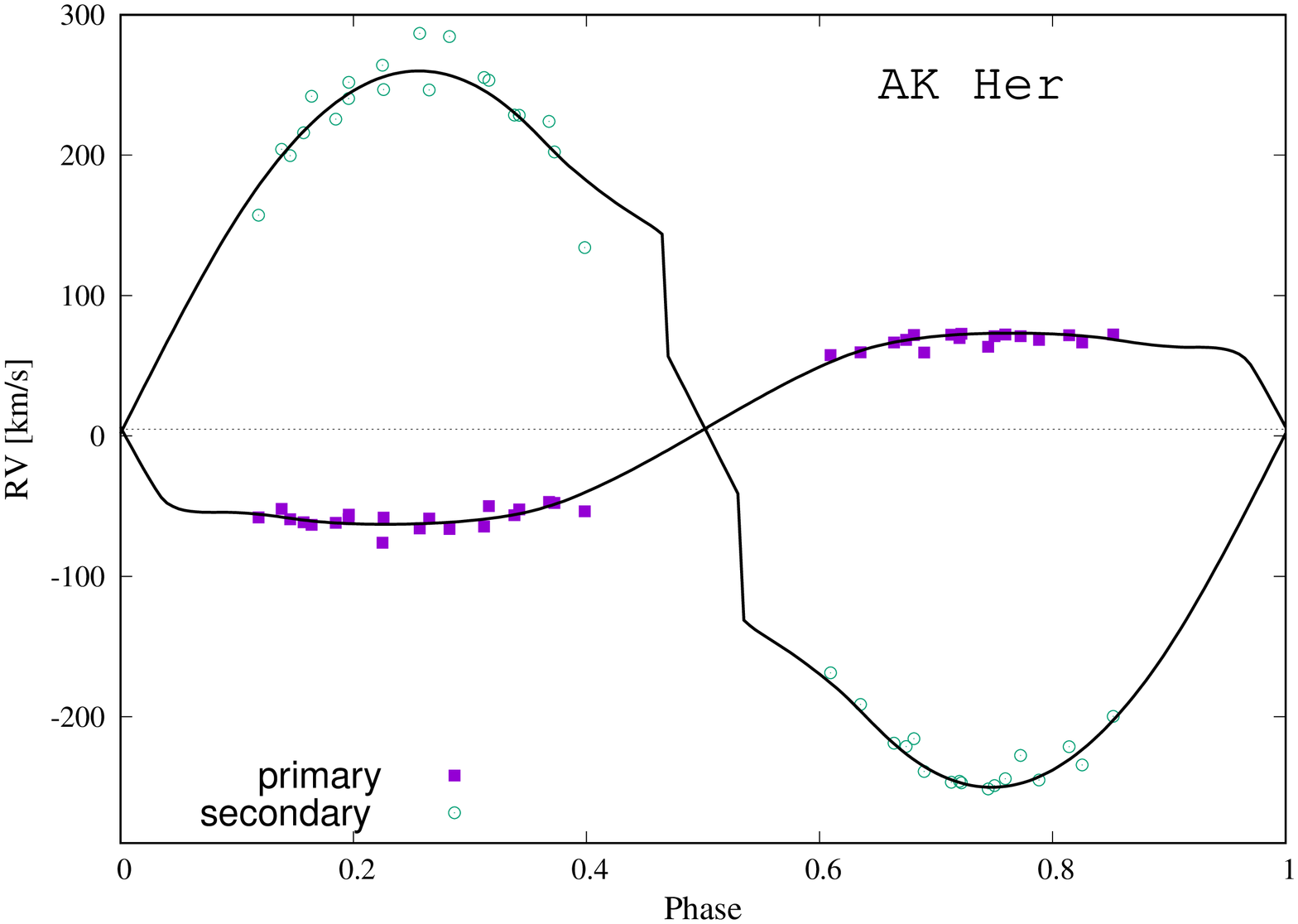}
\includegraphics[height=5.8cm,scale=1.0,angle=270]{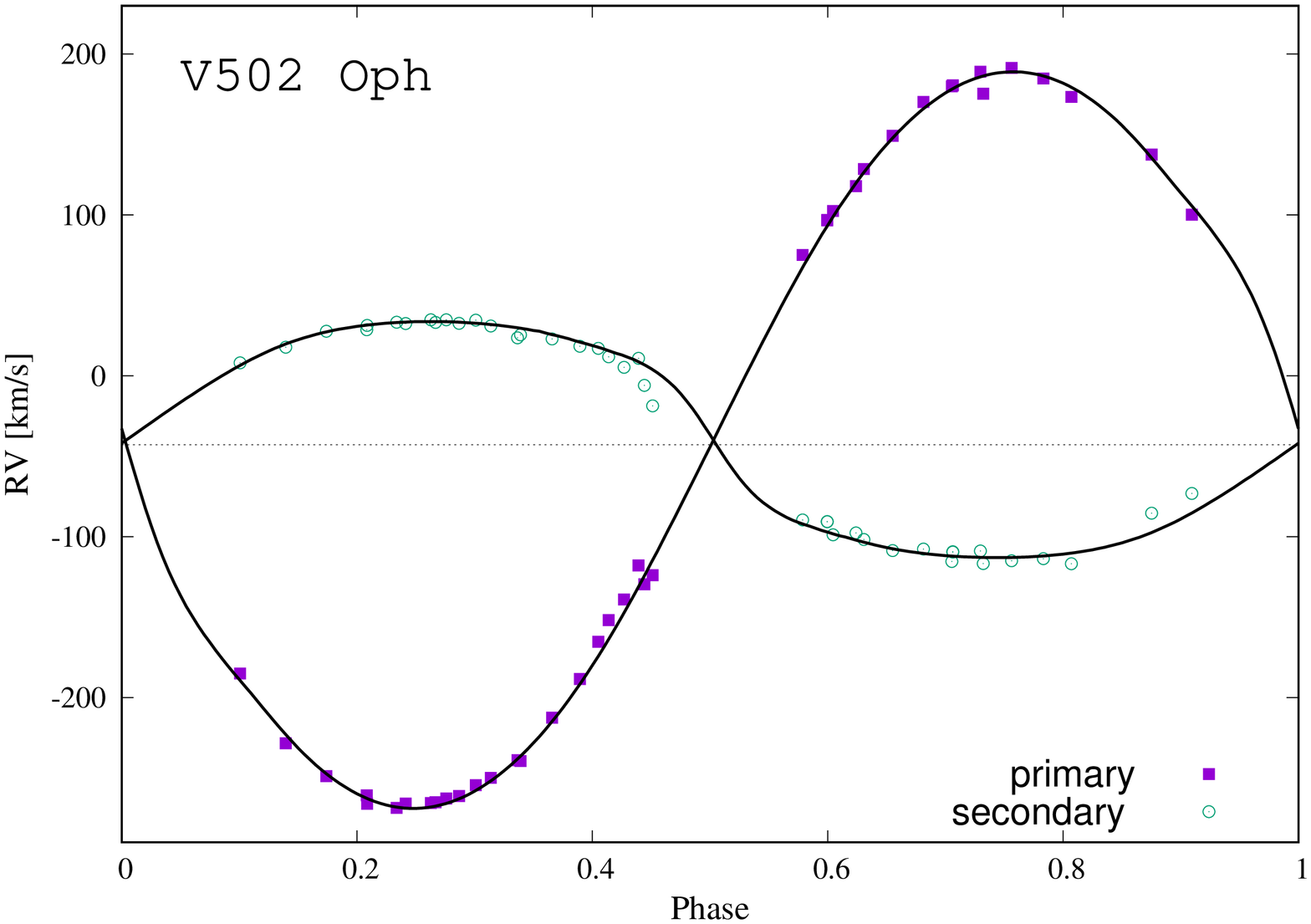}
 \caption{Radial velocity (RV) curves and observed data from the {\it DDO Program}
 for HV Aqr, OO Aql, FI Boo, TX Cnc, OT Cnc, EE Cet, RW Com, KR Com, V401 Cyg, V345 Gem, AK Her and V502 Oph,
 plotted versus orbital phase. Symbols represent individual observations, while theoretical
 curves are shown as continuous lines, and include the circular orbit with the proximity effects. The systemic velocity
 ($\gamma$) is shown as horizontal dashed line.}
\label{FigRV1}
\end{figure*}

\begin{figure*}
\includegraphics[height=5.8cm,scale=1.0,angle=270]{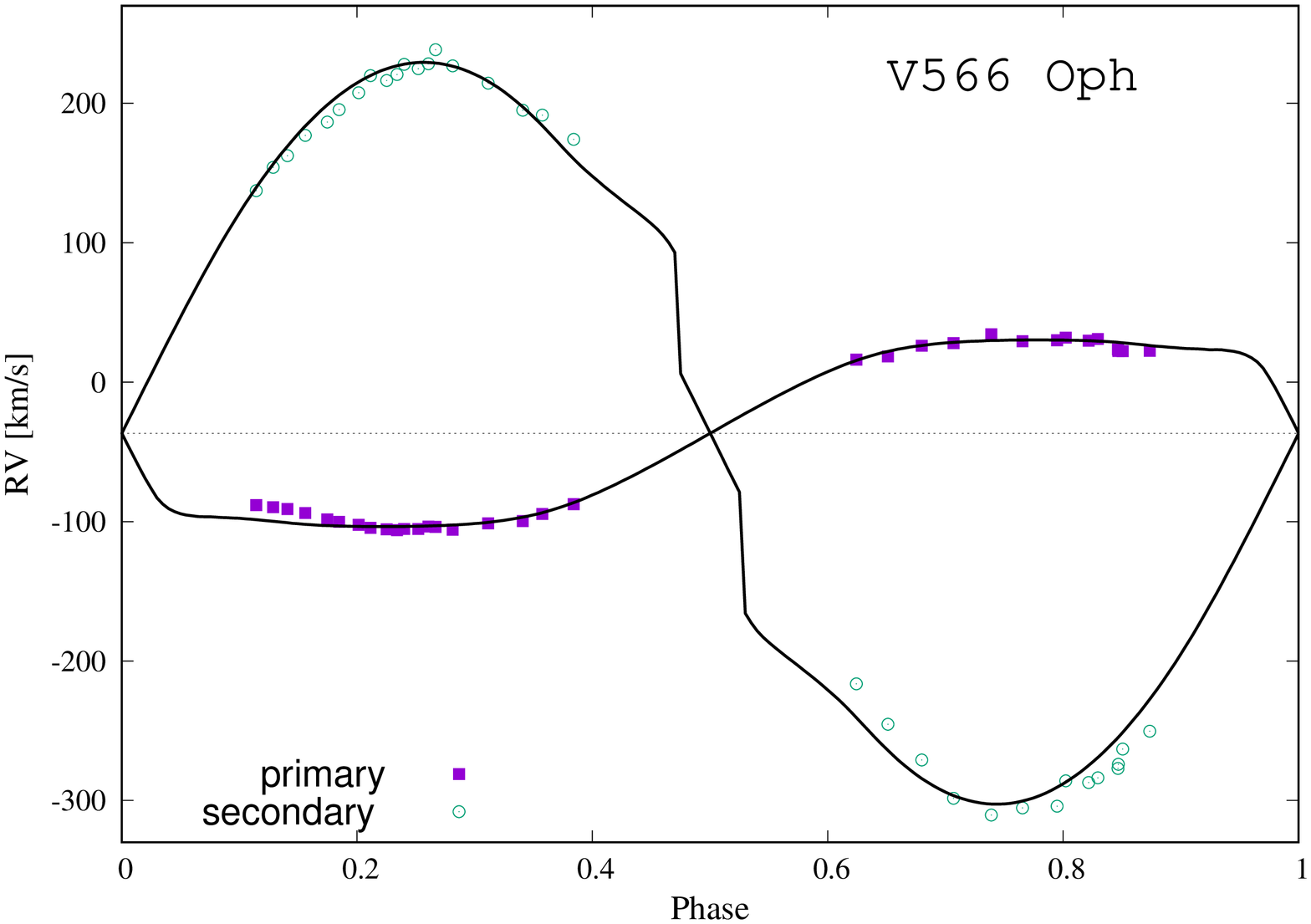}
\includegraphics[height=5.8cm,scale=1.0,angle=270]{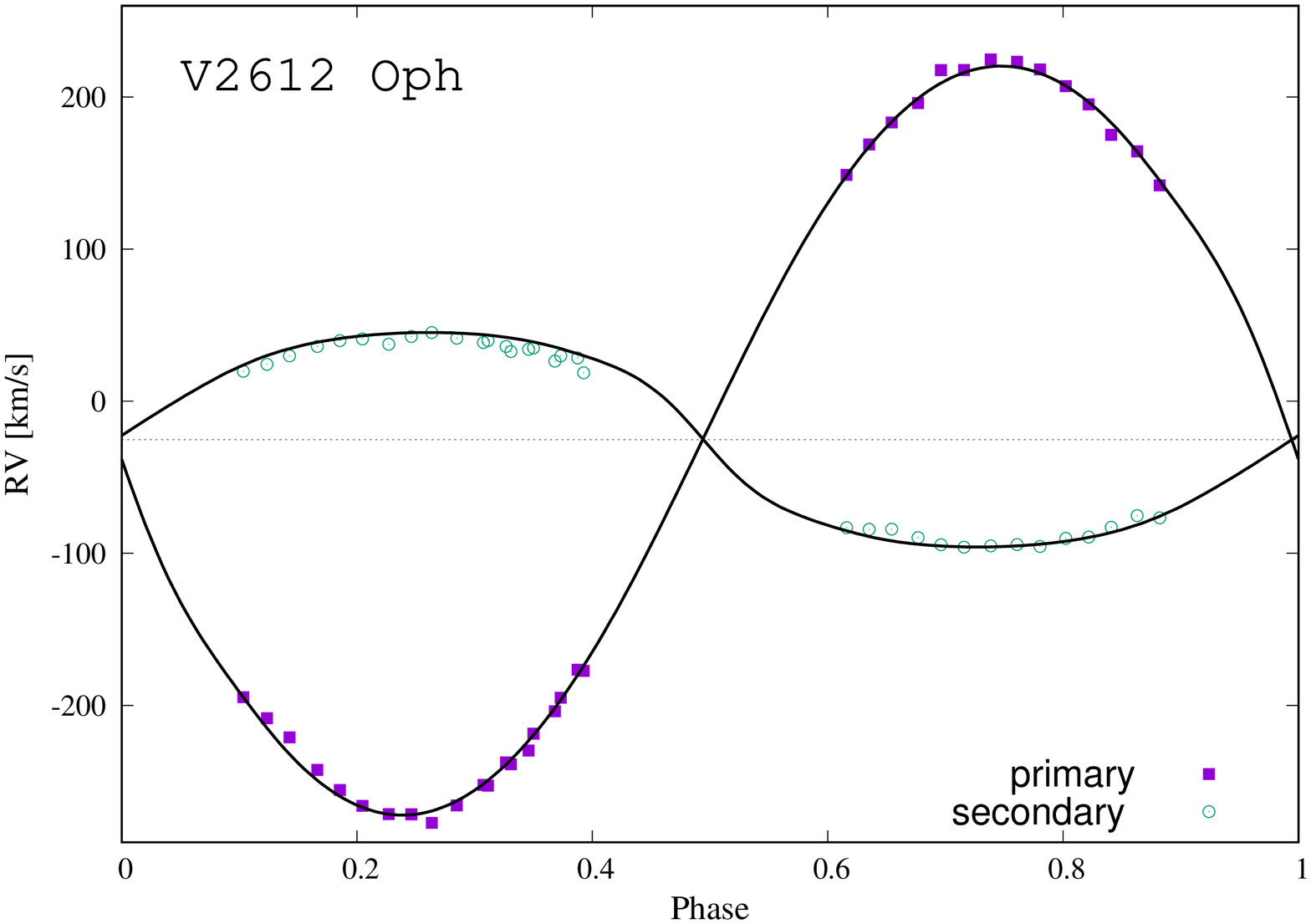}
\includegraphics[height=5.8cm,scale=1.0,angle=270]{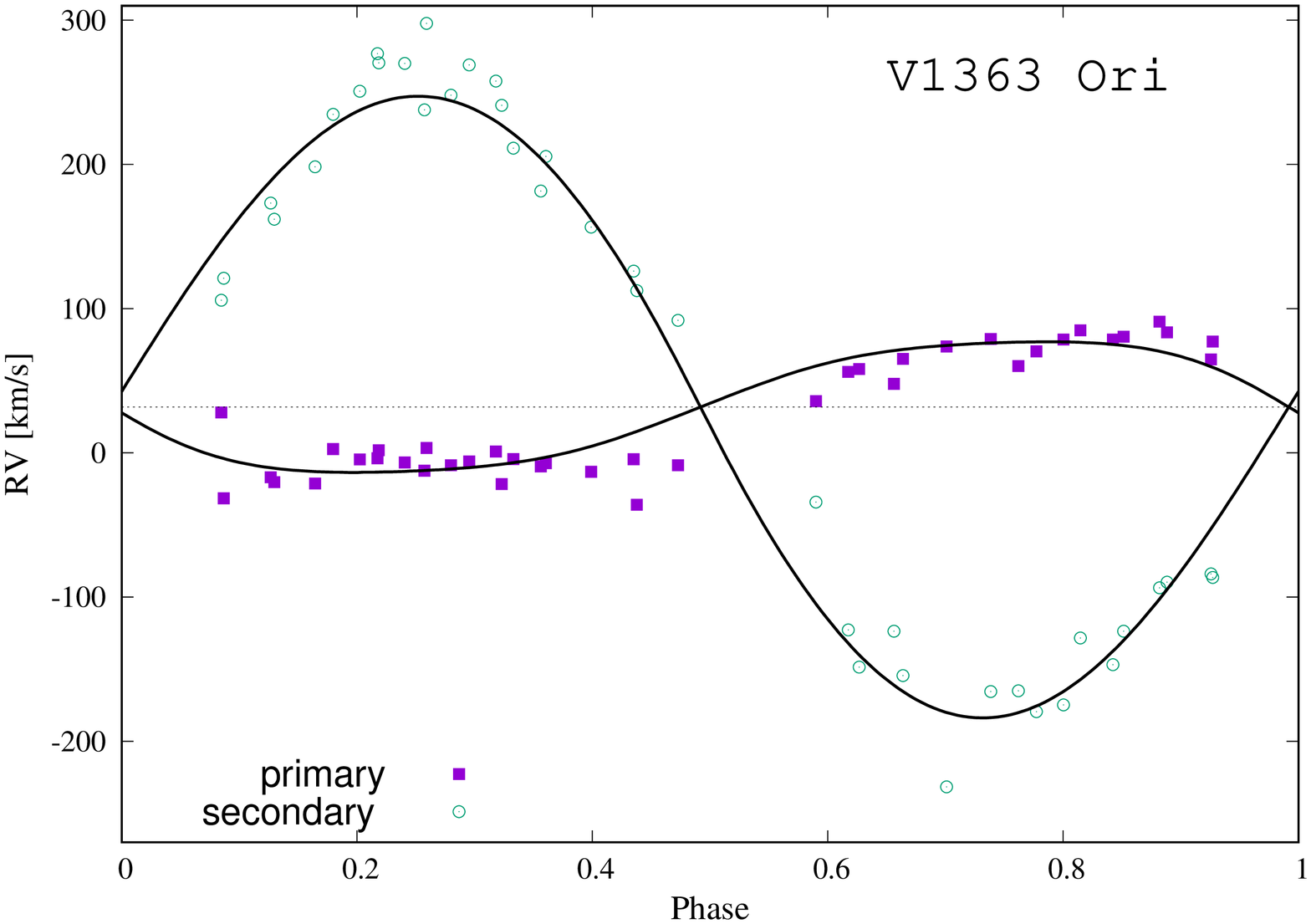}
\includegraphics[height=5.8cm,scale=1.0,angle=270]{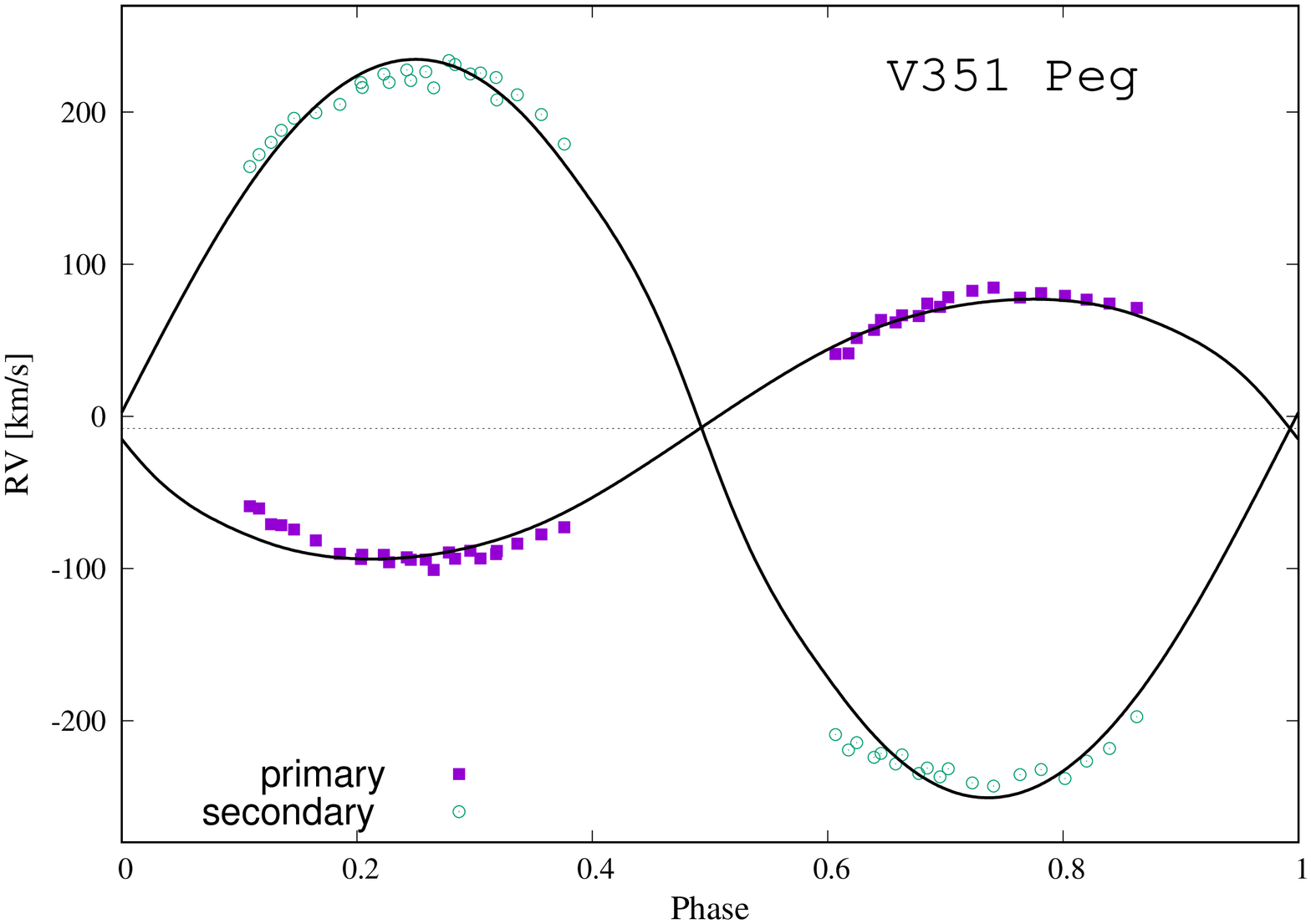}
\includegraphics[height=5.8cm,scale=1.0,angle=270]{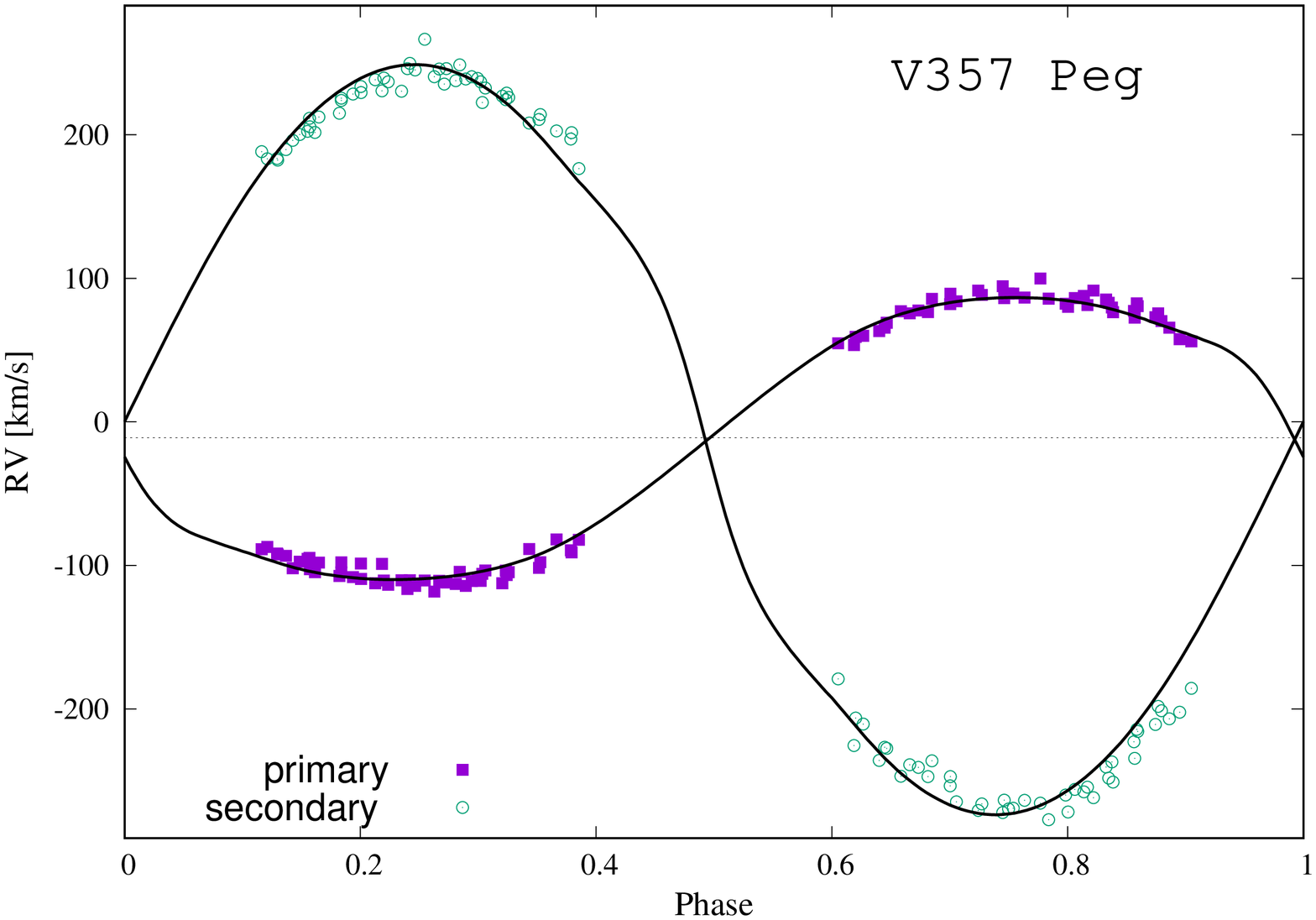}
\includegraphics[height=5.8cm,scale=1.0,angle=270]{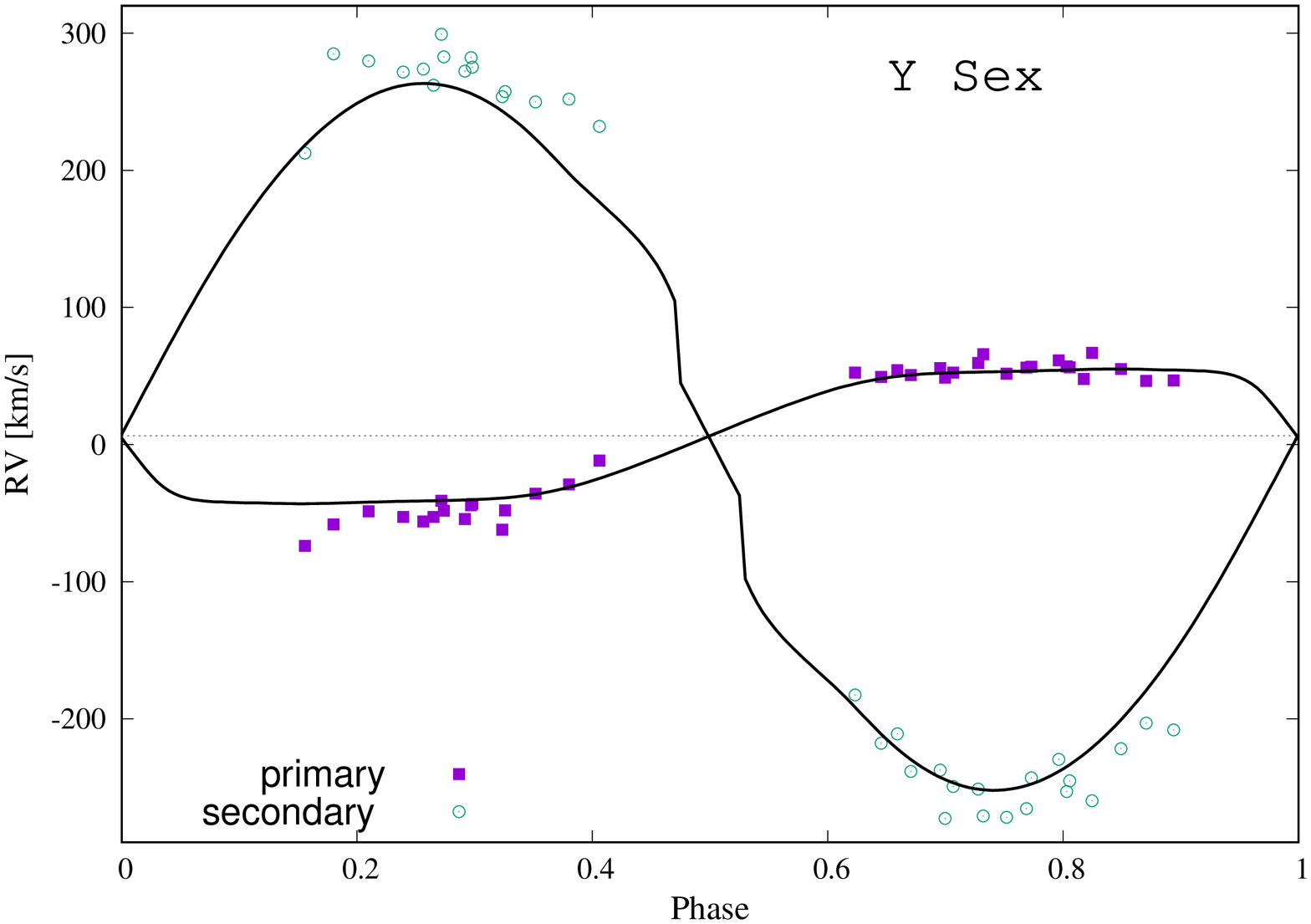}
\includegraphics[height=5.8cm,scale=1.0,angle=270]{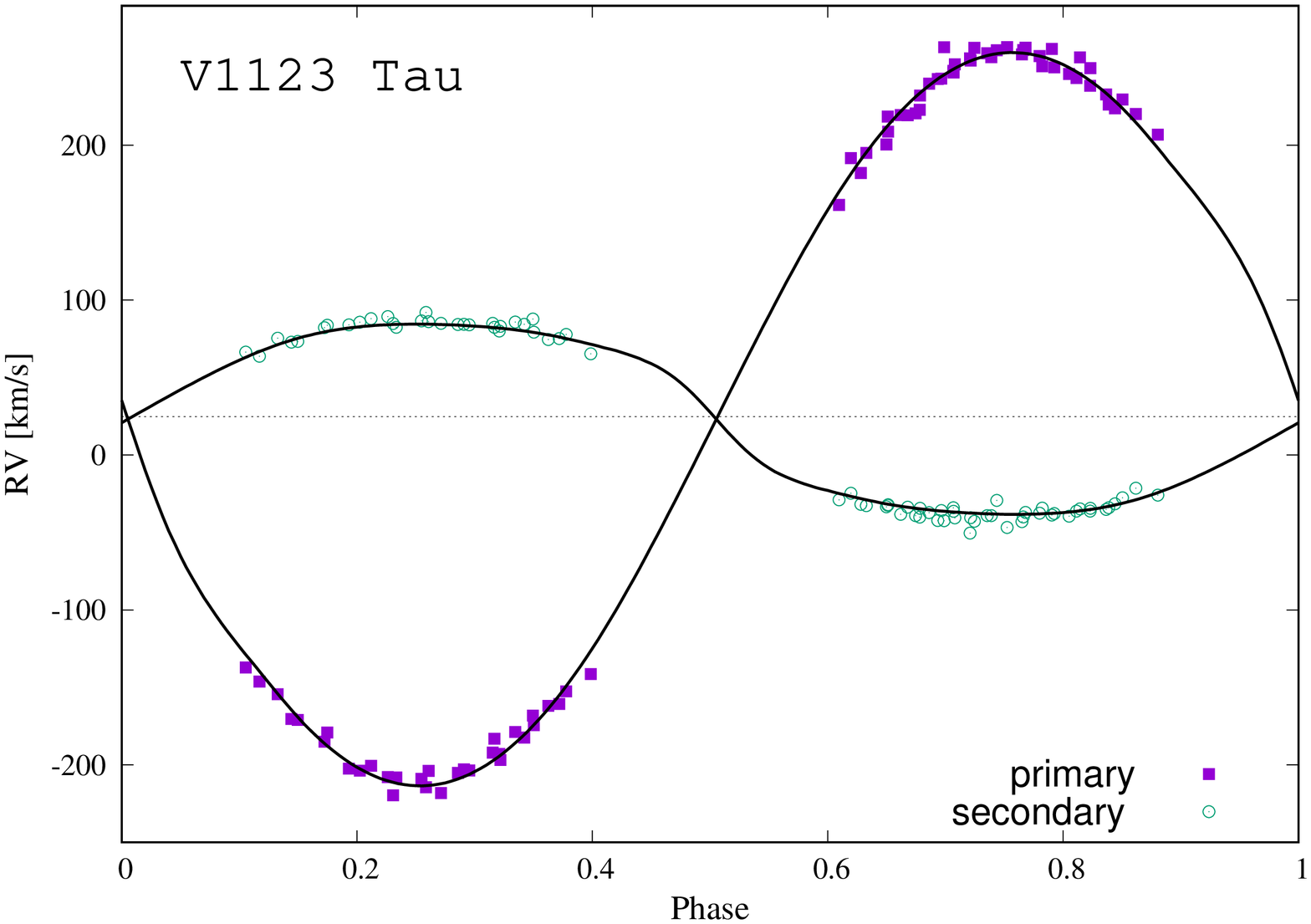}
\includegraphics[height=5.8cm,scale=1.0,angle=270]{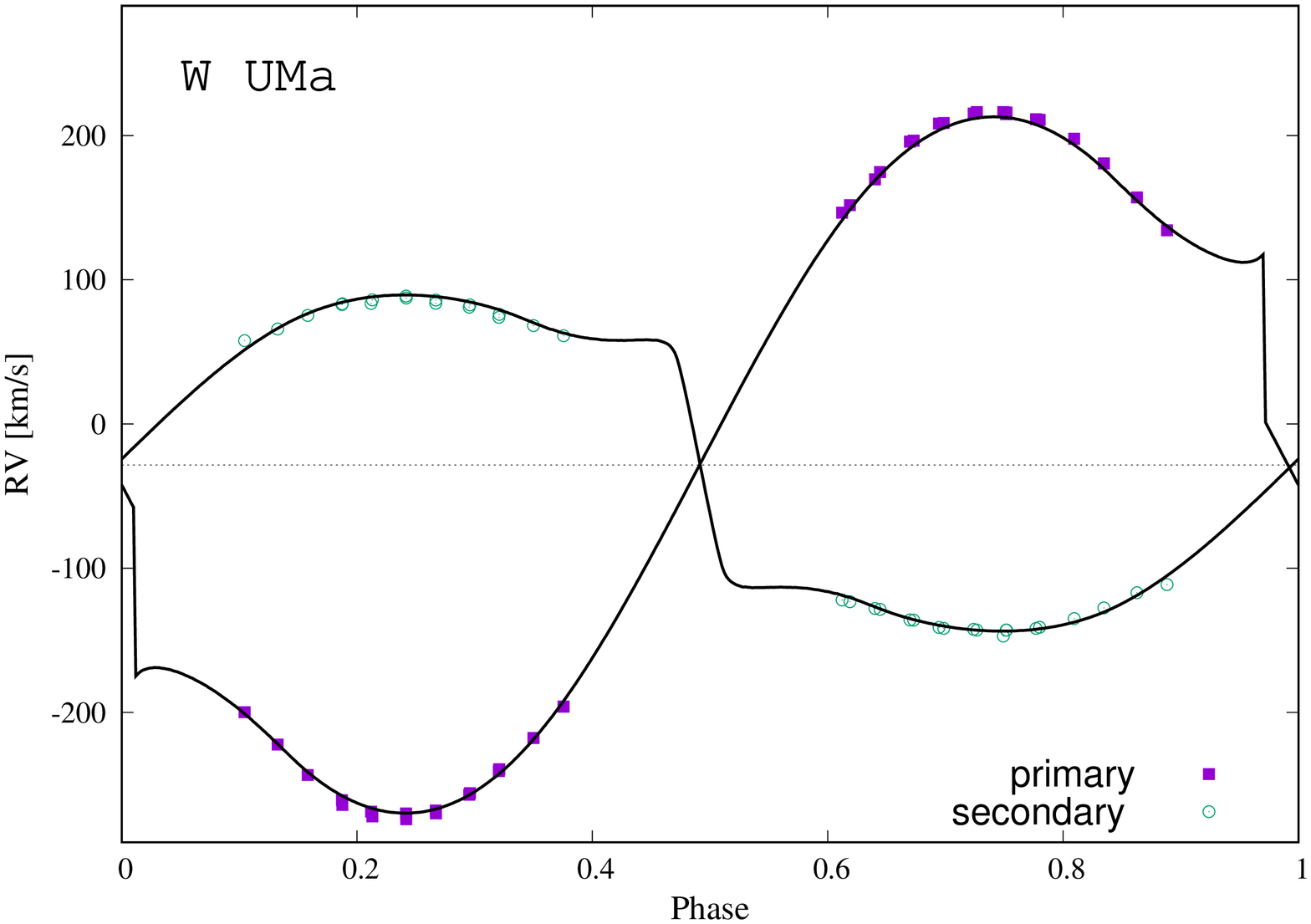}
 \caption{The same as in Fig. \ref{FigRV1} but for V566 Oph, V2612 Oph, V1363 Ori, V351 Peg, V357 Peg, Y Sex, V1123 Tau, and W UMa.}
\label{FigRV2}
\end{figure*}

Once the best solution was found, we proceeded with the next step in which we used the W-D code with
the original differential correction search method to fit the radial velocities alone. In this second
step only parameters relevant to the orbit and the mass ratio were adjusted. This procedure allows
correction of the mass ratio ($q$) for proximity effects. With the new value of $q$, we performed the
light curve modelling again and determined an improved set of system parameters. This procedure was
repeated, usually two or three times, until the correction to the mass ratio  became smaller
than its error. The changes of the original $q$ values are usually small, in some cases negligible,
and there is no noticeable difference in the RV curves. However, the corrected $q$ values represent
the observed radial velocity curves more realistically and are in accordance with the physical approach.
The observed light curves are plotted together with the resulted models, as shown in
Figs. \ref{FigLC1} and \ref{FigLC2},  while the radial velocity curves and their fits are
shown in Figs. \ref{FigRV1} and \ref{FigRV2}.

The parameters resulting from  the modeling are given in Tables \ref{TabRes1}, \ref{TabRes2},
\ref{TabRes3} and \ref{TabRes4} and they were used for determination of the physical parameters
of the components of the studied systems.


\section{Results for individual systems}

In the search for the best solutions, we initially tried to limit the number of free parameters
to a minimum. Due to asymmetries in the observed light curves, we were not able to
obtain sufficiently good fits for half of the systems analysed in this paper.
For those showing obvious asymmetries due to the O'Connell effect \citep{oco1951},
we included a cool spot on the surface of the more luminous component.
By adding a single spot, the number of free parameters increases by four (the spot
longitude, latitude, radius and temperature factor) and, as a consequence, these models resulted in a
significant improvement of the fit compared to the non-spotted solutions.
For all ten systems, a single spot was sufficient to reproduce the observed
light curves well.

Typically, a cool spot with a radius of from $20$ to $46^\circ$ and a temperature factor
of about 0.9 was needed to account for the observed asymmetries. For three systems
the models resulted in somewhat larger spotted areas, while in  V401~Cyg,
the best solution resulted in a spot which covers half of the hemisphere of the primary star.


\subsection{HV~Aqr}
The mass ratio of HV~Aqr is one of the smallest found so far,
even among the class of contact systems with extremely small mass ratios, e.g.
AW~UMa, FG~Hya, NSV~223 (DZ~Psc), GR~Vir, V410~Aql, KR~Com (current study).
In general, a small mass ratio together with a high orbital inclination produces
total eclipses when the less massive component is obscured,
which is also the case for HV~Aqr. This allows an accurate determination of the parameters
of the system \citep{gaz2007}.
Since there is no obvious difference in the amplitude of the maxima in our light curve,
we initially tried a non-spotted model. However, the resulting solution was not satisfactory.
While theoretical light curves fitted the observed ones quite well in $V$ and $R$ bands, the depths
of the minima were not correctly reproduced in $B$ and $I$ bands. Additionally, there were also small
discrepancies outside the minima. Therefore, we first added a cool spot to the model.
As a result, the fit was improved mostly in the parts of the light curve outside of the minima.
The depth and shape of the synthetic light curve in both minima were still not as observed.
In the next step, two independent trials were made: a fit with two spots and another one with one spot
and a third light as a free parameter. The two spot solution failed to reproduce the shape of the $I$-band light curve.
The solution with a third light \citep[][also justified by the adaptive optics findings]{ruc2007b},
produced the best match of the model to the observed light curve.
Since the tertiary component is of K2/3V spectral type, the level of third light is negligible in
the $B$ and $V$ bands, turning out be significant only in the infrared. Our solution gives a large
fill-out  factor of $f=74$~per cent, which is in agreement with the very small mass ratio,
indicating that the system could be a rather evolved one.


\subsection{OO~Aql}
Earlier studies have claimed a wide range of colour indices and base on these, the resulting spectral
types given for OO~Aql were G5V \citep{rom1956} or K0V \citep{hil1975}. In this study
we adopted the most recent estimation by \citet{pri2007}, who gave an earlier spectral
type of F9V, with the conclusion that the system was affected  by significant reddening.
Our analysis of OO~Aql produced a reasonable fit of the observed LCs with a non-spotted model.
We also checked a solution with a cool spot, however this did not improve much the quality of the fit.
Therefore, the small discrepancies, more obvious in $B$ band, may be due to atmospheric effects,
which are not fully corrected for.
We derived a contact configuration for this binary and the resultant fill-out factor
is small, the common envelope fills only about 17~per~cent of the space between the
Roche and the outer critical lobes.

\subsection{FI~Boo}
The low amplitude of FI~Boo indicates either a low orbital inclination or a significant
contribution from a third light source. Indeed, \cite{DAngelo2006} found a companion to this system,
but its light contribution was estimated to be very small, not larger
than about 1~per~cent of the total light.
Our first solution did not include a third light. We arrived at a model with an
inclination lower than 50$^{\circ}$ and a marginally contact configuration with a fill-out
factor less than 1~per~cent. This fact may indicate that this system is a relatively new contact binary.
We also made another trial model, adding a third light parameter to account for a possible companion.
The best solution resulted in a rather high third light contribution of about 30 per cent,
contradicting the previous results \citep{DAngelo2006}.
Obtaining a unique solution with a third light turns out to be very difficult using ground based
observations due to the very low amplitude of the light curve variations and correlation of the third
light with other crucial parameters, particularly the orbital inclination and the geometry of the system.
Therefore, we adopted a solution without a third light and argue that this is the best approximation that
can be made with current data.
The fit is perfect in the $V$ band, with only a small discrepancy near the primary minimum
in the $R$ band. Only the fit in $B$ band, in which the atmospheric
effects are more significant, does not closely resemble the observations.
The configuration of the system may be near-contact (with components very close
to filling the Roche lobe) or in marginal contact.


\subsection{TX~Cnc}
The light curve of TX~Cnc exhibits an obvious O'Connell effect (the first maximum
is higher than the second one). Therefore, a cool spot was included into the model.
A good fit was obtained requiring no third light, in agreement with
conclusions published by \citet{zha2009}.
We arrived at a contact configuration with an intermediate fill-out factor
of 19~per cent and  with a spot located near the equatorial region.

\subsection{OT~Cnc}
The very small amplitude of variability (0.09~mag) of OT~Cnc (originally designated as GSC~1387-0475) is a combination of the low
orbital inclination and the presence of an additional component orbiting the contact system.
\citet{ruc2008b} from spectroscopic observations estimated its contribution to be approximately
33~per cent. With this evidence a third light parameter was included into the model.
We completed the multicolour light curve of this system within four nights distributed over
about one and a half months.
Nonetheless, the light curve does not exhibit an O'Connell effect and we performed the modelling
assuming that there are no spots on the surfaces of the components. Our analysis resulted in a low
inclination of about $38^\circ$, and a contact configuration with a fill-out factor of 43~per cent.
The resulting third light contribution at the quadratures is between 30--40~per cent (from $B$ to $I$ bands,
respectively), in a good agreement with the value indicated by spectroscopy. Due to strong correlation
between orbital inclination and the third light parameter, we estimate the uncertainties of derived
parameters of the system to be large. OT~Cnc has very similar components to CC~Com, only
slightly less massive. However, a detailed comparison of these two systems can be hampered by
large uncertainties of parameters of the former.

\subsection{EE~Cet}
The light curves of EE~Cet analyzed in this study are rather symmetric without
a noticeable O'Connell effect and did not require the addition of a spot into
the model. Since during the reduction of our photometric data  we excluded
the nearby visual companion, we derived a light curve solution without
a third light.
This simplified the search procedure and allowed for quick convergence.
As a result, a good fit to the observed light curves and a reliable
model for EE~Cet were achieved.

\subsection{RW~Com}
There was some disagreement concerning  the spectral type of RW~Com in the literature in
the early years of observations. The 2MASS infrared colour  ($J-K$ = 0.618~mag)
corresponds to a spectral type of K2V. This is inconsistent with
the G5--G8 spectral type estimated earlier by \citet{mil1985}. It is, however, in better
agreement with the very short orbital period of the binary and the estimation of
\citet{pri2009a}, who assigned it as K2/5V spectral type.
A spectral type of K3V was adopted in this study, with an uncertainty or 200~K.
The observed light curve of RW~Com shows obvious asymmetry: the first maximum is
lower than the second one. A non-spotted model was initially tried but, due to the
asymmetry, the resulting fit was poor. After adding a dark spot placed on
the secondary, there was a significant improvement of the fit. The resulting  model
required a large polar spot of about $46^\circ$ in radius.
We arrived at the marginally contact configuration with a fill-out factor
of only 3~per~cent. The small fill-out factor indicates that the system could be
a relatively young contact binary, at a stage similar to that of FI~Boo.

\subsection{KR~Com}
KR~Com is another binary with a very low amplitude of variations. This is due to
both strong light contamination by a close companion and a low orbital inclination.
We encountered severe problems in deriving a unique solution for this binary,
as rather similar fits were obtained for third light contributions between 30 and 60~per cent
differing in inclination by just few degrees.
This system was classified as G0IV \citep{ruc2002b}, however, considering that the
components of contact binaries belong to the MS, we adopted a G0V spectral type in this analysis.
As the final model we present one for which the derived third light contribution is
around 60~per~cent, close to the value determined by \cite{ruc2002b}. In our
model, the secondary star is slightly hotter and the system is in very deep contact,
fully filling in the space between the Roche and the outer critical lobes.
This agrees with the very small mass ratio, which indicates that the more
massive component ``swallows'' the less massive one, leading towards a merger.
With such parameters the system seems to be well-evolved, similarly to HV~Aqr.
In this study we found a temperature difference between the components of $\Delta$$T$ = 210~K,
contrary to significantly  larger one  found by \citet{zas2010}. We confirmed the orbital inclination
to be small ($i$ = 54.4$^\circ$) in accordance with the earlier studies. The slightly
higher value of the inclination derived in this study resulted in a smaller derived mass for both components.

\subsection{V401~Cyg}
The light curves of V401~Cyg exhibit a clear O'Connell effect.
Since there is a difference in height between the maxima, we considered a solution with one
cool spot on the surface of the primary component. We also included a third light due
to the spectroscopic evidence for the presence of a faint companion.
Its contribution to the total light indeed turned out to be low (3--4~per~cent), in agreement
with the  spectroscopic observations.
A very large spot, covering one hemisphere was obtained as a result. However, its temperature
factor is very close to 1, which is equivalent to a larger number of
smaller spots with a lower temperature factor but scattered over a large area.

\subsection{V345~Gem}
The amplitude of light changes of V345~Gem is very small, just slightly exceeding the value of
0.05 in flux units. The derivation of differential magnitude between the comparison
and the target stars included the nearby brighter visual companion. Therefore, a third light was included
in the modelling and we derived its contribution to be around 60~per cent, in agreement with the
fact that the companion is brighter than the eclipsing binary itself. The resulting
system configuration is marginally in contact, with a fill-out factor of about $f=11$~per~cent.
Due to the very large contribution from the visual companion this result cannot be very reliable.
It is definitely a very close binary but the actual configuration may range from near contact to
marginally contact. The reliability of the solution may be increased only if the visual companion
light is excluded from future measurements.

\subsection{AK~Her}
AK~Her shows a clear asymmetry in the light curves: the amplitude of the light curve
in the first maximum is larger than in the second one. Therefore, our model included a dark spot.
We started our computations without including a third light, but the
derived light curve did not result in a flat secondary minimum, clearly seen in our new
observations. This feature can be reproduced with the inclusion of a third light with a
contribution in the range of 13--15~per~cent. The spot is again large
($R_{sp} = 66^\circ$) but cooler by a factor of only 0.946 than the surrounding
stellar surface.

\subsection{V502~Oph}
Comparison of solutions for V502~Oph with and without a third light indicates
that both fit the observations almost equally well. Therefore, we assumed no
third light in the model but we included a spot to account for a small difference
in the maxima. As a result, a cool spot located near the pole was derived.
The fit is almost perfect and the results indicate the system is in
a contact configuration with a 28~per cent fill-out factor. Small discrepancies
between the model and the observed light curves may be due to the rapid
intrinsic photometric variability which was previously reported by \citet{bin1969}.
The O$-$C diagram shows a parabolic trend with a linear period decrease \citep{kre2004}.
After removing this, the residuals show a semi-regular periodic behaviour which may
be attributed to the gravitational influence of a tertiary component orbiting the system.
However, the latter has not been confirmed by either spectroscopic observations \citep{DAngelo2006}
or by the AO astrometric and imaging study \citep{ruc2007b}.


\begin{table*}
\caption{Results for five systems derived from the light curve modelling.}
\label{TabRes1}
\begin{flushleft}
\begin{tabular}{lccccc}
\hline
parameter	            & HV~Aqr	        & OO~Aql	        & FI~Boo	        & TX~Cnc	        & OT~Cnc	        \\
\hline											
fill-out factor      	& 74\%           	& 17\%              & 0.6\%         	& 19\%	            & 43\%	            \\
phase shift             & 0.0004$\pm$0.0004	& -0.0003$\pm$0.0001& 0.0006$\pm$0.0006	& 0.0020$\pm$0.0005	& 0.0002$\pm$0.0011	\\
$i$ [$^\circ$]          & 78.5$\pm$0.5   	& 86.9$\pm$0.1      & 46.7$\pm$0.3   	& 63.5$\pm$0.2	    & 37.6$\pm$1.2	    \\
** $T_{\rm 1}[{\rm K}]$ & 6460	            & 6000	            & 5850	            & 6100	            & 4500	            \\
$T_{\rm 2}[{\rm K}]$    & 6655$\pm$23    	& 5845$\pm$10       & 5670$\pm$40    	& 5954$\pm$32	    & 4445$\pm$40	    \\
$\Omega_{\rm 1}=\Omega_{\rm 2}$        & 2.010$\pm$0.004	& 3.416$\pm$0.001   & 2.636$\pm$0.007	& 5.375$\pm$0.007	& 2.699$\pm$0.019	\\
q$_{corr}$              & 0.140	            & 0.847	            & 0.380	            & 2.170	            & 0.471	            \\
$L_{1}/(L_{1}+L_{2})~(B)$	&	0.819$\pm$0.005	&	0.569$\pm$0.001	&	0.737$\pm$0.005	&	0.360$\pm$0.004	&	0.676$\pm$0.030	 \\
$L_{1}/(L_{1}+L_{2})~(V)$	&	0.824$\pm$0.003	&	0.565$\pm$0.001	&	0.733$\pm$0.004	&	0.357$\pm$0.004	&	0.675$\pm$0.029	 \\
$L_{1}/(L_{1}+L_{2})~(R)$	&	0.826$\pm$0.006	&	0.561$\pm$0.001	&	0.730$\pm$0.003	&	0.354$\pm$0.003	&	0.672$\pm$0.028	 \\
$L_{1}/(L_{1}+L_{2})~(I)$	&	0.827$\pm$0.006	&	0.556$\pm$0.001	&	--			&	0.349$\pm$0.003	&	0.669$\pm$0.026	 \\
*** $L_{2}/(L_{1}+L_{2})~(B)$	&	0.181			&	0.431			&	0.263			&	0.640			&	0.324			 \\
*** $L_{2}/(L_{1}+L_{2})~(V)$	&	0.176			&	0.435			&	0.267			&	0.643			&	0.325			 \\
*** $L_{2}/(L_{1}+L_{2})~(R)$	&	0.174			&	0.439			&	0.270			&	0.646			&	0.328			 \\
*** $L_{2}/(L_{1}+L_{2})~(I)$	&	0.173			&	0.444			&	--			&	0.651			&	0.331			 \\
$l_{3}~(B)$             & 0.000$\pm$0.005	& --                	& --             	& --	                & 0.300$\pm$0.028	\\
$l_{3}~(V)$             & 0.000$\pm$0.003	& --                	& --             	& --	                & 0.365$\pm$0.026	\\
$l_{3}~(R)$             & 0.010$\pm$0.007	& --                	& --             	& --	                & 0.373$\pm$0.027	\\
$l_{3}~(I)$             & 0.045$\pm$0.007	& --                	& -             	& --	                & 0.394$\pm$0.026	\\
\hline										
r$_{1}~^{side}$         & 0.5921$\pm$0.0002 & 0.4032$\pm$0.0002 & 0.4671$\pm$0.0001 & 0.3179$\pm$0.0002	& 0.4745$\pm$0.0048	\\
r$_{2}~^{side}$         & 0.2443$\pm$0.0002 & 0.3724$\pm$0.0002 & 0.2907$\pm$0.0002 & 0.4603$\pm$0.0002	& 0.3358$\pm$0.0047	\\
\hline											
co-latitude [$^\circ$]  & 151$\pm$8     	& --                	& --              	& 84$\pm$6	        & --	                \\
longitude [$^\circ$]    & 33$\pm$7      	& --                	& --              	& 261.6$\pm$5.6	    & --	                \\
radius [$^\circ$]       & 30.7$\pm$5.3  	& --                	& --              	& 25.3$\pm$2.5	    & --	                \\
temp. factor            & 0.89$\pm$0.04 	& --                	& --              	& 0.96$\pm$0.05	    & --	                \\
\hline
\end{tabular}
\begin{small}
$*$~-~assumed,~~~~$**$~-~computed,
the subscripts 1 and 2 refer to the star being eclipsed at primary and secondary minimum, respectively. $l_{3}$ is the third light contribution to
the total light at phase 0.25. \\
\end{small}
\end{flushleft}
\end{table*}

\begin{table*}
\caption{Results for five systems derived from the light curve modelling.}
\label{TabRes2}
\begin{flushleft}
\begin{tabular}{lccccc}
\hline
parameter	            & EE~Cet	        & RW~Com	        & KR~Com	        & V401~Cyg	        & V345~Gem	         \\
\hline											
fill-out factor         & 19\%           	& 3\%	            & 99\%	            & 42\%	             & 11\%	             \\
phase shift             & 0.0001$\pm$0.0044 & 0.0010$\pm$0.0003	& -0.0105$\pm$0.0013& -0.0026$\pm$0.0003 & 0.0122$\pm$0.0008 \\
$i$ [$^\circ$]          & 73.7$\pm$0.2   	& 71.5$\pm$0.2	    & 54.4$\pm$1.2	    & 79.9$\pm$0.5	     & 61.2$\pm$0.9	     \\
** $T_{\rm 1}[{\rm K}]$ & 6100	            & 4600	            & 5920	            & 6980	             & 6200	             \\
$T_{\rm 2}[{\rm K}]$    & 5800$\pm$11    	& 4360$\pm$12	    & 6130$\pm$70	    & 6630$\pm$35	     & 5840$\pm$54	     \\
$\Omega_{\rm 1}=\Omega_{\rm 2}$        & 6.734$\pm$0.002	& 5.364$\pm$0.005	& 1.877$\pm$0.008	& 2.356$\pm$0.004	 & 2.148$\pm$0.006	 \\
q$_{corr}$              & 3.179	            & 2.093	            & 0.093	            & 0.284	             & 0.171	         \\
$L_{1}/(L_{1}+L_{2})~(U)$	&	--			&	--			&	0.858	$\pm$	0.085	&	--			&	--			 \\
$L_{1}/(L_{1}+L_{2})~(B)$	&	0.314$\pm$0.001	&	0.428$\pm$0.004	&	0.856$\pm$0.089	&	0.792$\pm$0.008	&	0.866$\pm$0.065	 \\
$L_{1}/(L_{1}+L_{2})~(V)$	&	0.307$\pm$0.001	&	0.416$\pm$0.003	&	0.858$\pm$0.088	&	0.781$\pm$0.007	&	0.863$\pm$0.063	 \\
$L_{1}/(L_{1}+L_{2})~(R)$	&	0.300$\pm$0.001	&	0.420$\pm$0.003	&	0.862$\pm$0.087	&	0.777$\pm$0.007	&	0.859$\pm$0.062	 \\
$L_{1}/(L_{1}+L_{2})~(I)$	&	0.292$\pm$0.001	&	0.393$\pm$0.002	&	--			    &	0.775$\pm$0.007	&	0.853$\pm$0.063	 \\
*** $L_{2}/(L_{1}+L_{2})~(U)$	&	--			&	--			&	0.650			&	--			&	--			 \\
*** $L_{2}/(L_{1}+L_{2})~(B)$	&	0.686			&	0.572			&	0.144			&	0.208			&	0.134			 \\
*** $L_{2}/(L_{1}+L_{2})~(V)$	&	0.693			&	0.584			&	0.142			&	0.219			&	0.137			 \\
*** $L_{2}/(L_{1}+L_{2})~(R)$	&	0.700			&	0.580			&	0.138			&	0.223			&	0.141			 \\
*** $L_{2}/(L_{1}+L_{2})~(I)$	&	0.708			&	0.607			&	--			&	0.225			&	0.147			 \\
$l_{3}~(U)$             & --              	& --	                & 0.598$\pm$0.041	& --	                 & --	             \\
$l_{3}~(B)$             & --              	& --	                & 0.621$\pm$0.041	& 0.041$\pm$0.009	 & 0.652$\pm$0.029	 \\
$l_{3}~(V)$             & --              	& --	                & 0.599$\pm$0.042	& 0.028$\pm$0.009	 & 0.619$\pm$0.031	 \\
$l_{3}~(R)$             & --              	& --	                & 0.606$\pm$0.042	& 0.035$\pm$0.009	 & 0.597$\pm$0.033	 \\
$l_{3}~(I)$             & --              	& --	                & --	                & 0.035$\pm$0.009	 & 0.585$\pm$0.034	 \\
\hline											
r$_{1}~^{side}$         & 0.2851$\pm$0.0004 & 0.3110$\pm$0.0048	& 0.6330$\pm$0.0043	& 0.5158$\pm$0.0013	 & 0.5492$\pm$0.0023 \\
r$_{2}~^{side}$         & 0.4956$\pm$0.0004 & 0.4464$\pm$0.0048	& 0.2211$\pm$0.0042	& 0.2869$\pm$0.0013	 & 0.2356$\pm$0.0021 \\
\hline											
co-latitude [$^\circ$] 	& --             	& 12$\pm$2	        & --	                & 35$\pm$2	         & --	             \\
longitude [$^\circ$]   	& --             	& 27$\pm$1	        & --	                & 147$\pm$3	         & --	             \\
radius [$^\circ$]      	& --             	& 46$\pm$2	        & --	                & 89.8$\pm$2.2	     & --	             \\
temp. factor           	& --             	& 0.86$\pm$0.02	    & --	                & 0.972$\pm$0.009	 & --	             \\
\hline
\end{tabular}
\begin{small}
$*$~-~assumed,~~~~$**$~-~computed,
the subscripts 1 and 2 refer to the star being eclipsed at primary and secondary minimum, respectively.
$l_{3}$ is the third light contribution to the total light at phase 0.25.\\
\end{small}
\end{flushleft}
\end{table*}

\subsection{V566~Oph}
The V566~Oph system shows negligible light curve asymmetry and therefore we decided
on a model without a spot. The first solution did not include a
third light and the depth and shape of the secondary minimum were not well reproduced.
After adding a third light we obtained an almost perfect fit.
The third light contribution is not large. The highest (about 10~per cent) was found
in the $B$ band, decreasing towards longer wavelengths and reaching
about 7~per cent in the $I$ band. The system shows no evidence of a third
component in the spectroscopic observations of \citet{DAngelo2006} or with
AO astrometric and imaging \citep{ruc2007b}. The O$-$C diagram does
show a parabolic trend with a linear orbital period increase,
which might imply that there is mass transfer ongoing between the components \citep{kre2004}.

\subsection{V2612~Oph}
V2612~Oph shows a highly variable light curve, likely due to magnetic activity,
which prevented obtaining a perfect fit. Due to different maxima amplitudes,
a spot on the primary component was added into the model. This improved
the fit significantly compared to a no-spot solution. We obtained
a contact configuration for this system with a fill-out factor of 33~per cent.

\subsection{V1363~Ori}
V1363~Ori is a low inclination system and a distorted light curve. Due to this, there
is some discrepancy between the theoretical and the observed light curves, especially
in the $B$ and $V$ bands. A high fill-out factor of $f$ = 85~per~cent was found.
The system shows large astrometric (parallax) scatter, which might indicate the existence
of a third body \citep{pri2006}.
However, no such body has been detected either from spectroscopic
observations \citep{DAngelo2006} or with the AO astrometric and imaging study \citep{ruc2007b}.
We tried to obtain solutions both with and without a third light contribution. Since the improvement
of the fit while using a third light was negligible, we adopted a final model without a third light.
Our result therefore is consistent with the previous non-detections of a tertiary component.

\subsection{V351~Peg}
The observed light curves for V351~Peg show only a small amplitude in the light curve
and it is difficult to find a unique solution.
We initially performed computations without a spot. However,
comparing the theoretical light curves with observations, one can notice
small asymmetries, possibly due to the O'Connell effect.
Due to this, we finally adopted a model with a spot. The size of the cool spot is not large,
its radius being about $20^\circ$.
V351~Peg belongs to the group of  large effective temperature (early type)
contact binaries.

\subsection{V357~Peg}
V357~Peg shows a small light curve asymmetry, and due to this, we introduced
a dark spot into the model.
The best fit required a relatively cool spot located near the pole of the primary.
Although a large disagreement regarding the physical parameters of this system exists in the literature,
our solution indicates that this is a contact system with rather small fill-out factor
of $f$ = 18~per~cent and about 500~K temperature difference between the components.
We also confirm that the orbital inclination of the system is $i$ = 73.4$^\circ$,
in agreement with the findings of \citet{deb2011}.

\subsection{Y~Sex}
The search for the best fit for Y~Sex included a third light.
We achieved convergence very quickly and the resulting contribution of the third light was
between 10--12~per~cent, in perfect agreement with the value derived from spectroscopy \citep{pri2009b}.
The fit is very good, with small discrepancies between the model and the observations
in the descending branch after the second maximum solely in the $I$ band.
This is a contact system with a medium fill-out factor of $f = 53$~per cent and
a small temperature difference (about 100~K) between the components.

\subsection{V1123~Tau}
V1123~Tau is a typical contact binary system with a small fill-out factor of $f = 16$~per cent.
We needed neither a spot nor a third light to obtain a solution which perfectly matches
the observed $BVR$ light curves. The more massive star (secondary) is about 150~K hotter
than the primary and the orbital inclination is about $i$ = 68$^\circ$.

\subsection{W~UMa}
The amplitude of the maxima of the observed light curves are not equal and therefore
our computations included a spot. Since this is a triple system \citep{whe1973},
we also set the third light as a free parameter.
The ~4 mag fainter visual companion was always present in the 20--30 arcsec photometer
aperture. The fit of the resulting model is very good with very small deviations
in the $B$ band only. The third light contribution ranges
between 8--11~per cent, being slightly larger than expected for the brightness of the companion.
The derived orbital inclination is high ($i = 88.4^\circ$) and as a result the light curve exhibits
a flat bottom primary minimum. Though not so obvious in our observations, this feature is clearly
visible in the more densely sampled light curve published by \cite{Linnell1991}.


\begin{table*}
\caption{Results for five systems derived from light curve modelling.}
\label{TabRes3}
\begin{flushleft}
\begin{tabular}{lccccc}
\hline
parameter               & AK Her            & V502 Oph          & V566 Oph          & V2612 Oph          & V1363 Ori          \\
\hline
fill-out factor             & 23\%              & 28\%              & 38\%              & 33\%               & 85\%               \\
phase shift             & 0.0019$\pm$0.0005 & 0.0028$\pm$0.0004 & 0.0002$\pm$0.0004 & -0.0026$\pm$0.0007 & -0.0082$\pm$0.0004 \\
$i$ [$^\circ$]              & 84.8$\pm$1.2      & 71.4$\pm$0.2      & 83.0$\pm$0.5      & 66.0$\pm$0.4       & 54.1$\pm$0.2       \\
** $T_{\rm 1}[{\rm K}]$ & 6530              & 5920              & 6530              & 6460               & 6700               \\
$T_{\rm 2}[{\rm K}]$    & 6060$\pm$33       & 5850$\pm$27       & 6480$\pm$11       & 5980$\pm$48        & 6364$\pm$28        \\
$\Omega_{\rm 1}=\Omega_{\rm 2}$        & 2.383$\pm$0.005   & 6.327$\pm$0.009   & 2.343$\pm$0.005   & 6.731$\pm$015      & 2.189$\pm$002      \\
q$_{corr}$              & 0.281             & 2.911             & 0.274             & 3.245              & 0.234              \\
$L_{1}/(L_{1}+L_{2})~(B)$	&	0.817$\pm$0.012	&	--			    &	0.763$\pm$0.007	&	0.342$\pm$0.006	&	0.807$\pm$0.004	\\
$L_{1}/(L_{1}+L_{2})~(V)$	&	0.809$\pm$0.012	&	0.290$\pm$0.003	&	0.762$\pm$0.005	&	0.331$\pm$0.005	&	0.799$\pm$0.003	\\
$L_{1}/(L_{1}+L_{2})~(R)$	&	0.803$\pm$0.012	&	0.289$\pm$0.003	&	0.761$\pm$0.007	&	0.321$\pm$0.004	&	0.795$\pm$	0.003	\\
$L_{1}/(L_{1}+L_{2})~(I)$	&	0.792$\pm$0.010	&	0.287$\pm$0.003	&	0.760$\pm$0.006	&	0.307$\pm$0.004	&	0.792$\pm$0.002	\\
*** $L_{2}/(L_{1}+L_{2})~(B)$	&	0.183			&	--			&	0.237			&	0.658			&	0.193			\\
*** $L_{2}/(L_{1}+L_{2})~(V)$	&	0.191			&	0.710			&	0.238			&	0.669			&	0.201			\\
*** $L_{2}/(L_{1}+L_{2})~(R)$	&	0.197			&	0.711			&	0.239			&	0.679			&	0.205			\\
*** $L_{2}/(L_{1}+L_{2})~(I)$	&	0.208			&	0.713			&	0.240			&	0.693			&	0.208			\\
$l_{3}~(B)$             & 0.131$\pm$0.013   & --                 & 0.100$\pm$0.008   & --                  & --                  \\
$l_{3}~(V)$             & 0.139$\pm$0.012   & --                 & 0.086$\pm$0.006   & --                  & --                  \\
$l_{3}~(R)$             & 0.152$\pm$0.012   & --                 & 0.080$\pm$0.008   & --                  & --                  \\
$l_{3}~(I)$             & 0.160$\pm$0.011   & --                 & 0.073$\pm$0.008   & --                  & --                  \\
\hline
r$_{1}~^{side}$         & 0.5085$\pm$0.0030 & 0.2973$\pm$0.0005 & 0.5187$\pm$0.0008 & 0.2911$\pm$0.0013  & 0.5577$\pm$0.0008  \\
r$_{2}~^{side}$         & 0.2782$\pm$0.0030 & 0.4924$\pm$0.0005 & 0.2836$\pm$0.0008 & 0.5056$\pm$0.0014  & 0.2949$\pm$0.0008  \\
\hline
co-latitude [$^\circ$]  & 158$\pm$5         & 169$\pm$8          & --                 & 122$\pm$10         & --                  \\
longitude [$^\circ$]    & 35$\pm$3          & 226$\pm$8          & --                 & 95$\pm$5           & --                  \\
radius [$^\circ$]       & 66.4$\pm$3.0      & 45.5$\pm$6.0       & --                 & 52.7$\pm$6.9       & --                  \\
temp. factor            & 0.943$\pm$0.027   & 0.475$\pm$0.146    & --                 & 0.90$\pm$0.04      & --                  \\
\hline
\end{tabular}
\begin{small}
$*$~-~assumed,~~~~$**$~-~computed,
the subscripts 1 and 2 refer to the star being eclipsed at primary and secondary minimum, respectively. $l_{3}$ is
the third light contribution to the total light at phase 0.25.\\
\end{small}
\end{flushleft}
\end{table*}

\begin{table*}
\caption{Results for five systems derived from light curve modelling.}
\label{TabRes4}
\begin{flushleft}
\begin{tabular}{lccccc}
\hline
parameter               & V351 Peg           & V357 Peg           & Y Sex              & V1123 Tau         & W UMa             \\
\hline
fill-out factor             & 14\%               & 18\%               & 55\%               & 16\%              & 22\%              \\
phase shift             & -0.0054$\pm$0.0005 & -0.0074$\pm$0.0004 & -0.0011$\pm$0.0003 & 0.0053$\pm$0.0002 & 0.0044$\pm$0.0004 \\
$i$ [$^\circ$]               & 64.5$\pm$0.2       & 73.4$\pm$0.2       & 78.7$\pm$0.4       & 68.3$\pm$0.1      & 88.4$\pm$0.8      \\
** $T_{\rm 1}[{\rm K}]$ & 7500               & 6700               & 6400               & 5920              & 6450              \\
$T_{\rm 2}[{\rm K}]$    & 6524$\pm$76        & 6183$\pm$42        & 6294$\pm$15        & 6059$\pm$12       & 6170$\pm$21       \\
$\Omega_{\rm 1}\Omega_{\rm 2}$        & 2.580$\pm$005      & 2.636$\pm$005      & 2.152$\pm$004      & 7.293$\pm$0.005   & 5.072$\pm$0.008   \\
q$_{corr}$              & 0.368              & 0.401              & 0.195              & 3.597             & 1.967             \\
$L_{1}/(L_{1}+L_{2})~(U)$	&	--			&	0.751	$\pm$	0.004	&	--			&	--			&	--			\\
$L_{1}/(L_{1}+L_{2})~(B)$	&	0.821$\pm$0.008	&	0.769$\pm$0.005	&	0.815$\pm$0.008	&	0.221$\pm$0.001	&	0.404$\pm$0.004	\\
$L_{1}/(L_{1}+L_{2})~(V)$	&	0.804$\pm$0.006	&	0.757$\pm$0.005	&	0.813$\pm$0.008	&	0.223$\pm$0.001	&	0.398$\pm$0.003	\\
$L_{1}/(L_{1}+L_{2})~(R)$	&	0.793$\pm$0.006	&	0.749$\pm$0.004	&	0.811$\pm$0.008	&	0.225$\pm$0.001	&	0.391$\pm$0.003	\\
$L_{1}/(L_{1}+L_{2})~(I)$	&	0.783$\pm$0.005	&	--			    &	0.810$\pm$0.008	&	--			    &	0.382$\pm$0.003	\\
*** $L_{2}/(L_{1}+L_{2})~(U)$	&	--			&	0.249			&	--			&	--			&	--			\\
*** $L_{2}/(L_{1}+L_{2})~(B)$	&	0.179			&	0.231			&	0.185			&	0.779			&	0.596			\\
*** $L_{2}/(L_{1}+L_{2})~(V)$	&	0.196			&	0.243			&	0.187			&	0.777			&	0.602			\\
*** $L_{2}/(L_{1}+L_{2})~(R)$	&	0.207			&	0.251			&	0.189			&	0.775			&	0.609			\\
*** $L_{2}/(L_{1}+L_{2})~(I)$	&	0.217			&	--			&	0.190			&	--			&	0.618			\\
$l_{3}~(U)$             & --                 & --                  & --                  & --                 & --            \\
$l_{3}~(B)$             & --                 & --                  & 0.10$\pm$0.01      & --                 & 0.078$\pm$0.006       \\
$l_{3}~(V)$             & --                 & --                  & 0.11$\pm$0.01      & --                 & 0.089$\pm$0.006       \\
$l_{3}~(R)$             & --                 & --                  & 0.12$\pm$0.01      & --                 & 0.091$\pm$0.006       \\
$l_{3}~(I)$             & --                 & --                  & 0.11$\pm$0.01      & --                 & 0.115$\pm$0.007       \\
\hline
r$_{1}~^{side}$         & 0.4781$\pm$0.0010  &0.4724$\pm$0.0011   & 0.5566$\pm$0.0014  & 0.2739$\pm$0.0001 & 0.3286$\pm$0.0011 \\
r$_{2}~^{side}$         & 0.2952$\pm$0.0010  &0.3050$\pm$0.0010   & 0.2631$\pm$0.0013  & 0.5058$\pm$0.0004 & 0.4532$\pm$0.0011 \\
\hline
co-latitude [$^\circ$]  & 23 $\pm$7          & 162$\pm$3          & --                  & --                 &  15$\pm$3          \\
longitude [$^\circ$]    & 169$\pm$2          & 195$\pm$1          & --                  & --                 & 227$\pm$3          \\
radius [$^\circ$]       & 27.3$\pm$4.0       & 52.0$\pm$2         & --                  & --                 & 42.9$\pm$1.8       \\
temp. factor            & 0.871$\pm$0.048    & 0.440$\pm$0.12     & --                  & --                 & 0.423$\pm$0.12     \\
\hline
\end{tabular}
\begin{small}
$*$~-~assumed,~~~~$**$~-~computed,
the subscripts 1 and 2 refer to the star being eclipsed at primary and secondary minimum, respectively. $l_{3}$ is the third light contribution to
the total light at phase 0.25.\\
\end{small}
\end{flushleft}
\end{table*}

\section{Evolutionary status of the sample}

The combined photometric and spectroscopic models of the current sample of
contact binaries provide the means for an accurate determination of their orbital and physical
characteristics.
The uncertainties are smaller than 2--3~per~cent in most cases, following the requirements
set from the beginning of this project \citep{kre2003} (Paper I).
The physical parameters of all systems studied in this work are given in Table \ref{TabParam}.
For the sake of uniformity, in this table we use the designation '1' for
the more massive component, resulting in a mass ratio less than unity.
In this Table, the sub-types 'A' and 'W' denote whether the more massive component
is the hotter or cooler, respectively. Therefore, those
systems with $q>1$ (W-type) have reversed physical parameters for the
two components in Table \ref{TabParam} with respect to Tables \ref{TabRes1}--\ref{TabRes4}.
This convention follows the definition used in spectroscopic observations,
where the more massive component is usually referred to as the 'primary'.

The study of absolute physical parameters through correlation diagrams is a key tool for
understanding stellar evolution of single stars and components of close binary systems.
The relation between orbital and physical parameters shows that the components of close binary
systems are tightly correlated with each other, as a result of common evolution
\citep{mac1982, hil1988, gaz2006b, gaz2008, gaz2009, mic2019}.
Similar studies in the past included a large number of contact binaries,
aimed at extracting statistical information for the physical parameters
and evolution of binaries. However, the majority of these studies include
models determined solely from photometric data, without knowledge
of the spectroscopic mass ratio.
For contact and very close systems such models may lead to unreliable results,
since reliable physical parameters can be determined from photometry alone only for systems
which show light curves with total eclipses \citep{pri2003, ter2005}.

\begin{figure}
\includegraphics[width=9.5cm,height=7.2cm,scale=1.0,angle=0]{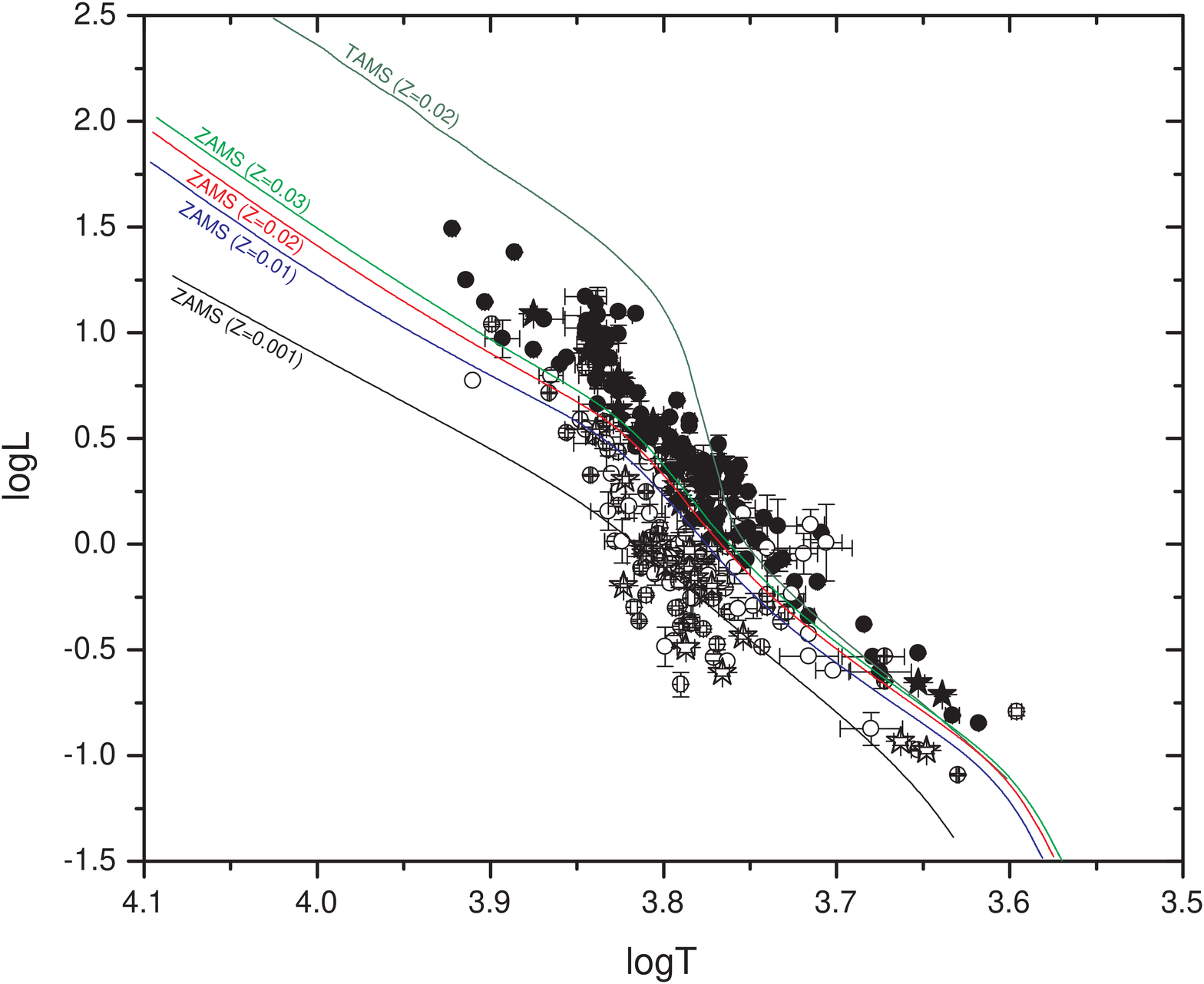}
\includegraphics[width=9.5cm,height=7.2cm,scale=1.0,angle=0]{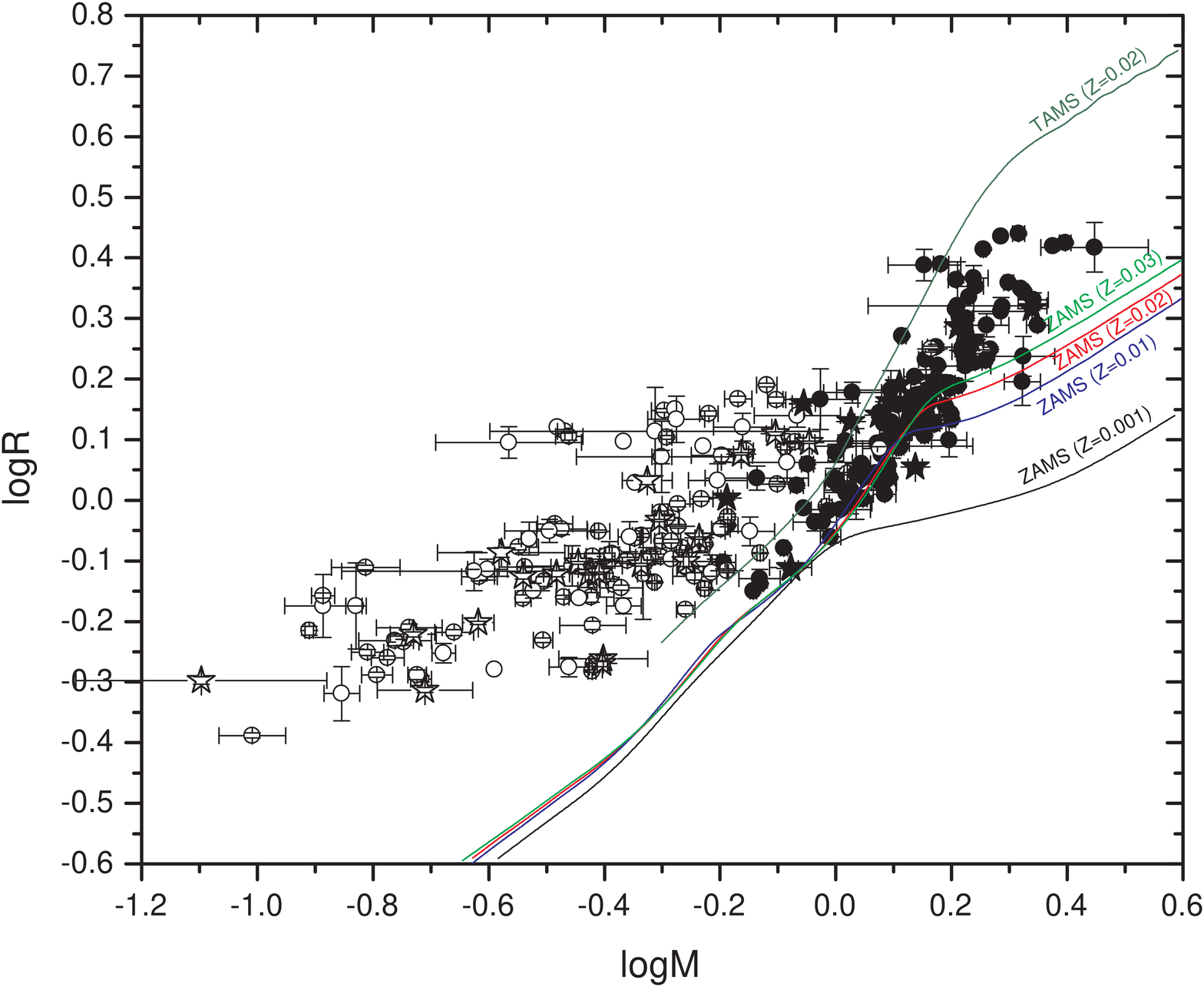}

\caption{Temperature-Luminosity (or H-R) and Mass-Radius diagrams for the best
models of contact binaries known to date. ZAMS and TAMS are calculated for
single MS stars and different metallicities, according to PARSEC models \citep{bre2012}.
The systems shown in these plots are taken from the sample of 51 systems from the W UMa Project,
together with 67 systems from the literature (filled and open black circle symbols),
and the 20 systems from the current study (filled and open black star symbols).}
\label{FigCD1}
\end{figure}

The evolutionary state of a sample of contact binaries can be studied in terms of the
$\log M-\log R$, $\log T-\log L$ diagrams (Figs. \ref{FigCD1}).
The diagrams show that the primary components (filled symbols in all plots) follow the Main Sequence (MS) trend
with a relatively high metallicity. This can be judged by the location of primaries with respect
to the MS zone. ZAMS and TAMS are calculated according to the PARSEC models \citep{bre2012}.
On the contrary, the secondary components (marked by open symbols in all plots) appear to be
oversized compared to single MS stars of the same mass.
The Hertzsprung-Russell (H-R) diagram ($\log T-\log L$) confirms the above findings.
Secondary components lie in a hotter (for their mass) region of the plot, likely as a result
of energy transfer between the components. It was also noticed that the primary
components are under-luminous for their location in the H-R diagram, while
the secondary components are over-luminous for their mass and appear to be evolutionarily more advanced.

The above properties arise as a result of the tight orbital configuration and significant interaction
between the components. The primary components show a 1:1 relation between mass and radius,
following the trend:

\begin{equation}
M_{1} \sim R^{0.91\pm0.04}
\end{equation}

The radius of secondary components seems to be
$\sim$30~per~cent larger than the size expected for single MS stars, following the trend:

\begin{equation}
M_{2} \sim R^{0.41\pm0.03}
\end{equation}

In addition, their temperature is higher than expected, due to thermal equilibrium with the primary
components, which co-exist within a common envelope. Taking into account the evolutionary model
described in \citet{ste2012}, the secondary components are probably stripped of their photosphere
due to mass-transfer processes, exposing their hotter internal layers. Secondary components are
evolutionarily more advanced, with hydrogen depleted (or nearly depleted) in their cores, hence they appear
oversized. As a consequence, the luminosities of the secondary components are larger than expected and
they are located above the MS on the H-R diagram.

Our study shows that the primary components follow a mass-luminosity relation:

\begin{equation}
M_{1} \sim L^{3.13\pm0.16}
\end{equation}

while the secondary components follow a less steep one:
\begin{equation}
M_{2} \sim L^{0.81\pm0.12}
\end{equation}

A few outliers exhibit slightly different behaviour divergent from that of the overall sample in
the $\log M-\log R$, $\log T-\log L$ diagrams (Fig. \ref{FigCD1}).
Systems like V345~Gem (in the current sample), V781~Tau and EF~Dra (from older studies) are located well below the ZAMS in the $\log M-\log R$ diagram, which is likely connected with their low metallicity.

Following careful examination, we see that these systems are either dominated by a large amount
of third light (due to a close companion, being members of triple or multiple systems), or they
are magnetically active stars with distorted light curves. In addition, the majority
of them have low inclination orbits. For the above reasons, their models may be less reliable
and therefore the uncertainties of the parameters could be higher than the formal estimates would suggest.

The same occurs for a few systems with large luminosity of their primary components,
located well above the TAMS:  MW~Pav, TY~Pup, XX~Sex, BV~Eri, TV~Mus, KR~Com, FI~Boo.
Furthermore, the majority of these outliers come from solutions based on the ASAS light curves,
showing a relatively large scatter and resulting in larger uncertainties of the physical parameters,
even when combined with the spectroscopically-determined mass ratio.

The physical parameters of binaries are constrained by their Roche geometry and they cannot be
directly compared to single stars with the same characteristics. Energy transfer between the two
components alters their evolution and therefore their parameters depend on this effect.
As a consequence we see primary components located close to the ZAMS (their luminosity and temperature
are lower than would result from the core luminosity), whereas the secondaries are located above the TAMS
(their luminosity and temperature are elevated compared to their core energy production).
\citet{moc1981} suggested a method of correcting the stellar parameters for the energy transfer,
assuming equal temperatures of the components and a simple mass-luminosity relation $L \sim M^{4.4}$,
which is a rather general assumption in this domain. Such assumptions are poorly fulfilled
in many systems. In addition, we have to deal with the energy transfer and the W-phenomenon
(where the secondary is apparently hotter than the primary), and with the O'Connell effect due to
magnetic activity, which results in surface luminosity modulation. Luminosity can also be affected by spots
due to magnetic activity and/or circumstellar material. All this shows that the observed luminosity and
temperature of each component cannot be solely connected with the physical (core) processes in these stars.

With the above in mind, we see that mass and radius are more reliable parameters
when deciphering  the evolutionary status of contact systems. An interesting result
of this study is that contact binaries are not low-metallicity objects. The majority of them
(about 93 per~cent  \citep{jon2010} and 96 per~cent \citep{aum2009}), are thin disc objects
(scale height~$<300$~pc) and have solar metallicity.
Our sample consists of the nearest binaries in our Solar vicinity, all within a radius of $\sim$500 pc.
Therefore Solar metallicity is a safe assumption to describe the current sample. \citet{ruc2013}
confirmed this spectroscopically, finding an average metallicity close to Solar,
although with a large scatter attributed partly to determination errors.

In this study, the ZAMS and TAMS fit much better for intermediate and high metallicity
and only a few systems are located in the low metallicity region of ZAMS and TAMS (systems
which resulted in large luminosity values are those based on ASAS data).

\begin{table*}
\begin{center}
\caption{Absolute parameters and their errors (in Solar units) for the systems which are described in the present paper. Type ``A'' or ``W'' denotes whether the more massive component is the hotter or cooler in the system, respectively.}
\label{TabParam}
\begin{tabular}{lccccccc}
\hline
System       & ${\cal M}_{\rm 1}$   & ${\cal M}_{\rm 2}$     & $R_{\rm 1}$     & $R_{\rm 2}$     & $L_{\rm 1}$    & $L_{\rm 2}$  & type     \\
\hline
HV~Aqr	     &$	1.240\pm0.028	$&$	0.186\pm0.017	$&$	1.456\pm0.012	$&$	0.601\pm0.005	$&$	3.326\pm0.213	$&$	0.638\pm0.044	$&A		\\
OO~Aql	     &$	1.063\pm0.015	$&$	0.899\pm0.021	$&$	1.350\pm0.006	$&$	1.247\pm0.006	$&$	2.128\pm0.143	$&$	1.635\pm0.113	$&A		\\
FI~Boo	     &$	0.648\pm0.016	$&$	0.241\pm0.015	$&$	1.009\pm0.007	$&$	0.628\pm0.004	$&$	1.074\pm0.075	$&$	0.362\pm0.028	$&A		\\
TX~Cnc	     &$	1.276\pm0.025	$&$	0.580\pm0.009	$&$	1.255\pm0.005	$&$	0.866\pm0.003	$&$	1.783\pm0.167	$&$	0.936\pm0.062	$&W		\\
OT~Cnc       &$	0.835\pm0.070	$&$	0.396\pm0.070	$&$	0.774\pm0.008	$&$	0.548\pm0.006	$&$	0.221\pm0.040	$&$	0.106\pm0.020	$&A		\\
EE~Cet	     &$	1.464\pm0.043	$&$	0.461\pm0.011	$&$	1.360\pm0.008	$&$	0.783\pm0.004	$&$	1.886\pm0.178	$&$	0.765\pm0.051	$&W		\\
RW~Com	     &$	0.838\pm0.022	$&$	0.395\pm0.008	$&$	0.772\pm0.006	$&$	0.538\pm0.006	$&$	0.194\pm0.025	$&$	0.117\pm0.010	$&W		\\
KR~Com	     &$	0.880\pm0.041	$&$	0.080\pm0.040	$&$	1.445\pm0.013	$&$	0.505\pm0.007	$&$	2.311\pm0.162	$&$	0.324\pm0.029	$&A		\\
V401~Cyg	 &$	1.628\pm0.080	$&$	0.472\pm0.047	$&$	1.938\pm0.031	$&$	1.078\pm0.018	$&$	8.032\pm0.527	$&$	2.023\pm0.168	$&A		\\
V345~Gem	 &$	1.371\pm0.040	$&$	0.195\pm0.037	$&$	1.134\pm0.008	$&$	0.486\pm0.004	$&$	1.712\pm0.113	$&$	0.248\pm0.019	$&A		\\
AK~Her	     &$	1.188\pm0.029	$&$	0.329\pm0.017	$&$	1.382\pm0.012	$&$	0.756\pm0.008	$&$	3.129\pm0.199	$&$	0.694\pm0.050	$&A		\\
V502~Oph	 &$	1.477\pm0.030	$&$	0.495\pm0.010	$&$	1.533\pm0.007	$&$	0.926\pm0.004	$&$	2.480\pm0.240	$&$	0.949\pm0.065	$&W		\\
V566~Oph	 &$	1.365\pm0.016	$&$	0.359\pm0.011	$&$	1.443\pm0.006	$&$	0.789\pm0.003	$&$	3.411\pm0.211	$&$	0.989\pm0.063	$&A		\\
V2612~Oph	 &$	1.304\pm0.024	$&$	0.374\pm0.005	$&$	1.315\pm0.005	$&$	0.757\pm0.003	$&$	1.992\pm0.177	$&$	0.899\pm0.056	$&W		\\
V1363~Ori	 &$	1.289\pm0.057	$&$	0.264\pm0.067	$&$	1.552\pm0.029	$&$	0.821\pm0.015	$&$	4.373\pm0.673	$&$	0.996\pm0.163	$&A		\\
V351~Peg	 &$	2.209\pm0.047	$&$	0.795\pm0.026	$&$	2.049\pm0.053	$&$	1.265\pm0.084   $&$11.968\pm1.140	$&$	2.612\pm0.464	$&A		\\
V357~Peg	 &$	1.713\pm0.019	$&$	0.686\pm0.021	$&$	1.847\pm0.007	$&$	1.193\pm0.005	$&$	6.193\pm0.373	$&$	1.874\pm0.126	$&A		\\
Y~Sex	     &$	1.476\pm0.045	$&$	0.288\pm0.033	$&$	1.586\pm0.018	$&$	0.750\pm0.008	$&$	3.802\pm0.253	$&$	0.795\pm0.059	$&A		\\
V1123~Tau	 &$	1.397\pm0.024	$&$	0.390\pm0.007	$&$	1.401\pm0.005	$&$	0.759\pm0.003	$&$	2.383\pm0.230	$&$	0.637\pm0.043	$&W		\\
W~UMa	     &$	1.139\pm0.019	$&$	0.551\pm0.006	$&$	1.092\pm0.016	$&$	0.792\pm0.015	$&$	1.557\pm0.166	$&$	0.978\pm0.071	$&W		\\
\hline
\end{tabular}
\end{center}
\end{table*}



\section{Overview of physical properties of contact binaries}

\begin{figure*}
\includegraphics[width=8.0cm,height=6.1cm,scale=1.0,angle=0]{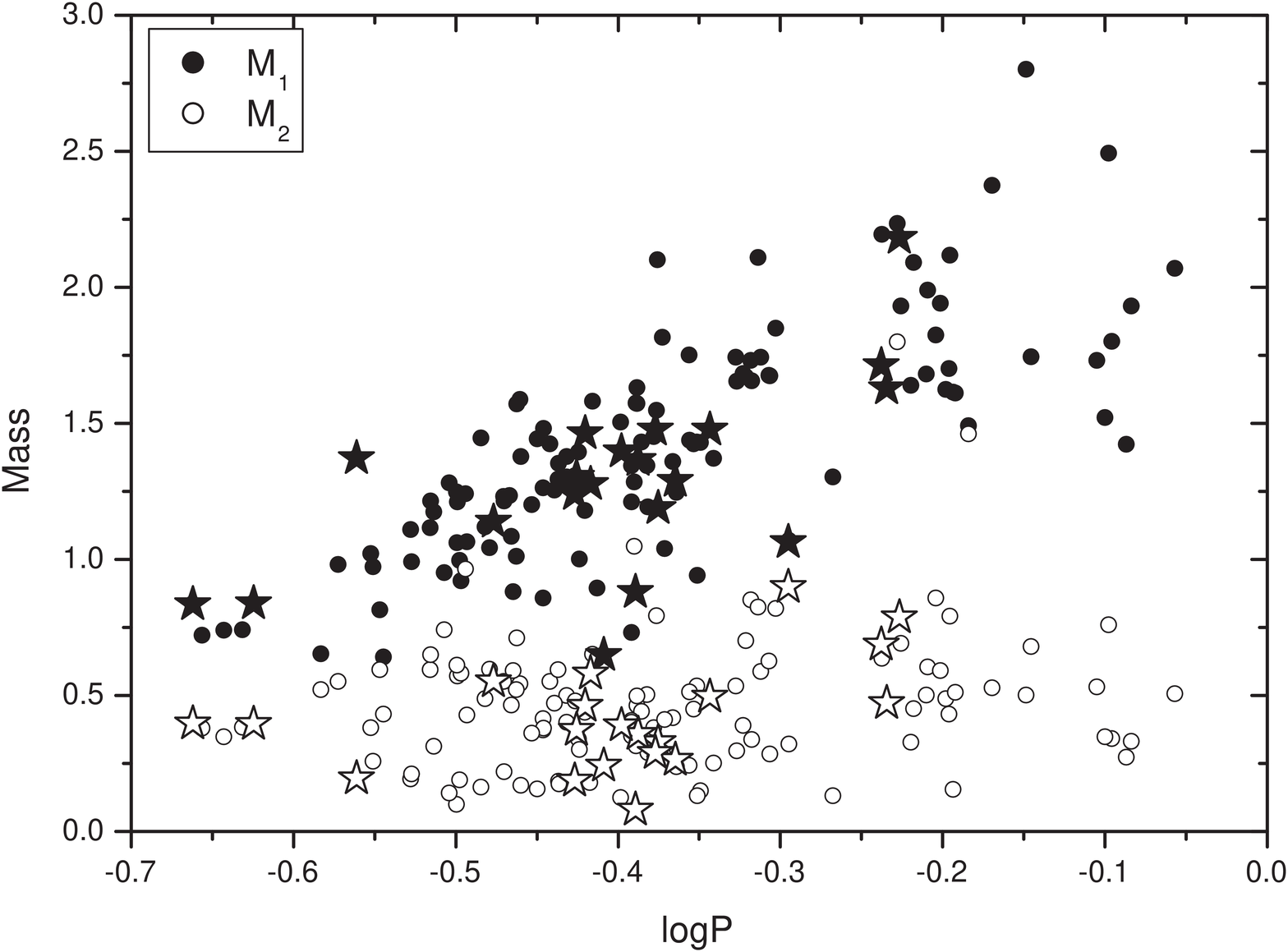}
\includegraphics[width=8.0cm,height=6.1cm,scale=1.0,angle=0]{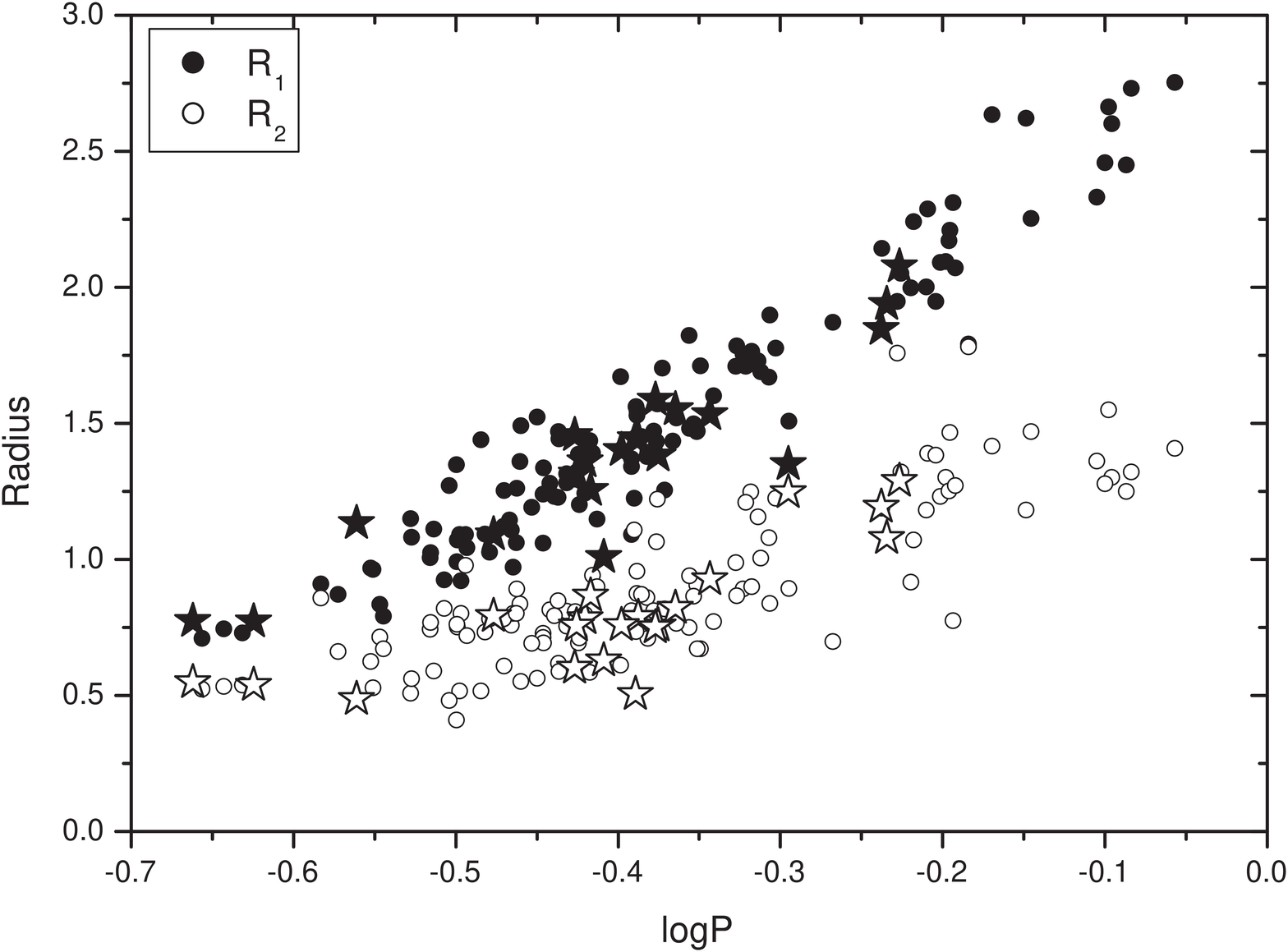}
 \caption{Correlation diagrams between orbital period and mass (left) and radius (right) for the
sample of 51 systems, observed in the frame of the {W~UMa Project} and 67 systems collected from
literature (black and white circles), together with the current 20 systems (black and white stars).}
\label{FigCD3}
\end{figure*}

\begin{figure*}
\includegraphics[width=8.0cm,height=6.1cm,scale=1.0,angle=0]{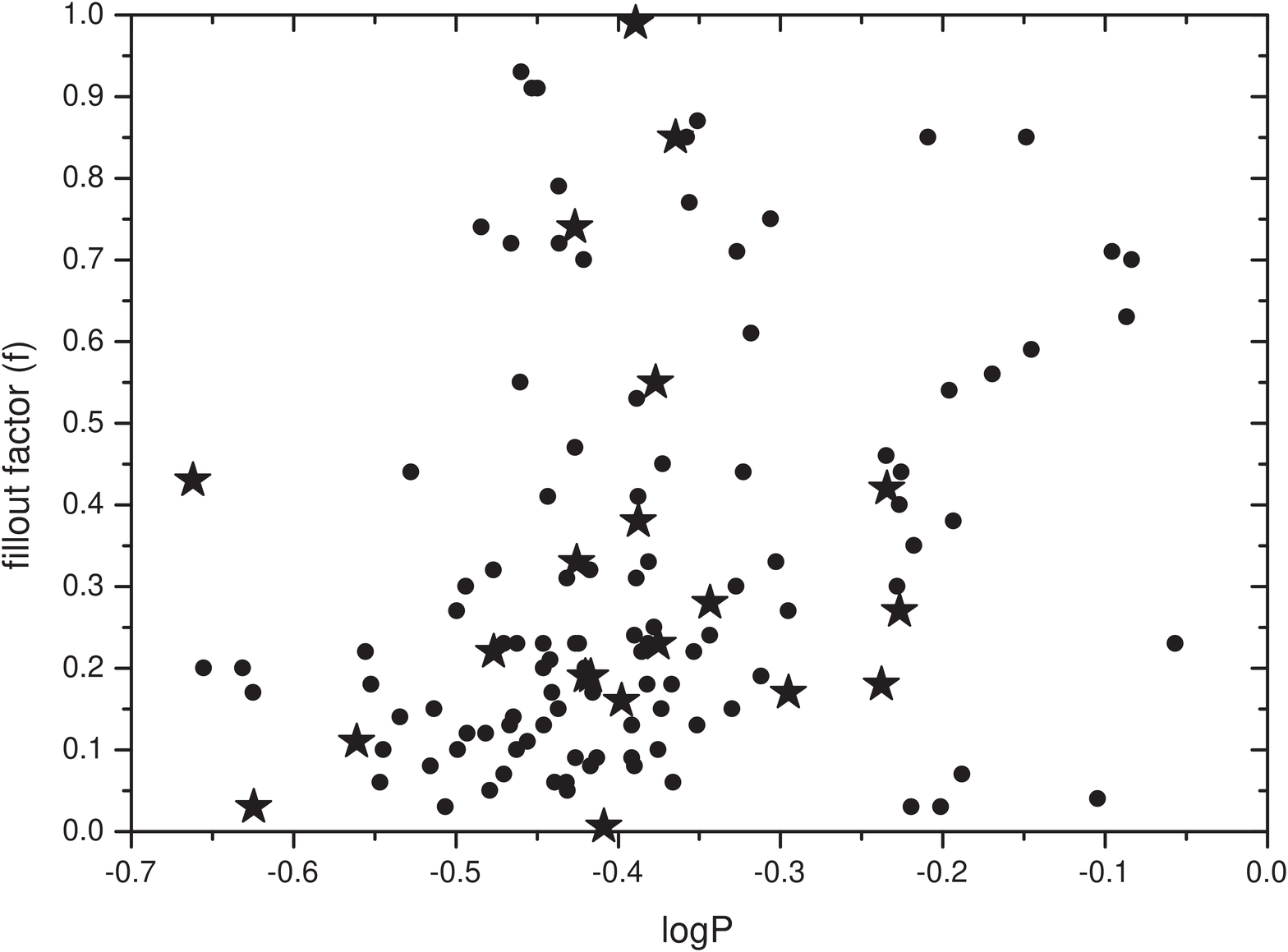}
\includegraphics[width=8.0cm,height=6.1cm,scale=1.0,angle=0]{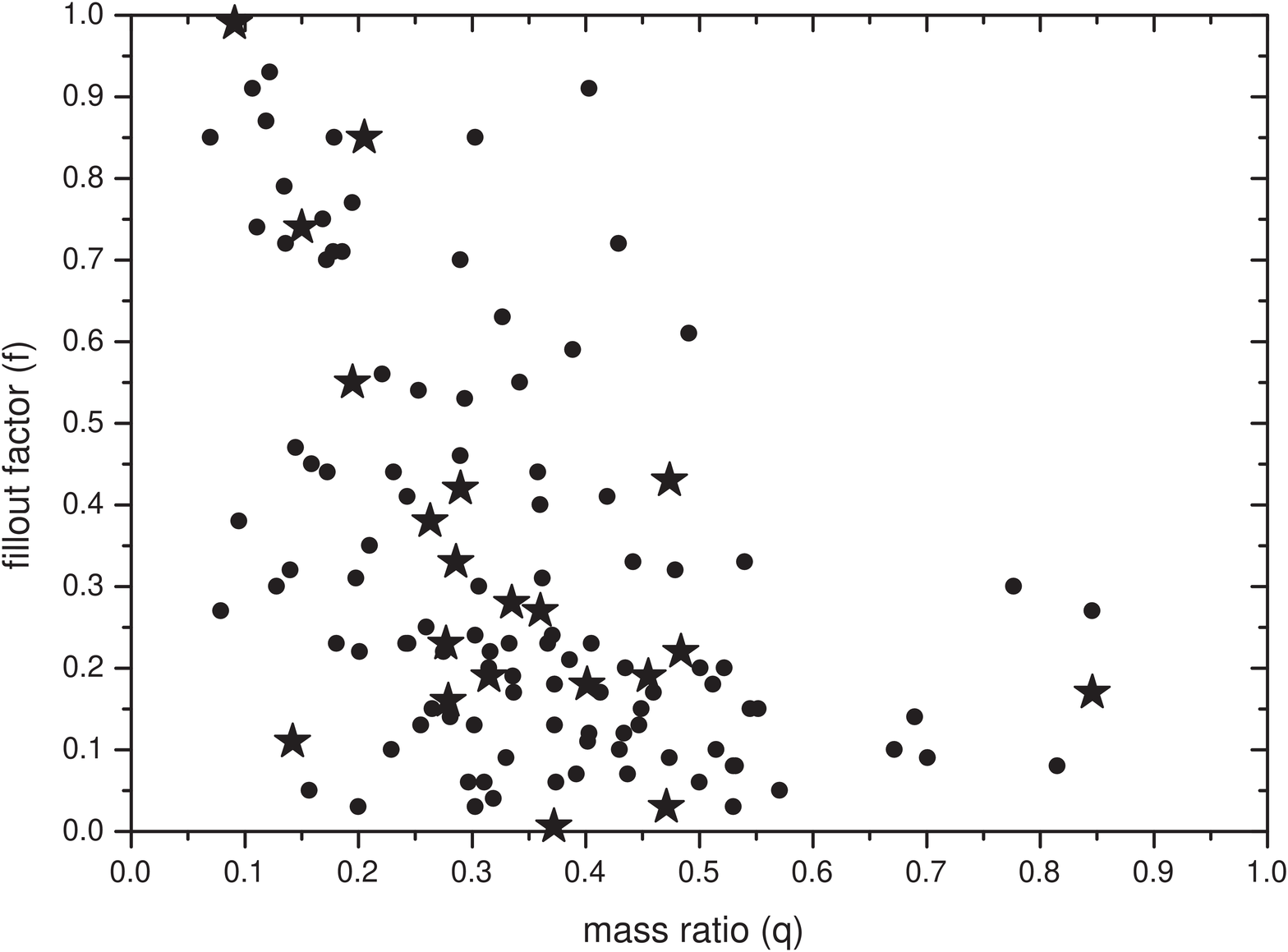}
 \caption{Correlation diagrams between orbital period and fill-out factor (left), and between mass
ratio and fill-out factor (right), for the sample of 51 systems observed in the frame of {W~UMa Project}
and 67 systems collected from literature (circles), together with the current 20 systems (stars).}
\label{FigCD4}
\end{figure*}

The mass, radius, luminosity and temperature of the components of the 20 additional close binary systems analysed in this study seem to follow the trends of the complete sample of
contact binaries with well defined parameters (Fig. \ref{FigCD3} and Fig. \ref{FigCD4}).
It is shown that contact binaries appear to be oversized for their ZAMS
and TAMS masses as compared with single MS stars from stellar evolution models.

The primary components in such systems more closely follow the mass-radius relation of MS stars \citep{gaz2008},
while the secondary components are severely oversized due to the energy transfer cause by the proximity of the primary.
This can lead to the conclusion that contact binary systems are old objects with components
in an advanced evolutionary state due to continuous mass and angular momentum loss. According to \citep{ste2006}, the secondary components were initially more massive and they evolved much faster. Mass was transferred to the smaller (at that time) component, which is now more massive than before. The final configuration results in young (MS) primary components, while the secondary ones are less massive, more evolved and possibly hydrogen depleted with a small helium core.

A similar study by \citet{yan2015}, based on 46 binary systems in contact configuration included
several systems from the W~UMa Project \citep{kre2003}.
However, in that work the authors also included systems whose physical parameters
were derived only from modelling of photometric light curves.

Contact binary systems with low mass and low temperature (LMCB and LTCB, respectively) were the main
sample in the study presented by \citet{ste2012}). These systems exhibit evolved secondaries and very short
orbital period ($<0.25$ d), often reaching the lower observed threshold of $\sim$0.22 days {\citep{ruc2007a}}.
These are some of the most rare contact binary systems, and only a handful of such systems are known to date.
Their components follow the MS trend and they are located within the ZAMS and TAMS limits.
Such systems, e.g. OT~Cnc, CC~Com, GSC~3807-0759, V523~Cas, RW~Com,
exhibit very short orbital periods, and three of them,
namely V345~Gem (0.274~d), RW~Com (0.237~d) and OT~Cnc (0.218~d), were studied in this work.
The current study adds three more objects to the same region of the plots with $P_{orb}<0.3$ d, doubling the sample.

Low temperature CBs are often characterized by strong magnetic activity, frequently
detected by ground-based and orbital observatories. Soft X-ray emission is common for systems
with components of G and K spectral type, which is a result of magnetic activity in the stellar convection zone.
Such activity is expressed in spotted surfaces, frequent flare events and
subsequent light curve asymmetries and shape variability.
An indication of soft X-ray emission and/or light curve asymmetries, which are common on low temperature
(G, K, M sp. Type), components can be used as an additional pointer towards the spotted solutions given in binary models.
The inclusion of one or more spots in the models would generally result in better fits.
However, such models are characterized by non-uniqueness, which is a well known issue in light curve modelling \citep{mac1993, eke1999, gaz2019}.

Computations of the solutions for each system in the current study was achieved using the Monte Carlo code,
which provides the best fit within a range of adjusted parameters.
This procedure, in the case of spotted solutions, may result in spots appearing unrealistically large or cool,
but this could be overcame by replacing them with several smaller spots in the same area.
The single spots allow the models to be kept as simple as possible, and indicate the presence of magnetic
activity of the stellar surface. They are added in order to explain the observed asymmetries,
without getting deeper into details which would require additional model parameters.


Systems with low mass primaries ($M_{1} < 0.7 M_{\sun}$) such as RW~Dor, FI~Boo, VZ~Psc, show more
evolved components, evidenced mostly by the primary components which are located
within the MS region above the ZAMS, in contrast to the systems with high mass primaries ($M_{1} > 2.3 M_{\sun}$)
such as HV~UMa, V376~And, FN~Cam which occupy the lower part of the MS, with primaries still being closer
to the ZAMS. The same occurs for systems with high temperature primaries ($T_{1} > 8000$ K) such as
V376~And (also with a high-mass primary), V535~Ara and V2150~Cyg.

Systems with low mass ratio ($q < 0.1$) always puzzle those who deal with theoretical models. Systems like
SX~Crv, V870~Ara, KR~Com, FP~Boo, XX~Sex host extremely small secondary components compared to the
primary ones. The primaries are somewhat evolved MS stars, located well above the ZAMS,
while the secondaries are much more evolved.

Systems with high mass ratio ($q > 0.8$) such as FT~UMa, OO~Aql, and HT~Vir do not show peculiarities.
They consist of two similar components, with both primaries and secondaries located within the MS.

Systems with low fill-out factors ($f < 5$~per cent) such as TW~Cet, V1128~Tau,
RW~Com, V402~Aur host rather young components, since  the secondaries are located very
close to the TAMS (in contrast to the overall sample, which has evolved secondaries)
and primaries close to the ZAMS. FI~Boo ($f = 0.6$~per cent) might be an exception,
and it seems to have evolved both components. The same applies to TT~Cet ($f = 8$~per cent) as
found by \citet{deb2011}. Both systems have very small fill-out factors, being marginally in contact
or near-contact (NCB) configuration.
TT~Cet is more likely a near-contact binary and hence its evolution is different from the rest
of the sample of contact binaries.
On the other hand, systems with high fill-out factor show the opposite. KR~Com has a very high
fill-out factor ($f = 99$~per cent), while UZ~Leo, GR~Vir, and CK~Boo follow with $f > 90$~per cent.
Being in such a deep contact, the systems may be rather evolved. The members of this group will probably
evolve quickly towards coalescence.

In Fig. \ref{FigCD3} (left), there is an interesting trend seen in the fill-out factor parameter.
It appears that systems with shorter orbital period tend to have shallow contact
between their components.

The majority of systems with orbital periods less than 0.3~d are indeed in shallow contact with fill-out factors less than 25~per~cent. This finding could plausibly be explained by assuming the systems to be evolutionary young.
This is also in accordance with the research by \citet{li2019}, which was conducted on contact binary systems at the short orbital period limit.

This effect partially explains the very low mass transfer and mass loss rates observed
in such systems, as predicted by \citet{ste2012}, assuming that these systems have just entered
the contact configuration.
In addition, systems with low mass ratio ($q < 0.15$) tend to have larger fill-out factors
(Fig. \ref{FigCD4}, right). These systems are in a deep contact configuration, where the
significantly larger star ``absorbs'' the tiny companion.
This concerns two systems, analysed in this paper: HV~Aqr ($q = 0.140$), and KR~Com ($q = 0.093$).
Both systems host a very small in size component, which most likely will be "swallowed" gradually
by the more massive star.

Low mass-ratio and short orbital period systems are worth being studied more thoroughly,
also in terms of the O$-$C diagrams, in order to see any orbital period variability which
could eventually lead to their merger into single fast-rotating stars. The evolution of such
systems appears to be slow \citep{gaz2008, ste2012} and this explains why these systems
still exist as MS stars after several Gyr of evolution.

It is known \citep{DAngelo2006} that more than 30~per~cent of contact binary systems belong to triple systems. It is therefore worth monitoring all systems over the long-term (several decades), not only to observe any intrinsic period changes, but also to indicate whether or not they belong to multiple stellar systems.

The current study also confirms that there is no fundamental difference between A- and W-type contact
binary systems. Both types smoothly blend with each other in the plots, as was shown by \citet{gaz2006b}.

It is widely considered that contact binary systems begin life as the components of close detached binaries and gradually approach each other, forming contact binaries. They are following evolutionary paths in a similar way but under slightly different circumstances, leading to the appearance of both the A-type and W-type. The W-type contact binary systems mainly occupy the low-temperature and low-mass regime in contrast to A-type. Systems with intermediate masses can be either A-type or W-type or, sometimes, both in turn, as it happens for example with V839~Oph \citep{gaz2006c}.
This trend is assumed to be a result of energy transfer, or that the A-type are in a more preliminary state of evolution than W-type ones \citep{gaz2008, gaz2006b}. However we cannot exclude other possible evolutionary pathways as there is a plethora of theoretical evolution models \citep{yak2005} which are trying to best describe the observational data, including all the mechanisms which could take place within the binary. Nonetheless the literature has not yet provided a clear answer on the evolutionary scenarios, making the continued research on contact binary systems essential.
Stellar evolution studies showed that a binary star evolves significantly differently from a single star if the two components are not too far apart. The evolution is controlled by loss of angular momentum, nuclear evolution and by mass loss and mass transfer between the components (e.g. \cite{yak2005}, \cite{ste2006}). All mechanisms run in parallel and gradually bring the components closer to each other. The question of whether there is an evolutionary sequence among different types of contact binary systems is therefore still open.


\section*{Acknowledgments}

This work was partially supported by the NCN grant No. 2016/23/N/ST9/01218.
The authors wish to thank J.~Papamichael J. and S.~Varrias
for their contribution in observations of the systems AK~Her and V566~Oph, respectively.
DM acknowledges partial support by Shumen University Science Fund.
The authors wish to thank the anonymous referee for the valuable comments that improved the manuscript.


\section*{Data Availability}

The data underlying this article are available upon request to the corresponding author.

\bibliography{biblio}
\bibliographystyle{mnras}

\bsp
\label{lastpage}

\end{document}